\documentclass[lettersize,journal]{IEEEtran}
\usepackage{amsmath,amsfonts}
\usepackage{algorithmic}
\usepackage{algorithm}
\usepackage{array}
\usepackage[caption=false,font=normalsize,labelfont=sf,textfont=sf]{subfig}
\usepackage{textcomp}
\usepackage{stfloats}
\usepackage{url}
\usepackage{verbatim}
\usepackage{graphicx}
\usepackage{cite}
\usepackage{lipsum}
\usepackage{tikz,xcolor,hyperref}
\usepackage{orcidlink}
\usepackage{cleveref}
\usepackage{acronym}
\usepackage{xcolor}
\usepackage{bbding}
\usepackage{pifont}
\usepackage{todonotes}
\usepackage{fixme}
\usepackage{enumitem}
\usepackage{adjustbox}
\usepackage{amssymb}  
\usepackage{pifont}
\usepackage{colortbl} 
\usepackage{subcaption}  
\usepackage{float}       
\usepackage{mathtools} 
\usepackage[most]{tcolorbox}
\usepackage[normalem]{ulem}  

\usepackage{color}
\usepackage{multirow}
\usepackage{multicol}

\newcommand{\MC}[1]{\begin{tabular}[c]{@{}c@{}}#1\end{tabular}}

\newcommand{\subsubsubsection}[1]{\paragraph{#1}\mbox{}\\}
\newcommand{\subsubsubsubsection}[1]{\paragraph{#1}\mbox{}\\}

\newcommand{\orcidauthorMA}{0000-0002-1798-7003} 
\newcommand{\orcidauthorWJ}{0000-0002-8807-0673} 
\newcommand{\orcidauthorDP}{0000-0002-8879-5076} 

\usepackage{tikz}
\newcommand*\emptycirc[1][1ex]{\tikz\draw (0,0) circle (#1);} 
\newcommand*\halfcirc[1][1ex]{%
  \begin{tikzpicture}
  \draw[fill] (0,0)-- (90:#1) arc (90:270:#1) -- cycle ;
  \draw (0,0) circle (#1);
  \end{tikzpicture}}
\newcommand*\fullcirc[1][1ex]{\tikz\fill (0,0) circle (#1);} 

\begin{document}
\title{Unmodulated Visible Light Positioning: A Deep Dive into Techniques, Studies, and Future Prospects}
\author{Morteza~Alijani~\orcidlink{\orcidauthorMA},
		Wout~Joseph~\orcidlink{\orcidauthorWJ},~\IEEEmembership{Senior Member,~IEEE},
        and~David~Plets~\orcidlink{\orcidauthorDP},~\IEEEmembership{Member,~IEEE}%
\thanks{\uline{This work has been submitted to the IEEE for possible publication. Copyright may be transferred without notice, after which this version may no longer be accessible.} This work was supported by the imec ICON Project “PULSAR” (Flanders Innovation \& Entrepreneurship) through the Agentschap Innoveren en Ondernemen Project under Grant HBC.2023.0606. (\emph{Corresponding author:} Morteza.Alijani@UGent.be).}
\thanks{The authors are affiliated with the research group imec-WAVES, part of the Department of Information Technology (INTEC), Ghent University, 9052 Ghent, Belgium (e-mail: Morteza.Alijani@UGent.be).}}
%

\markboth{JOURNAL OF LATEX CLASS FILES, VOL. xx, NO. xx, July 2025}%
{Shell \MakeLowercase{\textit{et al.}}: A Sample Article Using IEEEtran.cls for IEEE Journals}

\maketitle
\begin{abstract}
Visible Light Positioning (VLP) has emerged as a promising technology for next-generation indoor positioning systems (IPS), particularly within the scope of sixth-generation (6G) wireless networks. Its attractiveness stems from leveraging existing lighting infrastructures equipped with light-emitting diodes (LEDs), enabling cost-efficient deployments and achieving high-precision positioning accuracy in the centimeter-to-decimeter range. However, widespread adoption of traditional VLP solutions faces significant barriers due to the increased costs and operational complexity associated with modulating LEDs, which consequently reduces illumination efficiency by lowering their radiant flux. To address these limitations, recent research has introduced the concept of unmodulated Visible Light Positioning (uVLP), which exploits Light Signals of Opportunity (LSOOP) emitted by unmodulated illumination sources such as conventional LEDs. This paradigm offers a cost-effective, low-infrastructure alternative for indoor positioning by eliminating the need for modulation hardware and maintaining lighting efficiency. This paper delineates the fundamental principles of uVLP, provides a comparative analysis of uVLP versus conventional VLP methods, and classifies existing uVLP techniques according to receiver technologies into intensity-based methods (e.g., photodiodes, solar cells, etc.) and imaging-based methods. Additionally, we propose a comprehensive taxonomy categorizing techniques into demultiplexed and undemultiplexed approaches. Within this structured framework, we critically review current advancements in uVLP, discuss prevailing challenges, and outline promising research directions essential for developing robust, scalable, and widely deployable uVLP solutions.

\end{abstract}


\section{Introduction}
\label{sec:Introduction}
\subsection{Background and Motivation}
\label{subsec:BackgroundandMotivation}

\IEEEPARstart{I}{ndoor} localization and navigation have been extensively researched in the last decade, within the context of a growing \ac{IoT} and the plethora of potential use cases in a wide range of sectors (e.g., industry, healthcare, retail, etc)\cite{Zafari2019,Farahsari2022,Yassin2017}. Notable and popular enabling technologies are \ac{BLE}, \ac{RFID}, \ac{UWB}, \ac{mmWAVE}, or \ac{LiDAR}\cite{Zafari2019,Farahsari2022,Yassin2017}. However, no single dominant solution like outdoor \ac{GPS} has yet come forward for indoor environments\cite{Zafari2019,Farahsari2022,Yassin2017}. As such, one has to deal with the ever-recurring trade-off between the desired accuracy and the total system cost (e.g., hardware, installation, maintenance, etc), making an optimal solution very use-case specific. \ac{VLP}, first coined in 2004\cite{Horikawa2014}, has recently gained research attention due to its low cost and high precision (centimeter-to-decimeter level accuracy) \cite{Hassan2015, Do2016, Luo2017, Zhuang2018, Pathak2015, Wang2024VLP, Afzalan2019, Kouhini2021, Mapunda2020, Bastiaens2023AnEA, Maheepala2020, Bastiaens2024}. It capitalizes on existing lighting infrastructure by employing \acp{LED} as positioning beacons, does not require \ac{LED} synchronization, is less affected by reflections, provides room-level certainty due to the inability of light to penetrate walls, avoids using the scarce \ac{RF} spectrum, and scales efficiently since light functions as a broadcast signal~\cite{Hassan2015, Do2016, Luo2017, Zhuang2018, Pathak2015, Wang2024VLP, Afzalan2019, Kouhini2021, Mapunda2020, Bastiaens2023AnEA, Maheepala2020, Bastiaens2024, Bastiaens2022}. Recent advancements in \ac{VLP} research have successfully demonstrated sub-decimeter accuracy, further underscoring its potential for precise indoor localization applications~\cite{Du2019, Bastiaens2022}.

However, despite claims of leveraging existing lighting infrastructure, \ac{VLP} often necessitates additional \ac{LED} drivers for modulation (e.g., \ac{OOK}, \ac{PWM}, etc.), to distinguish individual \ac{LED}'s contributions at the receiver\cite{Bastiaens2024}. A primary challenge in \ac{VLP} with modulated \acp{LED} is that \ac{LED} modulation reduces illumination levels by lowering the \ac{LED}’s radiant flux\cite{Bastiaens2020}. This reduction usually requires deploying additional \acp{LED} to maintain sufficient illumination, significantly increasing deployment costs\cite{Bastiaens2020, Bastiaens2022, Bastiaens2024}. Given the large number of light fixtures in indoor environments, this cost factor becomes even more critical\cite{Bastiaens2020,Bastiaens2022}. For example, in the \ac{FDMA} scheme proposed by De Lausnay et al. \cite{DeLausnay2015}, the radiant flux is effectively reduced by half. Similarly, polarization-based \ac{VLP} reduces radiant flux by at least 50\% due to the use of linearly polarized light \cite{Bastiaens2020}. To mitigate modulation loss, three approaches are considered within the framework of \ac{FDMA} using amplitude- and \ac{PWM}-compatible light sources: \textit{(i)} reducing the amplitude swing of $I_{LED}/P_k(t)$ (where $P_k(t)$ represents the transmitted power of \ac{LED}$_k$, and $I_{LED}$ denotes the driving current), \textit{(ii)} increasing the duty cycle to improve radiant flux, and \textit{(iii)} introducing non-interfering signal components, known as 'return-to-one coding.' These methods have been explored in \cite{Berton2020}. However, the first approach lowers the \ac{SNR}, thereby affecting positioning range and accuracy \cite{Zhang2014,BastiaensPhdBook}. The second approach modifies the waveform, leading to transmitter non-idealities \cite{Zhang2014,BastiaensPhdBook,Bastiaens2018}. Finally, although the third approach enhances the average radiant flux by superimposing a square wave, it also weakens key frequency components, ultimately degrading system performance \cite{Zhang2014,BastiaensPhdBook,Berton2020}. To address these challenges, \ac{uVLP} has emerged as a cost-effective alternative~\cite{Bastiaens2020,Bastiaens2022,Xu2015}. It leverages unmodulated or unmodified light sources (i.e., \acp{LED} used in their original form, without any additional control circuitry such as an \ac{FPGA}\cite{Liu2019}, synchronization mechanisms like shutter timing\cite{GPinto2012}, or dedicated modulation hardware~\cite{Bastiaens2023AnEA,Bastiaens2024}) by postprocessing either total light intensity (irradiance) measurements or using imaging techniques (see Section~\ref{sec:uVLPtechniques})~\cite{Amsters2019,Amsters2018,RaviFiatLuxF,Faulkner2020,Konings2020,Faulkner2019,Bastiaens2022,Bastiaens2020,Singh2023,Singh2022,Bastiaens2024}.

\begin{table*}[]  
  \centering
  \caption{A Comparative Analysis of Our Article on \ac{uVLP} with Existing Related Studies.}
  \small
    \hspace*{-0.20cm} 
    \setlength{\tabcolsep}{4.0pt} 
    \renewcommand{\arraystretch}{0.9} 
    \begin{tabular}{c c c c p{1.1cm}}
      \hline 
      \textbf{Reference}
      & \MC{\textbf{Fundamental of \ac{uVLP}}\\(Tx-Rx, \ac{VLP} vs. \ac{uVLP}, etc.)} 
      & \MC{\textbf{\ac{uVLP} Positioning Techniques}\\ (Demultiplexed \& Undemultiplexed)}
      & \MC{\textbf{\ac{uVLP} Investigations}\\ (Intensity- \& Image-based)}
      & \textbf{Year} \\
      \hline  \noalign{\vspace{0.1cm}} 
      Armstrong et al. \cite{Armstrong2013} & \halfcirc	 & \emptycirc &  \emptycirc &  2013\\
      Hassan et al. \cite{Hassan2015} & \halfcirc	 & \emptycirc &  \emptycirc &  2015\\
      Pathak et al. \cite{Pathak2015} & \halfcirc  & \emptycirc &  \emptycirc &  2015\\
      Do et al. \cite{Do2016} &  \halfcirc & \emptycirc & \emptycirc &  2016\\
      Luo et al. \cite{Luo2017} &  \halfcirc & \emptycirc & \emptycirc  &  2017\\
      Jiao et al. \cite{Jiao2017} &  \halfcirc & \emptycirc & \emptycirc  &  2017\\
      Zhuang et al. \cite{Zhuang2018} &  \halfcirc  & \emptycirc & \emptycirc &  2018\\
      Afzalan et al. \cite{Afzalan2019} & \halfcirc  & \emptycirc & \emptycirc  &  2019\\
      Chen et al. \cite{Chen2019} & \emptycirc  & \emptycirc  &  \emptycirc &  2019\\
      Rahman et al. \cite{Rahman2020} & \halfcirc  & \emptycirc  &  \halfcirc &  2020\\
      Maheepala et al. \cite{Maheepala2020} &  \halfcirc  & \emptycirc  & \emptycirc &  2020\\
      Liu et al. \cite{Liu2021} &  \emptycirc & \emptycirc & \halfcirc  &  2021\\
      Kouhini et al. \cite{Kouhini2021} &  \halfcirc  &  \emptycirc &   \emptycirc &  2021\\
      Tran et al. \cite{Tran2022}&  \halfcirc  &  \emptycirc &   \halfcirc &  2022\\
      Zhang et al. \cite{ZHANG2023}&  \halfcirc  &  \emptycirc &   \emptycirc &  2023\\
      Wang et al. \cite{Wang2024VLP} & \halfcirc  &  \emptycirc & \emptycirc  &  2024\\
      Bastiaens et al. \cite{Bastiaens2024} & \halfcirc  & \emptycirc & \emptycirc  &  2024\\
      Zhu et al. \cite{Zhu2024} & \halfcirc   & \emptycirc & \emptycirc  &  2025\\
      Alijani et al. (This article) &  \fullcirc & \fullcirc & \fullcirc  &  2025\\
      \hline
    \end{tabular}
    \label{tbl:SurveyComparison}
    \vspace{-0.1cm} 
    \begin{center}
    \footnotesize{The symbol $\fullcirc$ indicates that the topic is included, $\halfcirc$ denotes that it is partially included, and $\emptycirc$ signifies that it is not included.}
    \end{center}
\end{table*}

\subsection{Key Insights and Contributions}
\label{subsec:PaperContributions}
There is a rich body of literature on \ac{VLP}\cite{Jiao2017, Armstrong2013, Zhu2024, Hassan2015, Do2016, Luo2017, Zhuang2018, Pathak2015, Wang2024VLP, Afzalan2019, Kouhini2021, Maheepala2020,Alijani2025,ZHANG2023}, while research on \ac{uVLP} remains niche and relatively unexplored \cite{Bastiaens2020, Bastiaens2024}. Table~\ref{tbl:SurveyComparison} lists many comprehensive review papers on \ac{VLP}, but most either overlook \ac{uVLP} or focus only on \ac{VLP}. For example, \cite{Jiao2017, Armstrong2013, Zhu2024, Hassan2015, Do2016, Luo2017, Zhuang2018, Pathak2015, Wang2024VLP, Afzalan2019, Kouhini2021, Maheepala2020,ZHANG2023} provide in-depth analyses of \ac{VLP} but do not address \ac{uVLP}. Similarly, \cite{Bastiaens2024}, an extensive technical tutorial on \ac{VLP}, dedicates only a brief section to \ac{uVLP}. Studies \cite{Chen2019, Tran2022, Rahman2020} explore positioning algorithms, distinguishing between approaches that rely on modified versus unmodified light sources. However, these studies do not provide an in-depth examination of specific \ac{uVLP} implementation techniques (e.g., demultiplexed and undemultiplexed). Similarly, \cite{Liu2021} briefly reviews \ac{uVLP} studies but lacks a detailed analysis of \ac{uVLP} methods.

Unlike conventional \ac{VLP}, \ac{uVLP} eliminates the need for modulation, thereby simplifying deployment, reducing costs, and preserving illumination functionality \cite{Bastiaens2022,Bastiaens2020,BastiaensPhdBook}. These advantages make \ac{uVLP} a promising candidate for \ac{IPS} within the scope of \ac{6G}, where infrastructure-free, low-complexity, and energy-efficient technologies are highly desirable. This potential has been experimentally benchmarked in \cite{Bastiaens2022} (see Section~\ref{subsec:comparingvlpanduvlp} for additional advantages over conventional \ac{VLP}). Despite its potential, a comprehensive exploration of \ac{uVLP} techniques and an updated evaluation of existing studies remain lacking. To bridge this gap, our work provides a one-stop resource and structured overview of \ac{uVLP}, classifying its positioning techniques (i.e., demultiplexed and undemultiplexed) based on receiver types, including light intensity-based receivers (\ac{PD}, \acp{ALS}, solar cells, and spectral sensors) and imaging-based receivers (\ac{CMOS} and \ac{CCD}), while addressing practical deployment considerations. Furthermore, we trace the evolution of \ac{uVLP} from its early foundations (e.g., 1995~\cite{Facchinetti1995SelfPositioningRN}) to recent advancements, offering a thorough analysis of current methodologies and identifying future research directions.

\subsection{Structure of the Paper}
\label{subsec:Paperorganisation}
\begin{figure*}[]
  \centering
  \includegraphics[width=0.82\textwidth]{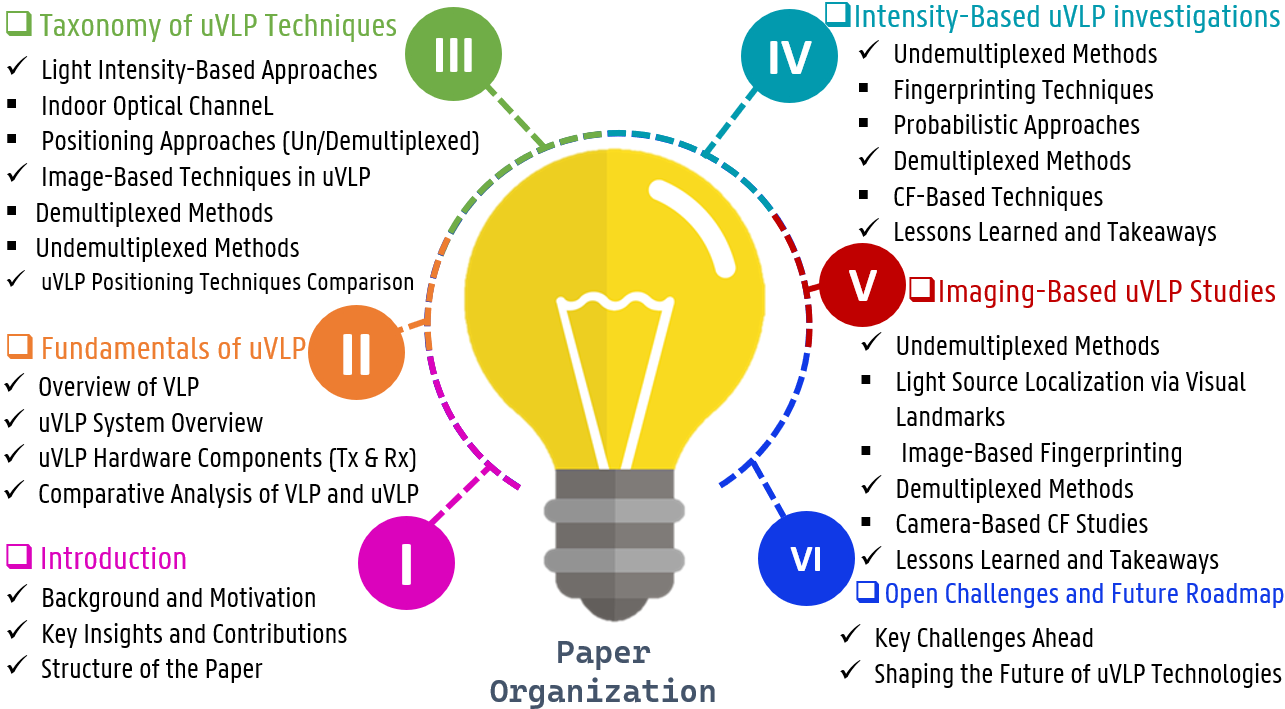} 
  \caption{The overall structure of the article.}
  \label{fig:SurveyStructure}
\end{figure*}

The remainder of this article is structured as follows (see Fig.~\ref{fig:SurveyStructure}). Section~\ref{sec:FundamentalsofuVLP} introduces the fundamentals of \ac{uVLP}, comparing it with conventional \ac{VLP} systems and discussing implementation considerations, hardware components, and underlying physical principles. Section~\ref{sec:uVLPtechniques} classifies \ac{uVLP} techniques based on receiver type into intensity/spectrum-based and imaging-based methods, further distinguishing between demultiplexed and undemultiplexed approaches. Sections~\ref{sec:Intensity-BaseduVLPTechniques} and~\ref{sec:imagbaseduVLP} respectively review intensity-based studies (e.g., fingerprinting, probabilistic, and \ac{CF}-based methods using \acp{PD}, solar cells, \acp{ALS}, and spectral sensors) and image-based investigations using smartphone cameras. Section~\ref{Sec:FutureDirections} highlights the key challenges and outlines potential directions for future research. Finally, Section~\ref{Sec:conclusion} presents the concluding remarks of the paper.

\section{Fundamentals of \ac{uVLP}}
\label{sec:FundamentalsofuVLP}
This section provides a brief overview of (u)VLP systems. For a more in-depth discussion, we refer the reader to \cite{Armstrong2013, Zhu2024, Hassan2015, Do2016, Luo2017, Zhuang2018, Pathak2015, Wang2024VLP, Afzalan2019, Kouhini2021, Maheepala2020, Alijani2025}. This paper specifically focuses on \ac{uVLP}, where (u)VLP denotes both \ac{VLP} and \ac{uVLP}.

\subsection{Overview of \ac{VLP}}
\label{subsec:OverviewofVLP}
\begin{figure*}[]
    \centering
    \includegraphics[width=0.85\textwidth]{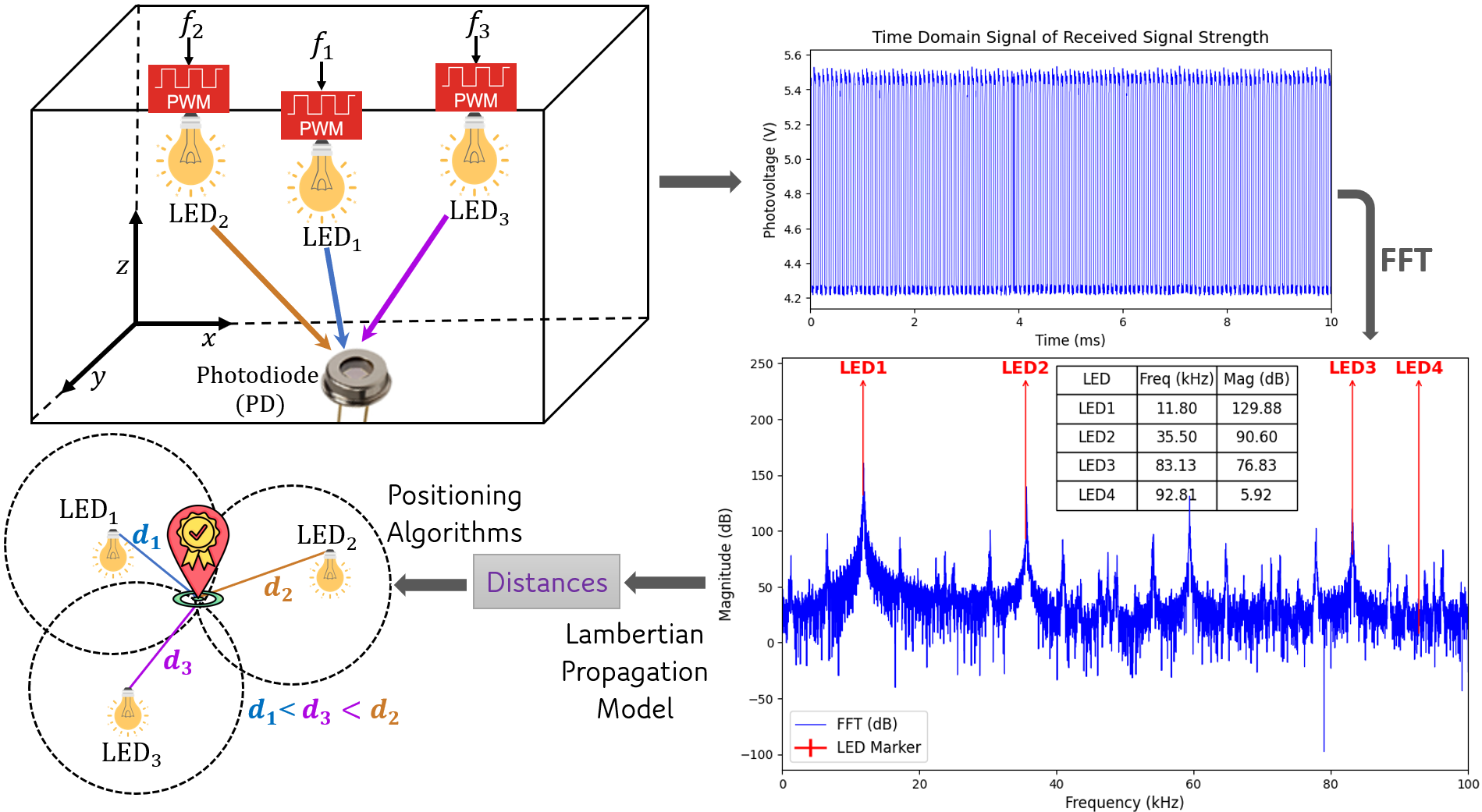} 
    \caption{Typical \ac{RSS}-Based \ac{VLP} Workflow: A photodetector receives light signals from multiple \acp{LED} with known locations, each modulated by an \ac{LED} driver (e.g., \ac{PWM}) and installed on the ceiling. A demultiplexing algorithm (e.g., \ac{FFT}\cite{oppenheim1999discrete,DeLausnay2015}) extracts the received power/\ac{RSS} (i.e., magnitudes) corresponding to each \ac{LED} based on their pre-assigned modulated frequencies (e.g., $f_1$, $f_2$, etc.). These magnitudes (i.e., \ac{RSS} values) are then converted into distances between the \ac{PD} and the corresponding \ac{LED} lamp using the Lambertian propagation model, assuming the \ac{LED} follows Lambertian propagation. Finally, a localization algorithm (e.g., trilateration\cite{BarcoAlvrez2024,Plets2019}) processes the extracted distances to estimate the position of the receiver.}
    \label{fig:VLP}
    \end{figure*}

With the increasing use of \acp{LED} for efficient lighting in indoor spaces (e.g., buildings, offices, etc.) and the emergence of \ac{VLC} technology, these \acp{LED} can piggyback on existing infrastructure to provide both lighting and communication services (e.g., sensing and positioning) at minimal cost\cite{Bastiaens2024,Pathak2015,Liang2022}. Typically, \acp{LED} are densely and uniformly placed on ceilings for consistent lighting\cite{Liang2022}. As solid-state devices, they can be modulated at high frequencies (e.g., above 200~Hz~\cite{Yang2025}) to transmit data via \ac{VLC}, while remaining safe for humans, causing no flicker or headaches, as the human eye cannot perceive frequencies above 60~Hz~\cite{Liang2022,Yang2025}. Although these high-frequency light variations are invisible to the human eye, they are detectable by \acp{PD} or cameras~\cite{Yang2025,Liang2022}. Modulated \acp{LED} at known, fixed locations broadcast unique identifiers, enabling precise \ac{2D}/\ac{3D} positioning through \ac{LoS} propagation~\cite{Liang2022}. This technique is widely recognized in the literature as \ac{VLP} (see Fig.~\ref{fig:VLP})~\cite{Liang2022,Bastiaens2024,Pathak2015,Zhuang2018}.

\ac{VLP} systems rely on various physical modalities (i.e., raw sensor data) such as \ac{RSS}\cite{Bastiaens2020,Bastiaens2023AnEA}, time of arrival (ToA) \cite{Wang2013}, time difference of arrival (TDoA) \cite{Jung2011}, phase difference of arrival (PDoA) \cite{Zhang2018,Xu2023}, and \ac{AoA} \cite{Steendam2018,Kuo2014}. These modalities do not directly yield position estimates but are processed by localization algorithms, which translate them into distance or angle measurements using channel propagation models (e.g., Lambertian model) \cite{EWLam2019, ZHANG2023, Armstrong2013, Bastiaens2024, Keskin20188, Jiao2017}. Common algorithms include trilateration, triangulation, fingerprinting, imaging, and proximity \cite{Bastiaens2024}. \ac{VLP} typically employs two types of receivers: cameras (also referred to as image sensors\cite{ElGamal2005,Yamazato2017,KAMAKURA2017}) and \acp{PD} (also referred to as photodetectors or light sensors)\cite{ZHANG2023}. Among these, \ac{RSS}-based \ac{VLP} using \ac{PD} receivers is the most extensively studied due to their simplicity, low cost, ultra-low power consumption ($\mu$W-range~\cite{Alijani2025}), and capability to support high modulation frequencies (from a few kHz to hundreds of MHz)~\cite{Bastiaens2021, Plets2017, Jin2025, Bastiaens2023AnEA, ZHANG2023, EWLam2019}. 

\subsection{\ac{uVLP} System Overview}
\label{subsec:uVLPConcept}
\begin{figure*}[]
    \centering
    \includegraphics[width=0.87\textwidth]{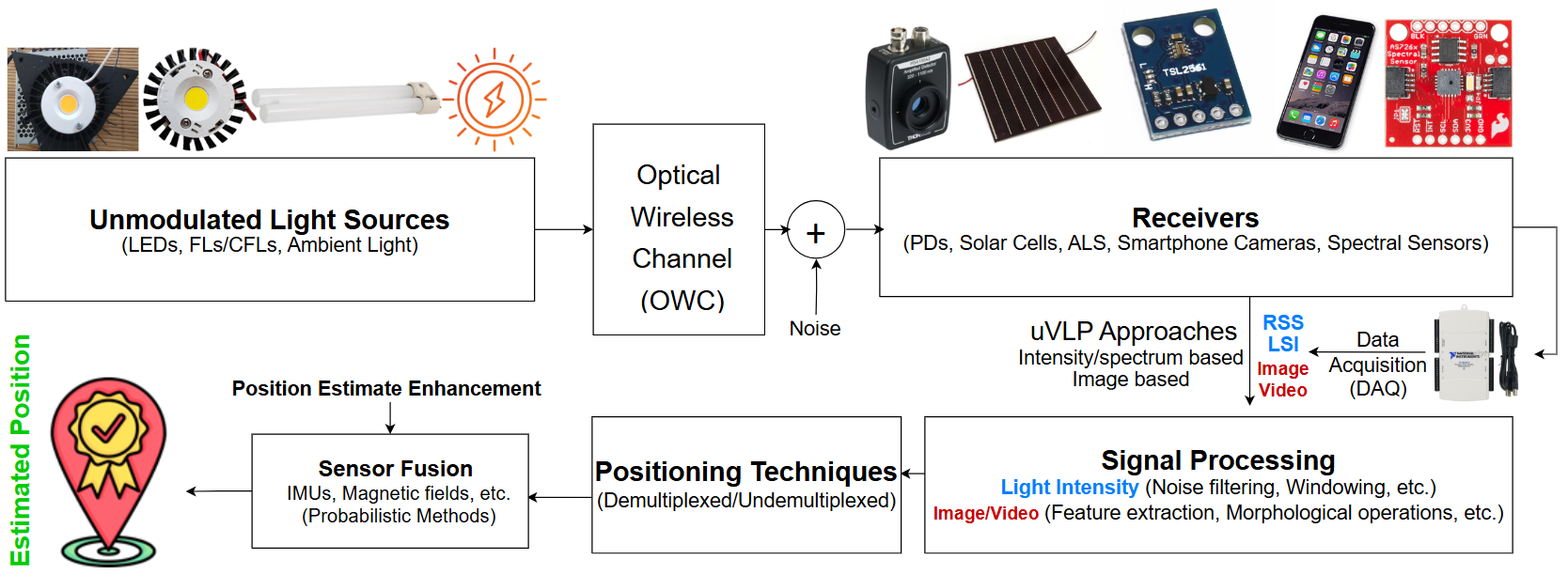}
    \caption{Illustration of the practical implementation of \ac{uVLP} systems, highlighting key hardware components. The diagram showcases various \ac{Tx} types, including \acp{LED} (e.g., COB \acp{LED}\cite{Bridgelux_BXRE-35E2000-C-73}), \acp{FL}/\acp{CFL}, and ambient light sources (e.g., sunlight, daylight). It also features \ac{Rx} components such as \ac{DAQ} system (e.g., USB-6212\cite{NI_USB6212_Specs}) for data acquisition, smartphone sensors (e.g., \ac{ALS}, \ac{IMU}, etc.), spectral sensors (e.g., AS7265x (Qwiic)\cite{SparkFun2022,ams_AS7265x_2018}), solar cells, and \ac{PD} (e.g., PDA100A2\cite{PDA100A2_Manual} or PDA36A2\cite{Thorlabs_PDA36A2}). Typically, captured signals/images undergo preprocessing before being fed into positioning algorithms, which can also be combined with data from other sensors (e.g., magnetic field, \ac{IMU}, etc.) to improve positioning accuracy.}
    \label{fig:uVLPPractical}
    \end{figure*}

The implementation of \ac{uVLP} has progressed from theoretical research to practical applications, with experimental validations conducted in relevant environments (e.g., office buildings)\cite{Zhang2016,Zhang2016LiTell2,Xu2015,Zhang2017,Bastiaens2022}. While many studies remain at the proof-of-concept stage, several investigations have demonstrated \ac{uVLP} in real-world settings, as detailed in Sections \ref{sec:Intensity-BaseduVLPTechniques} and \ref{sec:imagbaseduVLP}. Fig.~\ref{fig:uVLPPractical} depicts the practical implementation of an \ac{uVLP} system. The \ac{OWC} system inherently experiences environmental noise, and the transmitted signals are detected by light intensity receivers (e.g., \acp{PD}) interfaced with data acquisition (\ac{DAQ}) hardware, such as the USB-6212 device \cite{ni_usb6212}, or by imaging-based receivers. Depending on the receiver type, the captured optical signals can represent \ac{RSS}, \ac{LSI}, particularly in spectral-based (\ac{SB}) receivers \cite{Wang2023,Wang2022,Singh2022,Singh2023}, or image and video data from imaging-based receivers\cite{Zhang2016,Carver2017,Facchinetti1995SelfPositioningRN}. 

As illustrated in the signal processing block of Fig.~\ref{fig:uVLPPractical}, the acquired data, either from light \ac{IB} receivers such as \acp{PD}, \acp{ALS}, spectral sensors, or solar cells, undergo preprocessing steps including noise filtering, windowing, and other relevant operations. In the case of image or video data obtained from imaging sensors (e.g., \ac{CCD}, \ac{CMOS}), additional processing steps such as feature extraction and morphological operations (e.g., dilation and erosion) are applied~\cite{Bellini2002,Chen2014,Alves2013,launay2001fluorescent}. Subsequently, position estimation techniques such as demultiplexing or undemultiplexing can be employed (further detailed in Section~\ref{sec:uVLPtechniques}). Additional accuracy improvements can be achieved through sensor fusion with \acp{IMU}~\cite{Yang2025} or magnetometers combined with probabilistic methods~\cite{Wang20188,Wang2018DeepML,Amsters2018,Amsters2019} (see Section~\ref{sec:probabilistic}), which help compensate for environmental disturbances and enhance system robustness.

Research into \ac{uVLP} can be traced back over a decade~\cite{Facchinetti1995SelfPositioningRN}. Existing studies have predominantly investigated \ac{uVLP} using two primary methodologies: \ac{IB} receivers (e.g., \acp{PD}, solar cells, etc.) and imaging-based receivers, each with specific trade-offs concerning complexity, cost, latency, and implementation feasibility~\cite{Bastiaens2022,Bastiaens2024}. Detailed implementation aspects of these two approaches are presented in Section~\ref{sec:uVLPtechniques}, while a comprehensive review of the literature on \ac{IB}-\ac{uVLP} techniques is provided in Section~\ref{sec:Intensity-BaseduVLPTechniques}, and imaging-based methods are reviewed in Section~\ref{sec:imagbaseduVLP}.

\subsection{\ac{uVLP} Hardware Components (Transmitters \& Receivers)}
\label{subsec:uVLPHardware}
Fig.~\ref{fig:uVLPPractical} shows the general \ac{uVLP} hardware architecture. While there is well-established literature explaining how \acp{Tx} and \acp{Rx} operate \cite{Hassan2015, Do2016, Luo2017, Zhuang2018, Pathak2015, Wang2024VLP, Afzalan2019, Kouhini2021, Maheepala2020,Zhu2017,Zhang2016,Barwar2023,Zhang2016(2)}, this section highlights the key features relevant to their use in \ac{uVLP} systems.

\subsubsection{Transmitters (Light Sources)}
\begin{figure}[]
\centering
\includegraphics[width=0.50\textwidth]{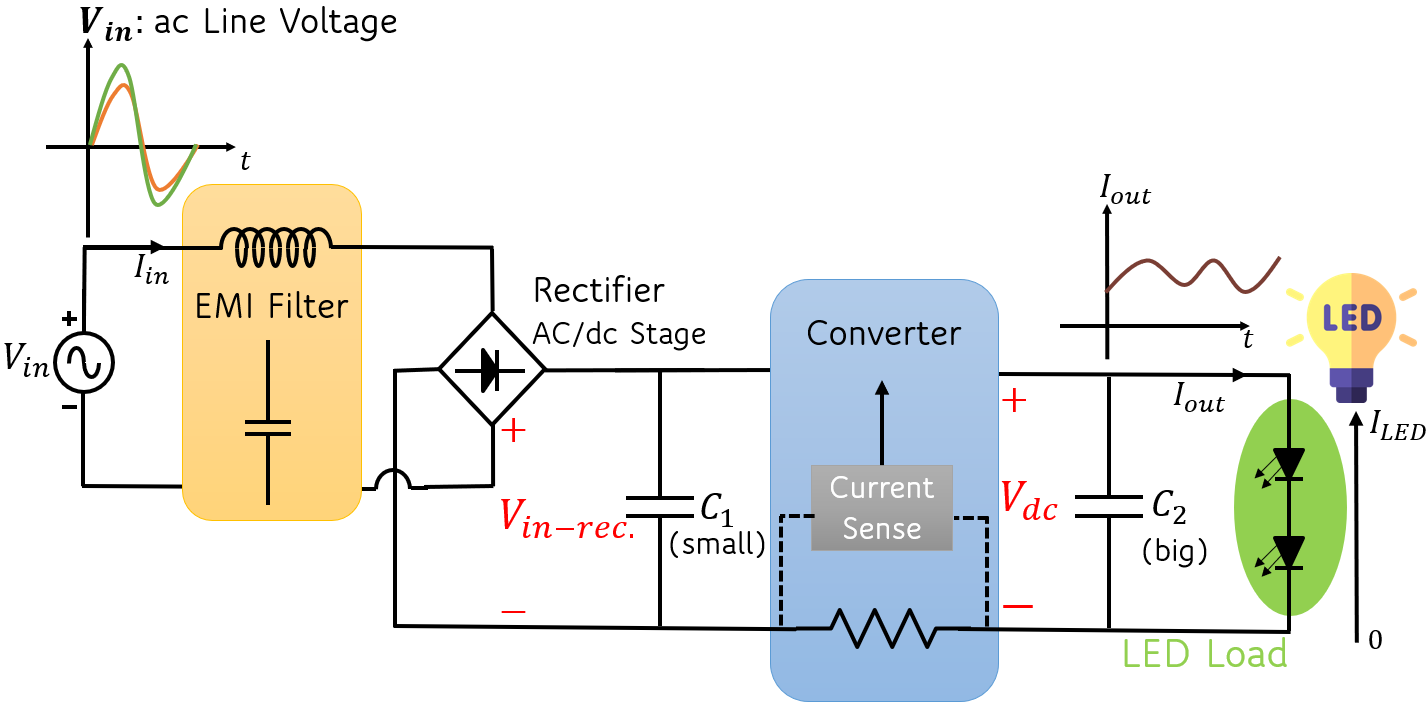} 
\caption{An illustration of a conventional single-stage \ac{LED} driver \cite{Munir2019,Barwar2023}.}
\label{fig:LEDWorkingPrinciple}
\end{figure}

Various visible light sources can serve as \acp{Tx} in \ac{uVLP} systems and are typically categorized as \acp{LED}, \acp{FL}/\acp{CFL}, and \acp{ICL} based on market availability~\cite{Wang2024, Wang2023}. Ambient sources such as sunlight or daylight can also be utilized \cite{BastiaensPhdBook,Bastiaens2024}. \acp{ICL} are being phased out due to energy inefficiency and have declined in use with the rise of \acp{LED}\cite{Lang2018}, which now dominate the lighting market—valued at USD 81.48 billion in 2023 and projected to grow at an 11.0\% CAGR through 2030 \cite{Pathak2015,statista_led_lighting,Bastiaens2024,Barwar2023,GrandViewResearch2025}. \acp{LED} are favored for their energy efficiency, long lifespan, and versatility, making them the primary choice for \ac{uVLP} systems \cite{Bastiaens2024,statista_led_lighting,Lang2018}. \acp{FL} remain in limited use \cite{Zhu2017,Zhang2016,Zhang2016(2)}. Accordingly, this section focuses on \acp{LED} and \acp{FL}/\acp{CFL}. Note that this study excludes infrared sources \cite{Yan2021}.
\subsubsubsection{Light-emitting diode (\ac{LED})}
\label{subsec:ledphysics}
The key takeaway is the underlying operating mechanism of \acp{LED}, which leads to their resonance frequency (i.e., \ac{CF}) and can be leveraged in \ac{CF}-\ac{uVLP} systems. As shown in Fig.~\ref{fig:LEDWorkingPrinciple}, \acp{LED} are unidirectional devices that operate on \ac{DC} power within a range of 3–3.6 volts \cite{Munir2019,Barwar2023}, while the primary power supply is typically \ac{AC} (110/220 volts, 50/60 Hz) \cite{Barwar2023}. This discrepancy necessitates the use of an internal power converter within the \acp{LED}. An electromagnetic interference (EMI) filter reduces input noise, the rectifier converts \ac{AC} to \ac{DC}, and the converter delivers a constant current for stable and efficient \ac{LED} operation. The diode bridge (i.e., the rectifier or \ac{AC}/\ac{DC} stage in Fig.~\ref{fig:LEDWorkingPrinciple}), being a nonlinear component, introduces harmonics and distortions due to its inherent characteristics such as forward voltage and resistance \cite{Munir2019}. These distortions can be further affected by duty cycling for brightness or temperature control \cite{Munir2019}. Although capacitors help to reduce voltage ripple, they do not eliminate it entirely \cite{Munir2019}. This process results in a unique resonance frequency (i.e., \ac{CF}) that can be used to distinguish individual \acp{LED} and quantify their contributions (e.g., amplitude) for positioning in \ac{uVLP} systems (see Section~\ref{sec:uVLPtechniques}) \cite{Zhu2017, Zhang2016, Barwar2023, Zhang2016(2)}.

\subsubsubsection{Compact Fluorescent Lights}
The fundamental principle behind \acp{FL}/\acp{CFL} for generating \ac{VL} is that they create an electric arc across a tube lamp using an electronic ballast, which excites the gas (e.g., mercury vapor) and fluorescent material inside, generating shortwave ultraviolet (UV) radiation. This UV radiation then strikes the phosphor coating, which fluoresces and emits \ac{VL}. In \acp{FL}/\acp{CFL}, the movement of electrons and mercury atoms within the vapor can lead to varying radiance levels and distinct spatial patterns along the tube. Moreover, non-uniformities in the phosphor coating and the glass wall contribute to differences in emission characteristics, even among \acp{FL}/\acp{CFL} of the same model. These factors result in diverse light outputs, influencing the overall performance and efficiency of the lighting system, and more importantly, generating different \acp{CF} through the \ac{AC} to \ac{DC} conversion by the \ac{FL}/\acp{CFL} driver\cite{Zhu2017,Zhang2016,Barwar2023,Zhang2016(2),Munir2019}. This unique feature and its application in \ac{uVLP} systems will be discussed in detail in Section~\ref{sec:uVLPtechniques}.

\subsubsubsection{Ambient Light}
In addition to primary light sources such as \acp{LED} and \acp{FL}, ambient light, including sunlight entering through windows or skylights, can also be utilized in \ac{uVLP}~\cite{Bastiaens2024,Ma2020,Liu2014}. Several \ac{uVLP} studies, particularly those focused on robotic applications, have employed ambient light as a source~\cite{Liu2014,Zhang2013,Sprute2017,Wang2018DeepML,Wu2022}.

\subsubsection{Receivers (Sensing Units)}
Similar to the \ac{Tx}, the \ac{Rx} in \ac{uVLP} systems can take on various forms. We categorize \ac{uVLP} \acp{Rx} into two main types: \textit{\textbf{light intensity-based}} receivers, including \acp{PD}, solar cells, \acp{ALS}, and spectral sensors; and \textit{\textbf{image-based}} receivers, such as smartphone cameras (cf. Fig.~\ref{fig:uVLPPractical}).

\begin{figure}[]
    \includegraphics[width=0.50\textwidth]{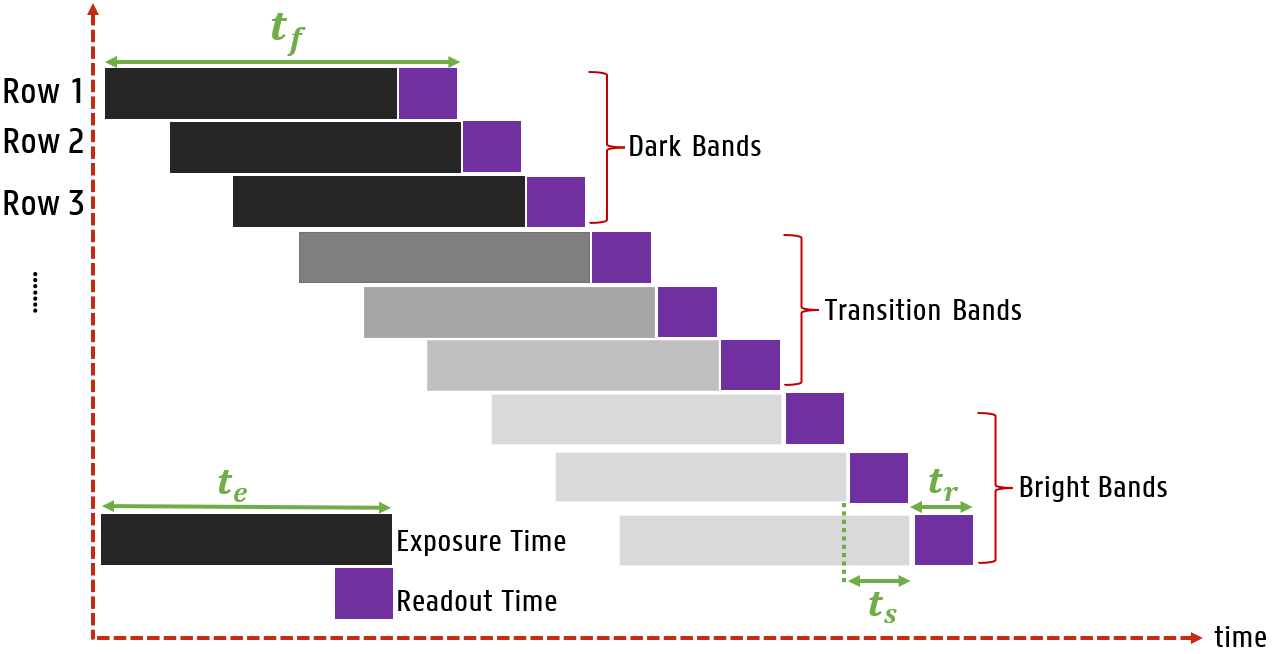} 
    \caption{Simplified illustration of a \ac{CMOS} image sensor using the rolling shutter effect. Transition bands result from overlap between the exposure time ($t_e$) and the \ac{LED}’s switching period within the total frame time ($t_f$) \cite{Zhang2016}.}
    \label{fig:rollingshutter}
\end{figure}

\subsubsubsection{Image Sensors}
Commercial cameras equipped with \ac{CMOS} image sensors can serve as receivers in \ac{uVLP} systems by leveraging the rolling shutter effect (illustrated in Fig.~\ref{fig:rollingshutter}) \cite{Zhang2016,Zhang2019,Carver2017}. These sensors are composed of a \ac{2D} matrix of \acp{PD} and employ a row-by-row scanning mechanism, sequentially reading out pixel data one row at a time. This sequential data acquisition technique, commonly referred to as the rolling shutter, introduces a systematic delay between consecutive row readouts. This delay is defined as the line time or readout time ($t_r$), and collectively results in the total frame capture duration or sampling interval ($t_s$)\cite{Jiao2017,Zhang2016,Teledyne2024,Yamazato2017,ElGamal2005}. While the rolling shutter effect can adversely affect image quality during high-speed motion capture \cite{Teledyne2024}, it simultaneously facilitates high-frequency sampling capabilities. Modern smartphone cameras can achieve readout frequencies ranging from 39.2 to 52.4 rows per millisecond~\cite{Lee2015}. This characteristic is particularly advantageous for demultiplexed \ac{uVLP} techniques (e.g., \ac{CF}-based \cite{Zhang2016,Zhang2019,Carver2017}). Section~\ref{subsec:CF-baseduVLPImage} discusses image-based \ac{uVLP} that utilize these properties of smartphone cameras for \ac{VL}-based positioning.

\subsubsubsection{Photodiode (PD)}
\acp{PD} operate as reverse-biased P--N junction semiconductors, converting optical photons into electrical current\cite{Rizk2022,Hui2020,Bastiaens2024,Paschotta2006}. They are integral components in optical communication systems (e.g., \ac{VLC}, \ac{uVLP}, and \ac{VLS}\cite{Alijani2024,Deprez2020,Alijani2025}) due to their compact size, rapid detection speed, and high efficiency, capable of sampling \ac{VL} at rates up to tens of MHz\cite{Rizk2022,Hui2020,Bastiaens2024,Paschotta2006,Jiao2017}. A widely used implementation of \acp{PD} in consumer electronics, particularly smartphones, is the \ac{ALS}~\cite{Zhuang2024,Wang2019,Liu2024}. The \ac{ALS} is primarily deployed to monitor ambient illuminance, enabling automatic screen brightness adaptation to enhance user readability and reduce energy consumption~\cite{Sato2022,Wang2019,Otsuka2025}. 

Functionally, the \ac{ALS} incorporates a \ac{PD} that converts incident light into an electrical current signal \( I(t) \), which is subsequently digitized by an \ac{ADC} to yield illuminance values, typically expressed in lux\cite{Wang2019}. Over a sampling window of duration \( T_s \), the output of the \ac{ALS}, denoted by \( O(t) \), can be modeled as:

\begin{equation}
O(t) = \frac{1}{T_s} \int_{t}^{t + T_s} I(\zeta) \, d\zeta,
\end{equation}

\noindent where \( T_s \) denotes the sensor-specific integration time, which defines the effective sampling interval and varies across different \ac{ALS} hardware implementations. The integration variable \( \zeta \) denotes the instantaneous time variable over the integration window \([t, t + T_s]\)\cite{Wang2019}. \ac{ALS} devices exhibit extremely low power consumption, typically below 1~mW~\cite{Sato2022}, which is orders of magnitude lower than the power usage of smartphone cameras (approximately 2000~mW) or dedicated high-speed \acp{PD} (around 150~mW)~\cite{Yang2018}. This energy efficiency makes \ac{ALS} highly suitable for power-constrained mobile applications~\cite{Sato2022,Otsuka2025,Wang2019}. However, a notable limitation is their significantly lower sampling rate (below 10 Hz\cite{Sato2022,Otsuka2025,Wang2019,Yang2025}) compared to high-speed \acp{PD}, which restricts their applicability in detecting fast-varying optical signals, such as \ac{CF}~\cite{Zhang2016}.

\subsubsubsection{Solar cells}
Solar cells generate a voltage or current when exposed to \ac{VL}, effectively converting optical energy into usable electrical power\cite{Rizk2022,Hui2020,Bastiaens2024,Paschotta2006}. This dual functionality is ideal for energy-efficient applications, such as \ac{IoT} devices powered by ambient light\cite{Rizk2022,Hui2020,Bastiaens2024,Paschotta2006}. Unlike conventional \acp{PD}, solar cells are optimized for power generation, which reduces the need for energy-intensive amplification circuits\cite{Rizk2022,Hui2020,Bastiaens2024,Paschotta2006}. However, they generally exhibit lower bandwidth and slower response times compared to \acp{PD}\cite{Elamrawy2025}, which can limit their suitability for high-speed communication scenarios. Despite this limitation, integrating solar cells into positioning and sensing applications can significantly reduce power consumption, thereby facilitating sustainable solutions in (u)\ac{VLP} systems\cite{Rizk2022,Hui2020,Bastiaens2024,Paschotta2006}.

\begin{table*}[]
    \centering
\caption{General comparison between \ac{VLP} and \ac{uVLP}.}
    \label{tab:vlp_comparison}
    \begin{adjustbox}{width=\linewidth} 
    \renewcommand{\arraystretch}{1.2}
    \begin{tabular}{|p{2.9cm}|p{6.2cm}|p{6.0cm}|}
        \hline
        \rowcolor[HTML]{D9D9D9} 
        \textbf{Features Taxonomy} & \textbf{Visible Light Positioning (VLP)} & \textbf{Unmodulated Visible Light Positioning (uVLP)} \\ \hline

        \textbf{Signal Type} & 
        \acp{LED} transmit modulated signals (coded) \cite{Bastiaens2024,Pathak2015} & 
        \acp{LED} emit constant, unmodulated light \cite{Bastiaens2020,Xu2015} \\ \hline

        \textbf{Accuracy} & 
        High (few cm to decimeters) \cite{Alam2019,Li2017,Plets20199,Almadani2021} & 
        Moderate to lower (few cm to meters) \cite{Xu2015,Yang2025} \\ \hline

        \textbf{Data Transmission} & 
        Can transmit data (e.g., \ac{LED} ID, location info) \cite{Pathak2015} & 
        Cannot transmit data (i.e., unmodulated \acp{LED})\cite{Bastiaens2020} \\ \hline

        \textbf{Implementation Cost} & 
        Higher (additional modulation hardware) \cite{Bastiaens2022,Almadani2021} & 
        Lower (existing infrastructure) \cite{Bastiaens2020,Bastiaens2022,Xu2015,Shao2018} \\ \hline

        \textbf{System Complexity} & 
        Higher complexity in \acp{Tx} and \acp{Rx} \cite{Bastiaens2024,Pathak2015,Zhuang2018} & 
        Lower complexity, simpler hardware \cite{Bastiaens2022,Bastiaens2020} \\ \hline

        \textbf{Lighting Performance} & 
        Potentially reduced by about 50\%\cite{DeLausnay2015,Bastiaens2020} & 
        Maintained (i.e., unmodified light sources) \cite{Bastiaens2020,Amsters2018} \\ \hline

        \textbf{Deployment Feasibility} & 
        Retrofit/modify lighting \cite{Bastiaens2024,Pathak2015,Sun2025,Bastiaens2022,Amsters2019} & 
        Easy integration (e.g., no light modification)\cite{Bastiaens2020} \\ \hline

        \textbf{Interference Resistance} & 
        Better resistance to ambient light and interference \cite{Bastiaens2024}& 
        More sensitive to ambient light noise\cite{Bastiaens2020,Bastiaens2022} \\ \hline

        \textbf{Maintenance Effort} & 
        Higher (complex components) \cite{Bastiaens2024,Armstrong2013} & 
        Lower (fewer components) \cite{Armstrong2013,Bastiaens2024,Bastiaens2022,Bastiaens2020} \\ \hline
    \end{tabular}
    \end{adjustbox} 
\end{table*}

\subsubsubsubsection{Spectral Sensors}
Spectral sensors (or spectrometers~\cite{Specim2024}) capture the reflectance spectrum of light from objects or scenes, unlike conventional \ac{IB} sensors (e.g., \acp{PD}) that only record a single light intensity\cite{Wang2023,Wang2022,Wang2024,Specim2024,Liu2014,Jiawei2024}. They provide \ac{LSI} by measuring light intensity across multiple wavelengths in the \ac{VL} spectrum~\cite{Wang2023,Specim2024}. Studies~\cite{Wang2023,Liu2014,Wang2024} show that even under identical lighting, indoor locations can produce distinct spectral signatures due to material-dependent reflectance. Spectral sensors include \ac{RGB} and multi-band types~\cite{Wang2022,Specim2024}. \ac{RGB} sensors use filters to capture three sub-bands, mimicking human vision after calibration~\cite{Liu2014,Kranjc2006}, while multi-band sensors (e.g., AS7265x~\cite{SparkFun2022,ams_AS7265x_2018}) measure radiant power across multiple sub-bands (e.g., 18 different wavelength frequencies)~\cite{Wang2023}. These sensors distinguish mixed and pure colors~\cite{Wang2022}, and are used in smartphones (e.g., Xiaomi~\cite{Xiaomi}) and virtual reality devices to enhance color accuracy~\cite{Wang2023}. Additionally, Hue sensors, which detect dominant wavelength powers, are useful for light feature extraction in white \acp{LED}\cite{Singh2022,Singh2023}.

\subsection{Comparative Analysis of \ac{VLP} and \ac{uVLP}}
\label{subsec:comparingvlpanduvlp}
Table~\ref{tab:vlp_comparison} presents a comparative overview of \ac{VLP} and \ac{uVLP} across various aspects. \ac{VLP} systems generally offer superior positioning accuracy due to their higher effective \ac{SNR}, achieved through controlled modulation. This makes them well-suited for high-precision applications such as Industry~4.0 (e.g., pallet tracking)~\cite{Danys2022, Sisinni2018} and industrial \ac{IoT} environments~\cite{Rekkas2023, Danys2022, Sisinni2018, Almadani2021}. However, this enhanced accuracy comes at the expense of increased cost, added system complexity, and potential impacts on illumination performance due to the integration of modulation hardware\cite{Bastiaens2020}. On the other hand, \ac{uVLP} systems leverage existing lighting infrastructure without requiring any physical modifications, making them a cost-effective and less complex alternative~\cite{Bastiaens2020,Bastiaens2022,Zhang2016,Zhang2016LiTell2,Zhang2019}. Nevertheless, they typically exhibit lower positioning accuracy due to reduced \ac{SNR} or limited discriminative capability when the contributions of individual \acp{Tx} are not demultiplexed. In contrast, demultiplexed approaches such as \ac{CF}-based \ac{uVLP} using \ac{FFT} or \ac{MUSIC} can achieve higher accuracy by leveraging the improved \ac{SNR} provided through the exploitation of the \ac{CF} of \acp{LED}, which enhances \ac{Tx} separability compared to undemultiplexed methods. Section~\ref{sec:uvlp_positioning} presents a detailed examination of \ac{uVLP} positioning methods, including both demultiplexed and undemultiplexed approaches.

\section{Taxonomy of \ac{uVLP} Techniques}
\label{sec:uVLPtechniques}

Fig.~\ref{fig:uVLPTechniquesClassification} illustrates the broad classification of \ac{uVLP} techniques based on the type of receiver, distinguishing between light \ac{IB} and image sensor-based receivers. Each category is further subdivided into \textit{\textbf{demultiplexed}} and \textit{\textbf{undemultiplexed}} methods. Demultiplexed methods (i.e., \ac{CF}-based \ac{uVLP}) assume that individual unmodulated light sources can be distinguished (or separated) based on the inherent features of each \ac{LED} (e.g., \ac{CF}\cite{Bastiaens2020,Bastiaens2022,Zhang2016}). Conversely, undemultiplexed approaches either treat the total received light signal in light \ac{IB} receivers (e.g., \acp{PD}, \ac{ALS}, etc.) as a composite of multiple indistinguishable sources or consider an image captured from an unmodulated light source in imaging-based receivers. These methods rely on fingerprinting~\cite{RaviFiatLuxF,Salem2021,Golding1999,Randall2007,Wu2022} or probabilistic models (e.g., the \ac{EKF}~\cite{Amsters2018,Amsters2019} and Bayesian estimation) to estimate the receiver’s position based on light intensity, without explicitly separating the contributions from individual light sources. In undemultiplexed approaches using image sensors, methods such as image-based processing (i.e., light source localization via visual landmarks) and fingerprinting can be employed. 

In this article, we address both \ac{IB} and imaging-based optical receivers (see Fig.~\ref{fig:uVLPTechniquesClassification}). Our focus is specifically on active (device-based) \ac{uVLP} systems, where the receiver (e.g., a \ac{PD} or camera) is carried by or attached to the object being localized. Passive \ac{uVLP} methods, which fall under the broader category of \ac{VLS}~\cite{Alijani2025}, are beyond the scope of this paper but are comprehensively reviewed in~\cite{Ullah2023,Singh2020,Wang2020,Alijani2025}.

\subsection{Light Intensity-Based Approaches}
\subsubsection{Indoor Optical Channel}
\label{subsec:channelmodel}
Indoor optical propagation is commonly modeled using the \textit{\textbf{inverse square law}}\cite{Murdoch1981,Taylor2000} and  \textit{\textbf{Lambert's cosine law}}\cite{Kahn1997,Taylor2000}. The inverse square law ($E = I/d^2$) relates the illuminance \(E\) on a surface to the luminous intensity \(I\) of a point source and the distance \(d\) from the source~\cite{Murdoch1981,Taylor2000}. It assumes a perpendicular receiving surface and a point-like emitter~\cite{QiangXu2015}. Lambert’s cosine law ($I_{\theta} = I_0\cos\theta$) describes the angular dependence of irradiance, applicable to Lambertian sources such as \acp{LED} and \acp{FL}~\cite{Taylor2000,Moreno2008,Bastiaens2024}. This model accounts for both distance and angular dependencies of the emitted and received light~\cite{Bhalerao2016,Bastiaens2020,Wang20188}.

\begin{figure}[]
    \centering
    \includegraphics[width=8.5cm]{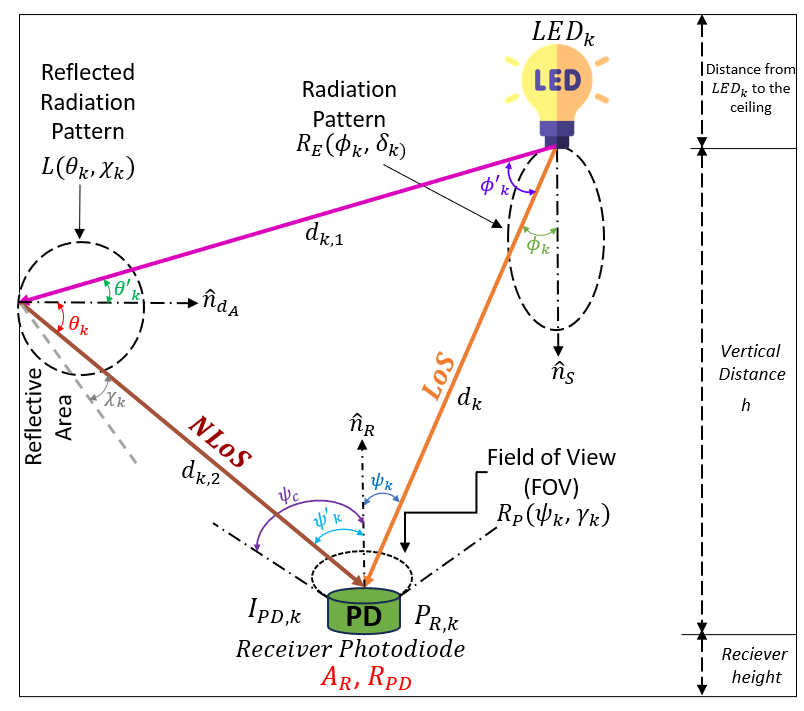}
    \centering
    \caption{Illustration of a typical \ac{PD}-based \ac{uVLP} geometry for optical channel modeling\cite{Bastiaens2024,Ghassemlooynew,Alijani2025,Bastiaens2023AnEA}.}
    \label{fig:VLPChannel}
    \end{figure}

\begin{figure}[]
\centering
\includegraphics[width=0.48\textwidth]{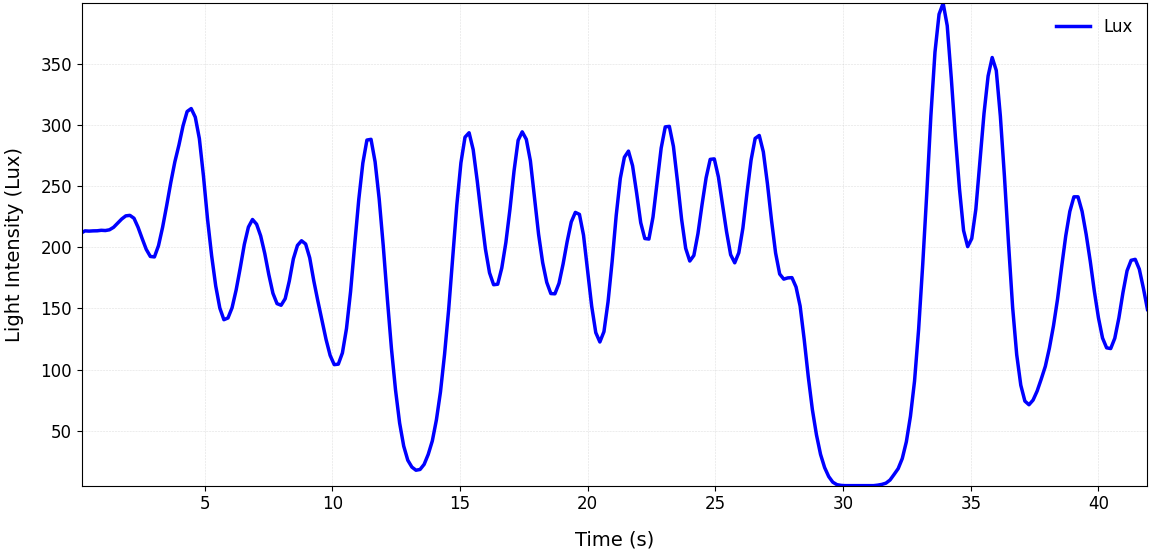} 
\caption{Light intensity captured by an \ac{ALS} sensor from a smartphone camera (e.g., Samsung Galaxy A52) while walking through a corridor with multiple \acp{LED} in the ceiling.}
\label{fig:ALSoutput}
\end{figure}

\subsubsubsection{Light Intensity}
Fig.~\ref{fig:VLPChannel} illustrates a typical light propagation model used in \ac{PD}-based \ac{uVLP} systems, assuming that \ac{LED}$_k$ follows a Lambertian radiation pattern within a flat fading channel (i.e., assuming no delay spread due to multipath\cite{Vatansever2017}) for indoor positioning purposes\cite{Verniers2017}. The total received power at the \ac{PD} from \ac{LED}$_k$ (i.e., $P_{r,k}$) can be expressed in terms of the transmitted optical power $P_{t,k}$ and the channel gain $H$, which includes the \ac{LoS} component $h_{c,LoS}^{(k)}$ and the \ac{NLoS} component $h_{c,\,NLoS}^{(k,\, dA)}$ within the receiver’s \ac{FoV} $\psi_c$ (semiangle). The general formulation is given by~\cite{Ghassemlooynew,Sander2021,Bastiaens2024,BastiaensPhdBook}:
\begin{figure*}[]
  \centering
    \includegraphics[width=0.80\textwidth]{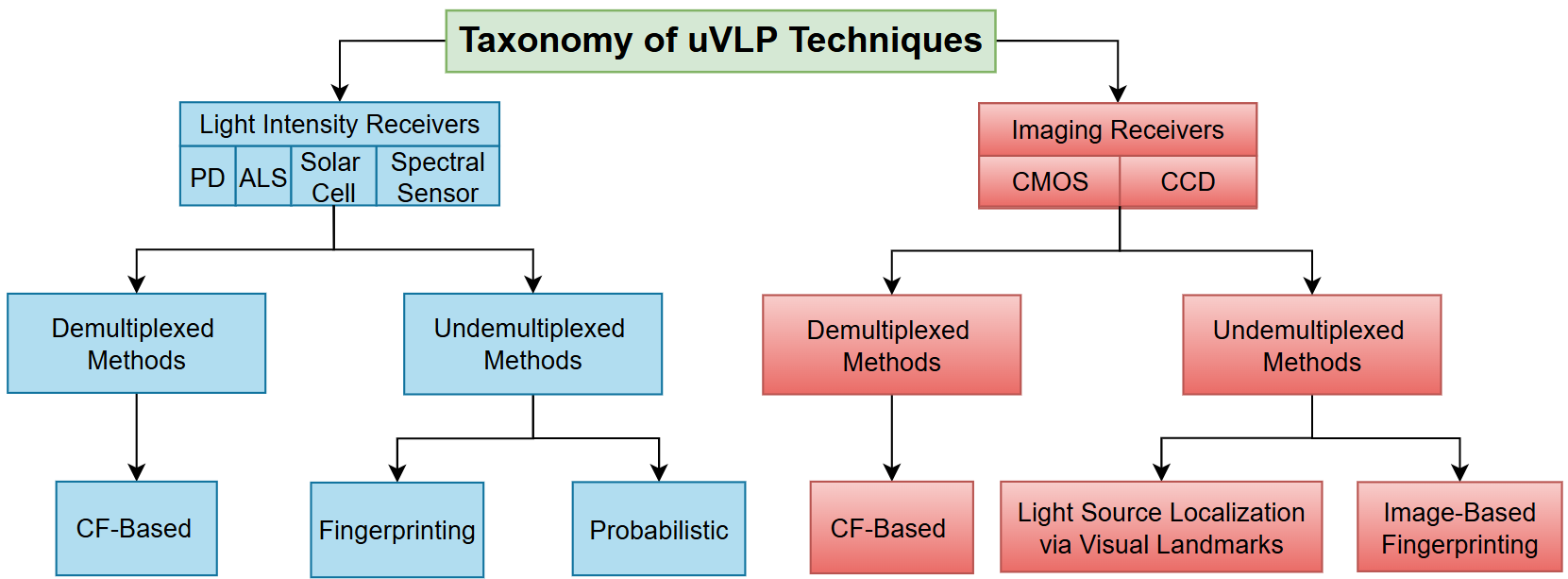} 
\caption{Taxonomy of \ac{uVLP} techniques categorized by receiver type and signal processing method.}
    \label{fig:uVLPTechniquesClassification}
\end{figure*}
\begin{equation}
    P_{r,k}=P_{t,k}\underbrace{\bigg[h_{c,LoS}^{(k)} + \sum_{A}^{}h_{c,\,NLoS}^{(k,\, dA)}\bigg]}_{H},
    \label{Main1}
\end{equation}
with:
\begin{equation}
    h_{c,LoS}^{(k)}=R_E(\phi_k,\delta_k)\frac{A_R}{d_k^2}\cos(\phi_k)T_s(\psi_k)\tau(\psi_k),
    \label{Loseq}
\end{equation}
\begin{equation}
    h_{c,\,NLoS}^{(k,\, dA)}=\frac{R_E(\phi'_k,\delta'_k) \cos(\theta'_k) dA \cos(\psi'_k)A_R L(\theta_k,\chi_k)}{d^2_{k,1} d^2_{k,2}},
    \label{NLoSeq}
\end{equation}
where $R_E(\phi_k,\delta_k)$ indicates the radation pattern of LED$_k$, $A_R$ is the \ac{PD} surface area, $d_{k}, d_{k,1}$, and $d_{k,2}$ represent the distances between the LED$_k$ and \ac{PD} (\ac{LoS} direction), the LED$_k$ and reflective area, and the reflective area and \ac{PD}, respectively\cite{Bastiaens2024,Bastiaens2023AnEA,BastiaensPhdBook,Ghassemlooynew,Sander2021}. $\phi_k$, $\phi'_k$, and $\theta_k$ designate the elevation irradiance angles for the \ac{LoS} and \ac{NLoS} components, and during reflection\cite{Bastiaens2024,Bastiaens2023AnEA,BastiaensPhdBook,Ghassemlooynew,Sander2021}. $\delta_k/\delta'_k$ symbolize the azimuthal irradiance angles, and $\psi_k/\psi'_k/\theta'_k$ represent the incidence angles at the \ac{PD} and at the reflective area, and finally $L(\theta_k,\chi_k)$ represents the reflected radiation pattern which is governed by Phong’s model\cite{Ghassemlooynew,Kahn1997,BastiaensPhdBook,Bastiaens2024,Sander2021}. Moreover, \( T_s(\psi) \) denotes the optical filter gain, and \( \tau(\psi) \) represents the optical concentrator gain~\cite{Ghassemlooynew,Kahn1997,Bastiaens2024,Sander2021,BastiaensPhdBook}. By considering $R_E(\phi_k,\delta_k)$ as the Lambertian radiation pattern of \ac{LED}$_k$, characterized by rotational symmetry (e.g., $\delta$-independence) and governed by a single parameter known as the Lambertian order $m_k$ \cite{BastiaensPhdBook,Bastiaens2023AnEA,Bastiaens2024}:
  \begin{equation}
  R_E(\phi_k,\delta_k)=R_E(\phi_k)=\frac{m_k+1}{2\pi}\cos^{m_k}(\phi_k).
  \label{Lambertain}
  \end{equation}
The Lambertian order $m_k=\frac{-ln(2)}{ln(cos \phi_{1/2})}$ is directly linked to the radiation pattern's single-sided half (or semiangle) angle $\phi_{1/2}\in \left[\frac{-\pi}{2}, \frac{\pi}{2}\right]
$\cite{Bastiaens2024,BastiaensPhdBook,Ghassemlooynew,Sander2021}. By substituting Eq.~\eqref{Lambertain} into Eq.~\eqref{Loseq}, the channel gains $h_{c,\text{LoS}}^{(k)}$ and $h_{c,\text{NLoS}}^{(k,\, dA)}$ (neglecting specular reflection) can be computed as follows~\cite{Bastiaens2024,BastiaensPhdBook,Ghassemlooynew,Sander2021}:
\begin{align}
\label{channelDC}
h_{c,\,LoS}^{(k)} = \begin{cases}
\dfrac{(m_k+1)A_{R}}{2\pi d_k^2} \cdot \cos^{m_k} (\phi_k) cos(\psi_k)T_s(\psi_k)\tau(\psi_k)\\
\quad \quad \quad \quad \quad \quad \quad \quad \quad \quad \quad 
,0\leq{\psi_k} \leq {\psi_c}\\
0\quad \quad \quad \quad \quad \quad \quad \quad \quad \quad \quad ,{0} \geq {\psi_c},
\end{cases}
\end{align}
\begin{align}
\label{channelDCnlos}
h_{c,\,NLoS}^{(k,\,dA)}  = \begin{cases}
\dfrac{(m_k+1)A_{R}}{2\pi^2 d_{k,1}^2d_{k,2}^2}\rho r_d dA \cdot \cos^{m_k} (\phi'_k)\cos(\theta_k^{'})\\
\cos(\theta_k) T_s(\psi_k)\tau(\psi_k)\cos(\psi'_k),\quad 
0\leq{\psi_k} \leq {\psi_c}\\
0,\quad \quad \quad  \quad \quad \quad \quad \quad \quad \quad \quad \quad \quad {0} \geq {\psi_c}.
\end{cases}
\end{align}
For a receiver chain with a linear response, the received optical power from \ac{LED}$_k$ (cf. Eq.~\eqref{Main1}) can be expressed in the form of an electrical voltage corresponding to the \ac{RSS}, typically measured in volts, as follows\cite{phdthesis,Bastiaens2024,Alijani2024,Alijani2025}:
\begin{equation}
RSS_{k} = G_{\text{PD}} \cdot 
\underbrace{
R_{\text{PD}} \cdot P_{r,k},
}_{I_{\text{PD}}(k)}
\label{eq:singleRSS}
\end{equation}
where \( R_{\text{PD}} \) is the \ac{PD} peak responsivity, \( G_{\text{PD}} \) is the transimpedance gain of the \ac{PD}, and \( I_{\text{PD}}(k) \) denotes the photocurrent generated by the \ac{PD} in response to the received optical power \( P_{r,k} \)~\cite{phdthesis,PDA100A2_Manual,AppliedVLP}. Note that any type of \ac{uVLP} receiver (e.g., camera) can be used instead of the \ac{PD} in Fig.~\ref{fig:VLPChannel}. For example, Fig.~\ref{fig:ALSoutput} shows the light intensity captured in lux by an \ac{ALS} sensor from a smartphone camera while walking through a corridor with several \acp{LED} in the ceiling. Finally, as noted in\cite{Moreno2008}, some \acp{LED} do not follow the traditional Lambertian model. To better represent these, a more general cosine-power function has been proposed\cite{Moreno2008}, which fits most commercially available \acp{LED} more accurately.
\subsubsubsection{Light Spectrum}
Indoor environments vary in layout, furniture, shadowing, scattering, and light source distribution (e.g., \acp{LED}, \acp{FL}, etc.), leading to non-uniform light intensity and spectrum\cite{Wang2022,Wang2023}. \ac{SB} localization based on \ac{LSI} is a feasible solution in \ac{uVLP}, thanks to three pillars, all of which have been experimentally verified in \cite{Wang2022,Wang2023}. First, \ac{LSI} is location-dependent, meaning that various indoor locations yield distinct \ac{LSI} when measured using spectral sensors\cite{Wang2023,Wang2022,Singh2022,Singh2023}. Second, \ac{LSI} is more reliable than single light intensity measurements because it captures richer spectral information, which results in better separability between locations. This is evidenced by t-distributed stochastic neighbour embedding (t-SNE)\cite{JMLR:v9:vandermaaten08a} visualizations showing that \ac{LSI} forms more distinct clusters with larger inter-cluster distances compared to intensity-only data, making it a more dependable indicator for accurate \ac{uVLP}. Third, \ac{LSI} remains stable over time under consistent lighting conditions, as shown by minimal variation in cluster distances across multiple days \cite{Wang2023,Wang2022,Singh2022,Singh2023}.

Consider the basic light propagation model depicted in Fig.~\ref{fig:VLPChannel}, where a spectral sensor (e.g., AS7265x~\cite{SparkFun2022,ams_AS7265x_2018}) is employed instead of a \ac{PD}. The spectral sensor possesses a wavelength-dependent responsivity denoted as \( R_{\text{SP}}(\lambda) \)\cite{SparkFun2022,ams_AS7265x_2018}. A single unmodulated \ac{LED}, labeled \( \text{LED}_k \), emits light characterized by a spectral distribution \( p_k(\lambda) \), where \( \lambda \in \{\lambda_1, \lambda_2, \ldots, \lambda_z\} \) defines the set of discrete wavelength bands corresponding to the sensor's spectral channels (e.g., 18 wavelengths~\cite{ams_AS7265x_2018}). The distribution \( p_k(\lambda) \) follows a Lambertian emission pattern with order \( m \), dependent on the specific \ac{LED} type, and is normalized over the visible spectrum from \( \lambda_1 = 380\,\text{nm} \) to \( \lambda_z = 780\,\text{nm} \)~\cite{Verniers2017}:
\begin{equation}
\sum_{i=1}^{z} p_k(\lambda_i) \, \Delta\lambda_i \approx 1,
\end{equation}
where \( \Delta\lambda_i \) represents the bandwidth of the \( i^{\text{th}} \) spectral channel. Assuming that \( \text{LED}_k \) emits a constant total optical power \( P_{t,k} \) across its spectral content, and given the optical channel gain \( H \) described in Eq.~(\ref{Main1}), the electrical \ac{RSS} at the \( i^{\text{th}} \) spectral band can be expressed as (cf. Eq.~\eqref{eq:singleRSS}):
\vspace{-\baselineskip}  
\begin{flushleft}
\begin{align}
\text{RSS}_{k,\text{SP}}(\lambda_i) =\; & H_k(\lambda_i) \cdot 
\underbrace{P_{t,k} \cdot p_k(\lambda_i) \cdot \Delta\lambda_i}_{\mathclap{\text{Optical power at } \lambda_i}} \cdot \notag \\
& \hspace{1.8cm} \underbrace{R_{\text{SP}}(\lambda_i)}_{\mathclap{\text{Sensor responsivity}}} \cdot \:\:\:\:
\underbrace{G_{\text{SP}}(\lambda_i)}_{\mathclap{\text{Sensor gain}}}
\end{align}
\end{flushleft}
where \( R_{\text{SP}}(\lambda_i) \) denotes the spectral responsivity of the sensor at wavelength \( \lambda_i \), and \( G_{\text{SP}}(\lambda_i) \) represents the internal gain (e.g., transimpedance or digital) of the spectral sensor at that wavelength~\cite{ams_AS7265x_2018}. The term \( H_k(\lambda_i) \) corresponds to the wavelength-dependent channel gain, which accounts for propagation effects such as the spectral reflectance of the scene (e.g., surface color and material), geometry, and ambient light interference, thereby enabling \ac{LSI} to discriminate between different positions.
\subsubsection{Positioning Approaches}
\subsubsubsection{Undemultiplexed Methods}
\vspace{-\baselineskip}  
\paragraph*{\textbf{(i) Fingerprinting techniques}}
Eq.~\eqref{eq:singleRSS} gives the \ac{RSS} from a single \ac{LED}$_k$, where the received light intensity depends on the receiver’s position relative to the source. In general, higher power is observed at shorter distances; however, this is not always the case, as large incidence or irradiance angles can significantly reduce the received power due to the angular dependence of the optical channel. In practice, the receiver (e.g., a \ac{PD}) collects light from all \acp{LED} (e.g., $k = 1, 2, \dots, M$) within its \ac{FoV}. Thus, the total received signal can be expressed as~\cite{Yang2025}.
\begin{equation}
\begin{aligned}
RSS_{\text{total}} &=  \sum_{k=1}^{M} RSS_{k}= RSS_1 + RSS_2 + \dots + RSS_M,
\end{aligned}
\label{totalRSS}
\end{equation}
where \( M \) is the number of \acp{LED} contributing to the total received signal. The main idea behind fingerprinting (or scene analysis methods) in \ac{uVLP} is to collect spatial (or spectral\cite{Hu2024,Hu2023,Wang2023,Wang2022}) intensity fingerprints at known reference points, creating a database (known as offline stage). In the online stage, the receiver captures current intensity/spectrum measurements, which are then matched against the database using classification~\cite{RaviFiatLuxF}, probabilistic methods\cite{Wang20188} (cf. Section~\ref{sec:probabilistic}), or \ac{ML} techniques\cite{Wang2023,Wang2022} to estimate the user’s position. In this category of \ac{uVLP} systems, light intensity is often combined with other sensor data to improve matching accuracy~\cite{Azizyan2009,Golding1999}. Section~\ref{subsec:finglightintenisty} reviews \ac{uVLP} studies that employ various light intensity receivers (e.g., \acp{PD}, spectral sensors, etc.) based on this approach.
\paragraph*{\textbf{(ii) Probabilistic methods}}
\label{sec:probabilistic}
There is well-documented evidence in the literature that probabilistic models such as the \ac{KF}\cite{Rahaim2012,Li2017}, the \ac{EKF}\cite{Amsters2018,Amsters2019,Saengudomlert2024,Vatansever2017,Li20177}, and Bayesian filters~\cite{Xu2015,Hu2015,Jimenez2013} can be effectively implemented in (u)\ac{VLP}. These filters (e.g., as post-processing) are typically applied on top of raw position estimates, which can be derived from geometric calculations, photometric models, or \ac{RSS}-based fingerprinting~\cite{Vatansever2017}, to account for motion dynamics and reduce noise. In such systems, the received power levels (cf. Eq.\eqref{Main1}) from each unmodulated \ac{LED}$_k$ serve as observation inputs for estimating the user's position over time\cite{Vatansever2017}. Prior implementations of \ac{KF}-based algorithms for \ac{VLP} generally rely on known parameters such as irradiance, angles of incidence, and distances to infer the user's position from received power measurements~\cite{Vatansever2017,Rahaim2012}. The discrete-time \ac{EKF} is a widely used tool for nonlinear state estimation problems~\cite{Vatansever2017}. Given a nonlinear discrete-time state-space model, the system evolves as follows~\cite{Vatansever2017}:
\begin{equation}
    \mathbf{x}_{k} = f(\mathbf{x}_{k-1}, \mathbf{u}_{k}) + \mathbf{q}_{k-1},
\end{equation}
\begin{equation}
    \mathbf{z}_{k} = h(\mathbf{x}_{k}) + \mathbf{v}_{k},
\end{equation}
where $\mathbf{x}_{k} \in \mathbb{R}^{n}$ is the state vector (e.g., the \ac{2D} or \ac{3D} position and possibly velocity of the user), $\mathbf{z}_{k} \in \mathbb{R}^{m}$ denotes the measurement vector, $\mathbf{u}_{k}$ is the control input (e.g., inertial data from \ac{PDR}), $f(\cdot)$ is the dynamic function model (i.e., nonlinear motion model), and $h(\cdot)$ is the observation (or measurement) function model which maps the user's position to the expected received optical power~\cite{Vatansever2017}. The vectors $\mathbf{q}_{k} \sim \mathcal{N}(\mathbf{0}, \mathbf{Q}_k)$ and $\mathbf{v}_{k} \sim \mathcal{N}(\mathbf{0}, \mathbf{V}_k)$ represent the process and observation noise, respectively, typically modeled as zero-mean Gaussian distributions with known covariance matrices $\mathbf{Q}_k$ and $\mathbf{V}_k$~\cite{Vatansever2017}.
\begin{figure*}[]
  \centering
  \includegraphics[width=0.90\textwidth]{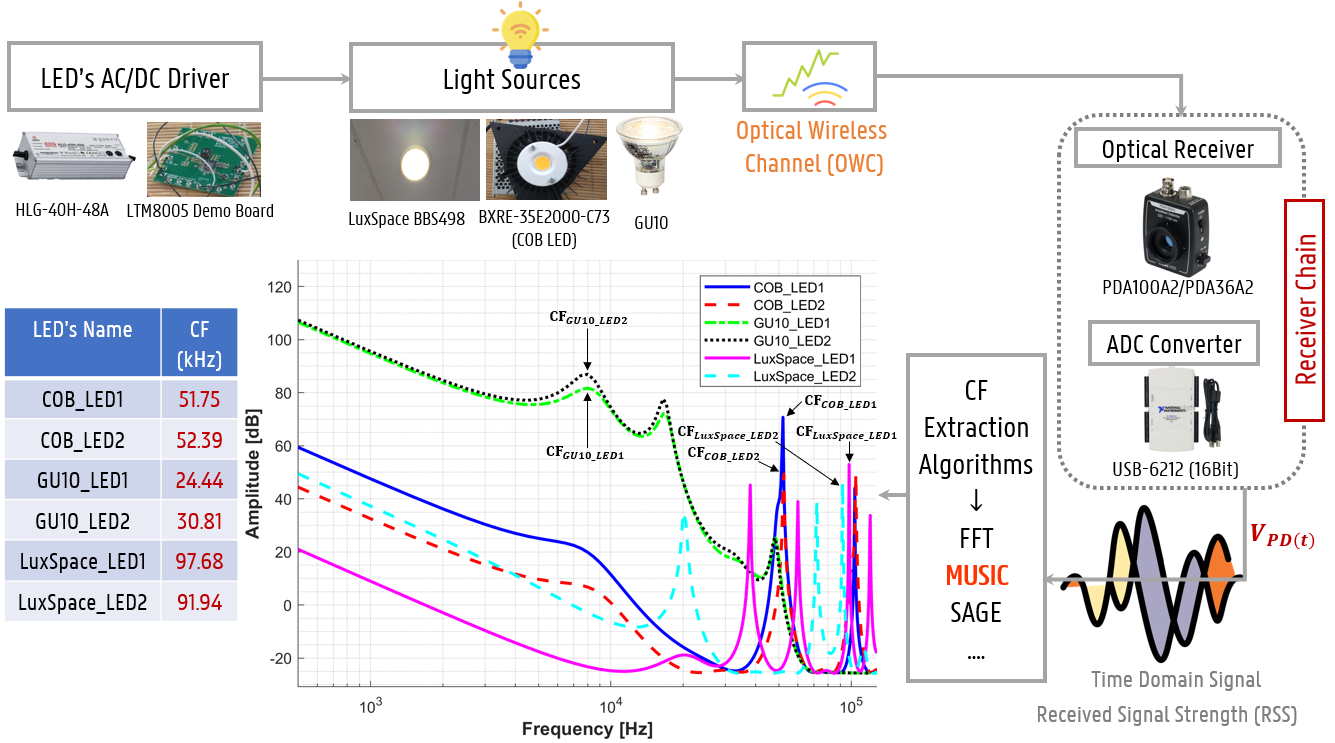} 
\caption{Illustration of a \ac{PD}-based \ac{CF} extraction workflow in \ac{uVLP}. A \ac{PD} (e.g., PDA100A2~\cite{PDA100A2_Manual}) connected to a \ac{DAQ} (e.g., USB-6212~\cite{NI_USB6212_Specs}) receives light signals from an \ac{LED}, sampled at 256~kHz. The \ac{CF} (i.e., the peak frequency) is extracted via peak detection on the 100-times averaged pseudospectrum obtained using the \ac{MUSIC} algorithm (subspace order 10)\cite{Bastiaens2020}. Alternative methods include \ac{FFT}\cite{oppenheim1999discrete,DeLausnay2015} and \ac{SAGE}\cite{Fleury1999}. After determining the \ac{CF}, the corresponding amplitude can be used for positioning purposes in \ac{uVLP}~\cite{Bastiaens2020,Bastiaens2022}.}
  \label{fig:cfbaseduVLPfft}
\end{figure*}

To estimate the user's position using this framework, the motion model $f(\cdot)$ can be informed by \ac{PDR} data, such as step counts, stride length, and heading direction from inertial sensors (e.g., \acp{IMU} and magnetic sensors~\cite{Yang2025}), providing a prediction of the user's movement~\cite{Yang2025}. The observation model $h(\cdot)$ uses known positions of the \acp{LED} and their emission characteristics (e.g., Lambertian pattern) to predict the expected received power at the estimated position~\cite{Vatansever2017}. During the correction phase, the \ac{EKF} compares the predicted received power to the actual sensor measurements (e.g., intensity or spectral \ac{RSS}) and updates the position estimate accordingly~\cite{Vatansever2017}. This probabilistic filtering framework fuses optical and inertial data for robust, continuous positioning, even with noise or partial observations. Section~\ref{subsec:probabllightintensity} reviews \ac{uVLP} systems using such techniques.
\subsubsubsection{Demultiplexed Methods (\ac{CF}-based \ac{uVLP})}
In a \ac{CF}-based \ac{uVLP} system, localization proceeds in two main stages: \textit{(i)} frequency‐domain demultiplexing of the total received signal (i.e., $\ac{RSS}_{\text{total}}$ describe in Eq.~\ref{totalRSS}), and \textit{(ii)} position estimation via a suitable positioning method (e.g., trilateration, etc).

\textbf{\textit{i) Frequency‐domain demultiplexing:}} As discussed in Section~\ref{subsec:uVLPHardware}, any \ac{LED} or \ac{FL}/\ac{CFL} type with a constant-current driver is suitable for \ac{uVLP}, as all exhibit \acp{CF}, typically in the tens of kHz range, due to their driver circuitry~\cite{vanderBroeck2007,Zhu2017,Bastiaens2020,Zhang2016}. These \acp{CF} can be effectively utilized for localization purposes. Initial studies~\cite{Bastiaens2020,Zhang2016,Zhang2016LiTell2,Zhang2019} demonstrated that the observed light power at the \acp{CF} maintains a sufficient \ac{SNR} and remains stable over time, enabling accurate (decimeter-level) positioning. Consider the total received signal strength, $\ac{RSS}_{\text{total}}$, computed in Eq.~\eqref{totalRSS}, which aggregates the contributions from all unmodulated \acp{LED} within the receiver’s (e.g., a \ac{PD}) \ac{FoV}. To enable positioning, these contributions must be demultiplexed (see Fig.~\ref{fig:cfbaseduVLPfft}). Owing to its simplicity and effectiveness, an \ac{FFT}-based method is typically employed to initially separate each \ac{LED}$_k$’s contribution by leveraging its unique \ac{CF}, $f_{c,k}$. The individual received power $P_{r,k}$ (or amplitude) is then estimated from the spectral content extracted from the received signal. In practice, however, the $f_{c,k}$ may exhibit slight temporal variations due to hardware imperfections or environmental fluctuations. To account for such variations, the method proposed in~\cite{Bastiaens2020} identifies each \ac{LED}’s spectral contribution by detecting the spectral peak within a frequency interval $\left[f_{c,k}-\Delta f,\; f_{c,k}+\Delta f\right]$, where $\Delta f = 100\,\mathrm{Hz}$ and the search step is $10\,\mathrm{Hz}$.

Note that current \ac{VLP} demultiplexing techniques rely on the modulation at known harmonics of a base frequency (e.g., \cite{DeLausnay2015}), demultiplexing is quite straightforward. In \ac{uVLP} however, the frequencies of interest (i.e., \ac{CF}) can be considered random. Hence, depending on the set of \acp{CF} of the observed \acp{LED}, optimal demultiplexing settings can change. For instance a \ac{MLE} algorithm for harmonic frequency estimation to determine the \acp{CF} could be an option and their variation with temperature, e.g., the expectation-maximization scheme in \ac{SAGE}\cite{Fleury1999}. Algorithms like \ac{SAGE} and \ac{MUSIC} are super-resolution, meaning that they can estimate the \ac{CF} with greater resolution than the simple Discrete Fourier Transform (DFT), which has a frequency resolution equal to the inverse of the time span of the total \ac{RSS} time series (cf. Eq.~\ref{totalRSS}). The advantage of \ac{MLE} algorithms over subspace-based algorithms such as \ac{MUSIC} \cite{Schmidt1986} is that \ac{RSS} estimation can be done with one total \ac{RSS} time series, allowing for faster localization and larger mobility of the \ac{uVLP} solution (e.g., embedded and mobile tags).

\textbf{\textit{ii) Position estimation:}}
\label{subsec:positioningcf}
After extracting the received-power contributions $\{P_{r,k}\}_{k=1}^M$ from the $M$ transmitting \acp{LED}, the user position $\mathbf{p}_u=(x_u,y_u,z_u)$ can be estimated by first converting each received power measurement into a distance estimate $d_k$ using the channel model described in Section~\ref{subsec:channelmodel} (e.g., Kahn’s model~\cite{Kahn1997}). During an offline site survey, each \ac{LED}$_k$ is characterized by its \ac{3D} spatial coordinates $\mathbf{p}_{S,k}=(x_{S,k},y_{S,k},z_{S,k})$, its characteristic frequency $f_{c,k}$, and the received optical power $P_{\text{ref},k}$ measured at a known reference position (typically on the floor directly beneath the \ac{LED}). Using the channel model, which includes the \ac{LED} radiation pattern and the receiver acceptance model, the measured power $P_{r,k}$ is mapped to a distance estimate $d_k$. These distance estimates ${d_k}$ are then used in established \ac{RSS}-based \ac{VLP} algorithms to estimate the user's location. Common approaches include multilateration (or trilateration)\cite{Bastiaens2020,ZHANG2023,Plets2019,Bastiaens2024}, the \ac{CMD} method\cite{Thomas2005,Bastiaens2020}, and \ac{MBF} techniques~\cite{Bastiaens2020,Sander2021}. These algorithms ultimately produce an estimate of the user's position, denoted as $\hat{\mathbf{p}}_u$, in \ac{3D} space.

\subsection{Image-Based Techniques in \ac{uVLP}}
\subsubsection{Undemultiplexed Methods}
\begin{figure}[htb!]
\centering
\includegraphics[width=0.50\textwidth]{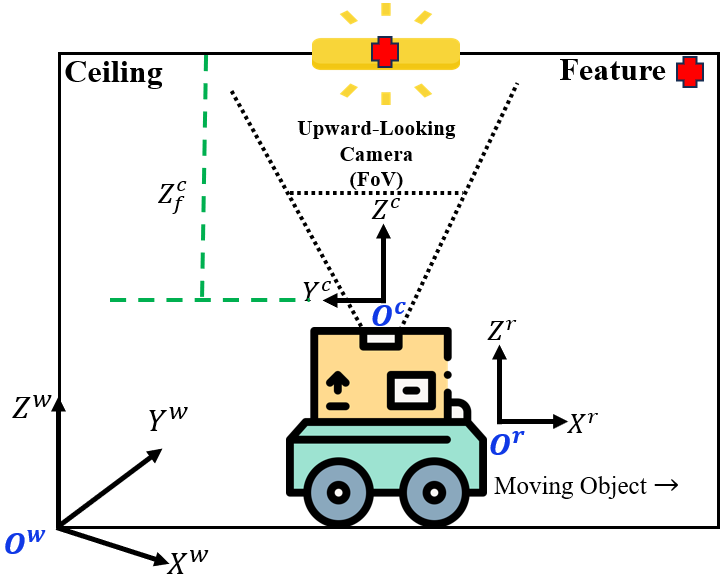} 
\caption{Image-based \ac{uVLP} system using an upward-facing camera to localize light sources (e.g., \ac{FL}) via image features within a \ac{SLAM} framework for a moving object (e.g., robot or wheelchair).}
\label{fig:imagebaseduVLP}
\end{figure}

\subsubsubsection{Light Source Localization via Visual Landmarks}
Localization of unmodulated (or unmodified) \ac{VL} sources (e.g., \acp{FL}), via ceiling fixtures is a well-established method for indoor \ac{SLAM} applications involving mobile agents like robots~\cite{Bellini2002,Chen2014,Alves2013,launay2001fluorescent} and wheelchairs~\cite{WANG1997,Wang2005} (see Fig.\ref{fig:imagebaseduVLP}). Ceiling-mounted light fixtures are robust landmarks due to their distinctive brightness and stable spatial arrangement when captured by upward-facing cameras\cite{lirias3591348,Alves2013,Chen2014}. Unlike natural landmarks such as doors or windows, ceiling fixtures are rarely obstructed, ensuring consistent detection and reliable feature tracking~\cite{lirias3591348}. Image processing techniques, including intensity thresholding and morphological operations (e.g., dilation and erosion), are commonly applied to extract relevant visual features such as brightness, geometric shapes, and spatial configurations of these fixtures~\cite{Chen2014}. For robust pose estimation, probabilistic state estimation methods (e.g., the \ac{EKF}), fuse these visual landmarks with odometry data to refine localization accuracy~\cite{Bellini2002,Panzieri2005,Alves2013,Folkesson2005,Hwang2011}.

Fig.~\ref{fig:imagebaseduVLP} illustrates a typical image-based \ac{uVLP} system utilizing an upward-facing camera mounted on a moving object (e.g., a robot or wheelchair) within a \ac{SLAM} framework. By defining the world coordinate frame $O^w=(X^w, Y^w, Z^w)$ attached to the floor, the robot coordinate frame $O^r=(X^r, Y^r, Z^r)$ fixed at the midpoint of the axis connecting its two driving wheels (with $X^r$ axis perpendicular to this axis), and the camera coordinate frame $O^c=(X^c, Y^c, Z^c)$ fixed at the optical center of the camera (with $Z^c$ axis aligned along the camera's optical axis), the stationary ceiling landmarks' positions denoted by $(x_f,y_f)$ can be determined in the global \ac{2D} world frame using coordinate transformations\cite{Chen2014}:
\begin{equation}
\label{eq.observationmodel}
\begin{aligned}
x_f &= x^r_f \cos\theta - y^r_f \sin\theta + x, \\
y_f &= x^r_f \sin\theta + y^r_f \cos\theta + y,
\end{aligned}
\end{equation}
where $(x, y)$ represents the position of the moving object (i.e., a robot) and $\theta$ is its orientation within the world frame\cite{Chen2014}. Here, $(x^r_f, y^r_f)$ denote the observed coordinates of the features in the robot's local frame\cite{Chen2014}. Typically, the odometry model of the moving object defines its state at time $t$ as $(x(t), y(t), \theta(t))$, and its subsequent state at time $t+1$ is estimated using a kinematic motion model as follows\cite{Chen2014}:
\begin{equation}
\label{eq.odemetry}
\begin{aligned}
x(t+1) &= x(t) + \Delta D \cos(\theta(t)), \\
y(t+1) &= y(t) + \Delta D \sin(\theta(t)), \\
\theta(t+1) &= \theta(t) + \Delta \theta,
\end{aligned}
\end{equation}
where $\Delta D$ represents the traveled distance and $\Delta \theta$ denotes the change in orientation between consecutive time steps\cite{Chen2014}. Finally, localization can be realized using probabilistic methods, typically with the \ac{EKF}, which accounts for nonlinearities in the observation (cf. Eq.~\eqref{eq.observationmodel}) and odometry (cf. Eq.~\eqref{eq.odemetry}) models\cite{Bellini2002,Panzieri2005,Alves2013,Folkesson2005,Hwang2011,Chen2014}.

\subsubsubsection{Image-Based Fingerprinting}
Image-based \ac{uVLP} fingerprinting techniques capture multiple images at predefined reference points under unmodulated \acp{LED} during an offline phase, extracting distinctive spatial or appearance features using either handcrafted descriptors or learned embeddings (e.g., \acp{CNN}\cite{Zhang20199}). In the online phase, the captured image is matched to stored fingerprints via classification (e.g., softmax-based\cite{Zhang20199}) or similarity-based methods to estimate the device position. Section~\ref{subsec:imagebasedfingerprint} provides a comprehensive review of \ac{uVLP} systems employing image-based fingerprinting approaches.

\subsubsection{Demultiplexed Methods (Camera-based \ac{CF})}
In a \ac{CF}-based \ac{uVLP} system using a smartphone camera, localization proceeds in two main stages: \textit{(i)} \ac{CF} extraction from captured images or videos, and \textit{(ii)} position estimation using a suitable positioning method (e.g., fingerprinting\cite{Zhang2016}, trilateration, etc.).

\textbf{\textit{i) \ac{CF} extraction:}}  
Extraction of the \ac{CF} from unmodulated light sources (e.g., \acp{LED} or \acp{FL}) using smartphone cameras is feasible with either the front-facing sensor or the higher-resolution rear sensor, despite their limited nominal frame rates~\cite{Zhang2016,Zhang2016LiTell2,Carver2017,Zhang2019}. The front-facing camera can capture MP4 video streams, which are preferred over JPEG images because JPEG compression suppresses high-frequency content (e.g., \ac{CF})~\cite{Carver2017}. Subsequent image processing steps, including grayscale conversion, thresholding, and morphological filtering, yield an intensity vector suitable for frequency-domain analysis using demultiplexing algorithms such as \ac{FFT}, \ac{MUSIC}, and others\cite{Zhang2016,Carver2017}. This capability is enabled by the rolling-shutter read-out mechanism (as illustrated in Fig.~\ref{fig:rollingshutter}), which provides a much higher effective sampling rate\cite{Zhang2019,Zhang2016}. For a \ac{CMOS} sensor with \(N\) rows and a frame period \(t_f\), the effective sampling interval and rate are given by \( t_s = \frac{t_f}{N} \) and \( f_s = N R_f \), respectively, where \(R_f\) is the frame rate (see Fig.~\ref{fig:rollingshutter}). For instance, a Full-HD sensor (1920\(\times\)1080) operating at \(R_f = 30\) frames per second yields a row sampling rate of \(f_s = 1080 \times 30 \approx 32.4\;\mathrm{kHz}\), nearly three orders of magnitude higher than the frame rate \cite{Zhang2019,Zhang2016}. However, this remains below the Nyquist rate required to sample typical \ac{CF}s from typical light sources (e.g., \acp{FL} (80--100~kHz)), which demands \(f_s > 200~\mathrm{kHz}\)~\cite{Zhang2019}. Early studies~\cite{Zhang2016LiTell2,Zhang2016,Zhang2019} leveraged the sparsity of the \ac{CF} (often a single spectral peak) and the fact that camera sensors, despite their low sampling rates, exhibit analog bandwidths up to several hundred kilohertz. As a result, high-frequency \ac{CF}s, when sampled at a low rate, are folded back to lower frequencies due to the aliasing effect\cite{oppenheim1999discrete}. By exploiting the rolling shutter effect, boosting \ac{SNR} via sequential imaging combination, and optimizing sampling parameters (e.g., exposure time), these works demonstrated the recovery of \ac{CF}s beyond 80~kHz using the \ac{MUSIC} algorithm\cite{Zhang2016}.

Particularly, the camera output at sampling time $t$ (measured from the start of exposure) can be modeled as~\cite{Zhang2016,Zhang2016LiTell2,Zhang2019}:
\begin{equation}
S(t) = \int_{-\infty}^{+\infty} I(\zeta) h(\zeta - t) \, d\zeta,
\label{eq:lightoutputcamera}
\end{equation}
where \(I(t)\) denotes the light intensity at time \(t\), and \(h(\zeta - t)\) is a rectangular exposure window of length \(t_e\) starting at \(t\), defined as~\cite{Zhang2019}:
\begin{equation}
h(\zeta - t) =
\begin{cases}
1, & \zeta \in [t, t + t_e] \\[4pt]
0, & \text{otherwise}.
\end{cases}
\end{equation}
Applying \ac{FFT} on both sides of the Eq.~\eqref{eq:lightoutputcamera} yields~\cite{Zhang2019}:
\begin{equation}
\begin{aligned}
\mathcal{F}(S) &= \int_{-\infty}^{+\infty} I(\zeta) e^{-j 2 \pi f \zeta} \, d\zeta \cdot \int_{-\infty}^{+\infty} h(-T) e^{-j 2 \pi f T} \, dT \\
&= -\mathcal{F}(I) \cdot \mathcal{F}(h),
\end{aligned}
\label{eq:freq_response}
\end{equation}
where \( T = t - \zeta \). Now, the frequency response of the sampling process is given by~\cite{Zhang2019}:
\begin{equation}
|H(f)| = \frac{|\mathcal{F}(S)|}{|\mathcal{F}(I)|} = |\mathcal{F}(h)| = |\text{sinc}(f t_e)|.
\end{equation}
However, actual photoelectron accumulation continues during the readout time ($t_r$), effectively increasing the integration interval to $t_e + t_r$. Therefore, the refined frequency response becomes~\cite{Zhang2019}:
\begin{equation}
|H(f, t_e)| = \left|\mathrm{sinc}\left[f (t_e + t_r)\right]\right|.
\label{eq:freqresponse}
\end{equation}
Experimental validation by LiTell~\cite{Zhang2016,Zhang2016LiTell2,Zhang2019} demonstrated that the frequency response of smartphone cameras (e.g., Nexus 5) follows Eq.~\eqref{eq:freqresponse}, revealing notch frequencies at multiples of \(x/(t_e + t_r)\), where \(x = 1, 2, \dots\). If an \ac{FL}'s \ac{CF} coincides with these notches, frequency components are attenuated, degrading detection performance. To mitigate this, LiTell optimizes the exposure time \(t_e^*\) to maximize camera sensitivity to the \acp{CF}~\cite{Zhang2016,Zhang2016LiTell2,Zhang2019}:
\begin{equation}
t_e^* = \arg\max_{i} \sum_{j=1}^M \left|H\left(f_j,t_e^i\right) \right|,
\end{equation}
where \(M\) denotes the number of \acp{LED} or \acp{FL}, and \(t_e^i\) represents the exposure time corresponding to the \(i^{\text{th}}\) exposure setting. It is worth noting that \(t_r\) is a camera-specific constant that can be determined through factory calibration\cite{Zhang2019}.

Aliased frequencies ($f_a$) are typically detected using \ac{FFT} or similar spectral analysis techniques. The relationship between the aliased frequency and the true \ac{CF} is given by~\cite{Zhang2016,Zhang2016LiTell2,Zhang2019}:
\begin{equation}
f_a =
\begin{cases}
CF - N f_s, & \text{if } 0 \leq CF - N f_s \leq \dfrac{f_s}{2}, \\[6pt]
(N+1) f_s - CF, & \text{if } \dfrac{f_s}{2} < CF - N f_s < f_s,
\end{cases}
\end{equation}
where \(f_s\) is the sampling frequency of the smartphone and \(N \in \mathbb{N}\). The candidate set for the true \ac{CF} is given by~\cite{Zhang2016,Zhang2016LiTell2}:
\begin{equation}
CF \in \Bigl\{ f_h \,\big|\, f_h = N f_s \pm f_a,\;N = 0,1,2,\ldots, \quad f_h \geq 0 \Bigr\},
\label{eq:origfreqcand}
\end{equation}
where \(f_h\) denotes a candidate estimate of the true \ac{CF}~\cite{Zhang2016}.
\textbf{\textit{ii) Position estimation:}}  
Once the true \acp{CF} from \acp{LED} are accurately estimated despite aliasing, they can be used for \ac{uVLP}. A practical approach involves measuring the ground-truth \ac{CF} of each light source (e.g., \acp{FL}) with a camera beforehand and storing the location–\ac{CF} pairs in a fingerprinting database. This fingerprinting overhead is bounded by the number of light sources and typically requires only a few seconds of measurement per source\cite{Zhang2016,Zhang2016LiTell2,Zhang2019}. During the online positioning phase, the captured image is processed to extract the \acp{CF}, which are then matched to identify the corresponding \acp{LED}. Proximity-based localization is performed by associating the receiver with the identified \acp{LED}' known positions. Alternatively, the extracted \acp{CF} and their corresponding signal amplitudes can be used in trilateration or multilateration algorithms.

\subsection{\ac{uVLP} Positioning Techniques Comparison}
\label{sec:uvlp_positioning}
As illustrated in Fig.~\ref{fig:uVLPTechniquesClassification}, \ac{uVLP} techniques can be broadly categorized into \emph{undemultiplexed} and \emph{demultiplexed} approaches. Each category can employ either light-intensity receivers (e.g., \acp{PD}, \acp{ALS}, etc.) or imaging-based receivers. The undemultiplexed method, which relies on the total received light intensity (cf. Section~\ref{subsec:finglightintenisty} and Section~\ref{subsec:probabllightintensity}), is conceptually the simplest and most cost-effective. It operates by measuring the aggregate illuminance from all \ac{VL} sources~\cite{RaviFiatLuxF,Golding1999,Randall2007,Jimenez2013,Amsters2019}. However, this approach presents several drawbacks. First, lighting infrastructures are typically designed to produce an illuminance that is as homogeneous as possible, whereas effective localization requires spatial gradients in illuminance to distinguish nearby positions. Second, this approach is highly vulnerable to ambient (\ac{DC}) light. Third, not separating the signals from different \acp{LED} makes it impossible to distinguish similar zones in repetitive \ac{LED} arrangements. As such, the recorded peak total illuminance can, at best, be successfully used as light fixes in an \ac{IMU}-based~\cite{Xu2015,Jimenez2013} or a \ac{2D} robot-based~\cite{Amsters2018,Amsters2019} dead reckoning (DR) system. The work in~\cite{Zhao2017} slightly enhances this technique by considering typical light intensity sequences when moving through a building, instead of single-shot measurements, as sequences offer a higher level of uniqueness. It is clear that the above approaches do not suffice to deliver accurate, large-scale, infrastructure-free positioning. \ac{SB} sensors help mitigate these limitations by partitioning the incident spectrum into multiple wavelength bins (e.g., 18 bins in~\cite{Hu2023,Wang2023}), thereby transforming a one-dimensional intensity measurement into a multi-dimensional observation space. While this improves localization accuracy, it also increases sensor cost and complexity (cf.\ Section~\ref{subsec:SB}). Demultiplexed \ac{uVLP} techniques, such as those based on \ac{CF}, identify individual light sources by exploiting subtle differences in their spectral or spatial emission patterns. Although these techniques yield lower \ac{SNR} compared to actively modulated \ac{VLP} systems and require more sophisticated signal processing, they preserve many of the functional benefits of conventional \ac{VLP} approaches~\cite{Alijani2025} (cf.\ Section~\ref{sec:CFbaseduVLP}).

Lastly, imaging-based receivers offer rich spatial information and enable pixel-wise light field analysis. Nevertheless, their adoption in the literature remains limited (cf. Section~\ref{sec:imagbaseduVLP}), primarily due to their higher computational and energy demands, increased algorithmic complexity, and inherent latency compared to \ac{PD}-based systems~\cite{lirias3591348}. For example, while a smartphone-based \ac{CF} method in~\cite{Zhang2016,Zhang2016LiTell2} achieves sub-decimeter accuracy, the reliance on a relatively expensive and power-hungry camera, along with complex image processing pipelines, makes it unsuitable for many practical applications. In summary, the choice of positioning techniques and sensing units as receivers in \ac{uVLP} involves trade-offs among hardware complexity, cost, and achievable localization accuracy.

\section{Intensity-Based \ac{uVLP} Investigations}
\label{sec:Intensity-BaseduVLPTechniques}
Based on the \ac{uVLP} techniques classification in Fig.~\ref{fig:uVLPTechniquesClassification}, this section reviews total light \ac{IB} methods, further categorized by receiver type: \ac{PD}, \ac{ALS}, solar cell, and spectral sensors (see Table~\ref{tbl:intensity-baseduVLPFingerprinitng}, Table~\ref{tbl:uVLPProbablisticMethods}, and Table~\ref{tbl:cf-baseduVLP})

\subsection{Fingerprinting Techniques}
\label{subsec:finglightintenisty}
\subsubsection{\ac{PD}-based}
FiatLux~\cite{RaviFiatLuxF} aims to localize users and identify the room they are in using only a light sensor, without relying on maps, lighting models, or sensor fusion. It does not assume prior knowledge of the user’s location and attempts to classify among any of the 20 rooms tested. Localization is performed by matching real-time light readings to previously collected fingerprints under static artificial lighting conditions (i.e., no windows). The system achieved up to 91.2\% average room-level classification accuracy using a Bayesian method (Range-Max achieved 71.6\%) with the sensor worn on a hat. However, performance dropped when the sensor was worn as a pendant due to body occlusion: the Bayesian method achieved 80.3\% accuracy, and the Range-Max method achieved 68.1\%\cite{RaviFiatLuxF}. While promising, this approach can struggle under dynamic lighting conditions (e.g., in the presence of sunlight).

Golding et al. \cite{Golding1999} used wearable sensor data combined with \ac{ML} and fingerprinting to infer a user’s room-level position and state during navigation tasks. Their approach relied on fingerprinting to collect sensor data patterns rather than modeling from maps or lighting information. Specifically, the system fused data from four sensor modalities: \ac{3D} accelerometers (capturing motion), \ac{3D} magnetometers (detecting orientation and magnetic fields), \ac{FL} detectors (which detect the 60 Hz flicker characteristic of \ac{FL} lighting), and temperature sensors (measuring ambient room temperature). This multimodal fusion enabled localization with 98\% accuracy in correctly identifying the user’s room-level location within a representative indoor environment, with an average latency of approximately 4.8 seconds \cite{Golding1999}. However, when operated without additional extracted features or range data and using only current sensor readings, the system’s error rate sharply increased to 55\%, averaging 52 errors per visit \cite{Golding1999}. This decline underscores the system’s reliance on temporal context and fused data for accurate navigation.

\subsubsection{Solar-based}
LuxTrace (see Table~\ref{tbl:intensity-baseduVLPFingerprinitng}) is a hybrid indoor localization system that combines wearable solar cells embedded in clothing with floor-level \ac{RFID} tags\cite{Randall2007}. As the user moves along a corridor beneath regularly spaced fluorescent ceiling lights, the solar cell tracks light intensity fluctuations (peaks and valleys) caused by passing under each light. These intensity patterns are mapped via a fingerprinting model trained using coarse \ac{RFID}-based absolute position fixes to estimate the user’s relative displacement along the path. In a 7.4~m test corridor with three \acp{FL} spaced 3.7~m apart, the system recorded 14 light-induced voltage peaks, achieving a displacement accuracy of approximately 21~cm (80\% quantile) and an average error of around 0.2~m. The \ac{RFID} component, which uses carpet-like tags read at key locations (four tag templates in the test), provides coarse yet essential anchors for model calibration. This coarse positioning allows LuxTrace to maintain high accuracy based primarily on light fingerprinting, with \ac{RFID} ensuring robustness but not serving as the main source of precision\cite{Randall2007}. While innovative, the method is sensitive to several environmental factors such as receiver orientation, natural light interference, variations in light power (e.g., different behavior for 30~W or 60~W lights), aging, dust, and surface reflections \cite{Jimenez2013}. Importantly, because the proposed approach yields only relative movement, the system must rely on \ac{RFID} tags embedded in the floor to provide coarse absolute positioning, thereby ensuring accurate and drift-free user localization over time\cite{Jimenez2013}.

\subsubsection{\ac{ALS}-based}
To date, the majority of research on \ac{uVLP} has focused on smartphone-based implementations using the \ac{ALS} sensor. We classify these investigations into two categories: \ac{ALS}-only approaches and fusion-based techniques.

\subsubsubsection{\ac{ALS}-only Approaches}
In \cite{Zhao2017}, NaviLight leverages light intensities from unmodulated sources (e.g., \acp{FL}, \acp{LFL}, \acp{LED}) captured by standard mobile phones. The study showed that single-point intensity values are insufficient as unique location signatures. Instead, a vector of multiple light intensity values along the user’s path, called LightPrint, provides better spatial discrimination and temporal stability. NaviLight captures LightPrint data throughout a user's walk and matches it with a pre-built \ac{LIF} map to predict the user's location. NaviLight has two important modules: coarse-grained localization and LightPrint matching. It scales by classifying a LightPrint obtained during a walk into one of the subareas of the building's floor plan using coarse-grained localization based on \ac{KNN}. Then, LightPrint matching using \ac{DTW} is used. Before matching, a LightPrint is separated into one or numerous light intensity vectors based on established walking directions, resulting in directed LightPrints. The preferred LightPrint is then compared with the \ac{LIF} map utilizing subsequent \ac{DTW}. Finally, the most suitable match is determined based on \ac{DTW} distances utilizing a clustering technique. Through the evaluation stage in three separate environments, including an office building, a shopping mall, and an underground parking lot (with a total area of over 1000 $m^2$ and random walking for building \ac{LIF} maps), LightPrints yielded localization accuracy of 85\% for indoor locations within 0.5 m (office), 0.35 m (shopping mall), and 3.9 m (underground parking lot). Moreover, for navigation, 85\% of errors were within 1.6 m, 2.2 m, and 43 m, respectively. While promising, NaviLight is limited to sun-free indoor use, sensitive to user behavior (e.g., phone height/angle), and requires frequent \ac{LIF} map updates due to environmental changes, making deployment labor-intensive \cite{Zhao2017}.

Liu et al. \cite{Liu2019Graph} developed the PILOT system for indoor navigation using peak intensities of unmodulated luminaires represented in a virtual graph (see Table~\ref{tbl:intensity-baseduVLPFingerprinitng}). The system consists of participants, a cloud server, and navigation users. Each participant uploads their pedestrian path, represented as vertices (peaks) and directed edges between consecutive peaks. The server integrates the uploaded paths by detecting overlapping segments and creates a virtual graph. During navigation, users are tracked based on the nearest vertex corresponding to detected peak intensities from light sources. PILOT was implemented on Android smartphones (e.g., Mate 8, Google Nexus 9) held in the user’s hand as they walk around indoor spaces. The system was tested in three scenarios: a 1000 m² one-story supermarket with 7 m high luminaires spaced 3 m apart; a 2000 m² six-story mall with 4 m high luminaires spaced 2 m apart; and an 800 m² four-story office building with 3.5 m high luminaires spaced 3.6 m apart. Peak detection accuracy was 100\%, 98.3\%, and 100\%, respectively, while overlap detection accuracy improved from 0.4 to 1 as the number of edges increased from 1 to 10\cite{Liu2019Graph}.

\subsubsubsection{Fusion-based \ac{ALS} Techniques} 
DeepML, developed in \cite{Wang2018DeepML}, combines bimodal magnetic field and ambient light data for indoor localization using a deep learning approach (e.g., \ac{LSTM}). Compared to traditional fingerprinting-based methods, the deep \ac{LSTM} network only requires one set of weights trained for all locations instead of creating a separate database for each training location. This feature accelerates location recognition and reduces data storage requirements. The proposed DeepML system includes a data preprocessing module for collecting magnetic field and light intensity data using a smartphone and creating bimodal data with a sliding window method. It also has an offline training phase that involves feature extraction, the deep \ac{LSTM} network, and a softmax classifier. In the online testing phase, DeepML incorporates an improved probabilistic approach for estimating the location of the target mobile device by feeding the bimodal images into the trained deep \ac{LSTM} network. DeepML was tested with a Samsung Galaxy S7 Edge smartphone running Android 7.0 in a 6~$\times$~12 $m^2$ laboratory with 12 training locations, where each pair of neighboring locations was separated by 1.6 m, and in a 2.4~$\times$~20 $m^2$ corridor with 10 training locations along a straight line, also spaced 1.6 m apart. The experimental results showed that DeepML achieved a maximum localization error of 3.7 m and 6.5 m, respectively, which is better than the maximum localization error of using only magnetic field data (5 m and 8.2 m, respectively)\cite{Wang2018DeepML}. 

SurroundSense~\cite{Azizyan2009} (see Table~\ref{tbl:intensity-baseduVLPFingerprinitng}) is a multi-sensor fusion system that combines ambient sound, light, and other environmental features for logical localization within a fingerprinting framework. A smartphone collects environmental data, preprocesses it to reduce the transmission load, and sends it to a remote server, where a fingerprinting module matches the sensed data to known fingerprints. Evaluated across 51 stores, SurroundSense achieved an average accuracy of 87\% using all sensing modalities. However, this approach raises concerns about energy consumption and real-time responsiveness due to sensor convergence delays.

LiLo\cite{Wu2022} is a smartphone-based, sensor fusion system that combines \ac{ALS} data with smartphone (e.g., Google Nexus 5 running Android) orientation to estimate user location and recognize \ac{ADL} in key indoor areas. The system constructs a luminance field map by measuring light intensity at various positions and device orientations. It then determines the user's frequency orientation, which refers to the most frequently observed set of orientation angles (e.g., azimuth, pitch, and roll) during typical phone usage. This information is used to align new \ac{ALS} readings with the luminance map. Although the device contains only one light sensor, LiLo approximates \ac{AoA} by combining light measurements with real-time orientation data to infer the direction of incoming light relative to the phone. These features are fed into a Bayesian network classifier to produce location and activity estimates\cite{Wu2022}. Experimental validation was conducted in an 800 ft$^2$ apartment, where LiLo achieved over 94\% accuracy for both \ac{ADL} and location classification and reached an average localization error of 0.4~m. The study highlights the method’s effectiveness using only off-the-shelf smartphone sensors, though it notes that managing changes in lighting conditions throughout the day remains a challenge.

Finally, GraphSLAM, proposed in \cite{Liang2020}, is a multi-sensor fusion system that combines ambient light intensity data (through peak illuminance detection using \ac{ALS}), magnetic field, and \ac{WiFi} signals captured by a smartphone (Samsung Galaxy S5) within a fingerprinting framework. Enhanced by a \ac{KF} using \ac{PDR} data, it exploits observations of ceiling lights distributed linearly along narrow corridors to further constrain \ac{PDR} drift. The system was evaluated in two environments, including an office area (21~\acp{FL}) and a public shopping mall, through walking trajectories lasting 10 to 20 minutes and covering areas of 2000 to 4000~m$^2$. It achieved real-time localization on a smartphone in the office area, with 50th percentile accuracy of 2.30~m and 90th percentile accuracy of 3.41~m. However, the localization performance was evaluated with only one user and a single device; the usability of the signal maps on other devices and for other users has not been demonstrated\cite{Liang2020}.

\subsubsection{Spectral Sensor-based \ac{uVLP}}
\label{subsec:SB}
We categorize \ac{SB}-\ac{uVLP} investigations into two groups: those that use only spectral data for positioning, and those that fuse spectral data with other sensor modalities (e.g., \ac{BLE}, etc.).
\subsubsubsection{Spectral-Only Approaches}
Salem et al.\cite{Salem2021} developed a \ac{uVLP} system using a helmet-mounted setup with an \ac{IMU} and a \ac{VL} sensor (a single \ac{RGB} \ac{PD}) to localize a person and recognize activity. The study involved an offline phase to collect light intensity data (i.e., \ac{RSS}) in the \ac{RGB} spectral bands across 110 points in a 44.7m\textsuperscript{2} room illuminated by 10 sets of \acp{FL}, creating a calibration database. In the online phase, a participant wore the helmet to gather data at selected points~\cite{Salem2021}. Using K-means clustering, the system achieved 91.8\% accuracy in position estimation, defined as the percentage of correctly classified location labels among the 110 predefined positions~\cite{Salem2021}. However, the experiments were conducted in a controlled environment with minimal ambient light interference due to sunshades.

Wang et al.\cite{Wang2023,Wang2022} investigated the potential of using \ac{LSI} as a \ac{SB}-\ac{uVLP} system, named Spectral-Loc. Their study covered two indoor environments: a meeting room and an open office, each illuminated by \ac{ICL}, \ac{CFL}, and \ac{LED}. In their study, the object of interest (e.g., a user) wore eight spectral sensors (e.g., AS7265x) and stood at different places about one meter apart, with each location measured for 30 seconds. They first collected fingerprint data during an offline phase over five days. This data was then used to train a \ac{CNN}, which was later validated during the online measurement phase. They compared the performance of Spectral-Loc with the EHASS approach\cite{Umetsu2019}, which relies on single light intensity. The results revealed that while EHASS performed well in the small meeting room, it struggled in the larger, more complex office environment, even with eight sensors worn in various positions. Particularly, EHASS accomplished localization accuracy within one meter in the meeting room, but in the office, the 90$^{\text{th}}$ percentile error reached up to five meters. In contrast, Spectral-Loc demonstrated sub-meter accuracy in both environments, achieving this level of precision even at the 90$^{\text{th}}$ percentile\cite{Wang2023,Wang2022}.

Jagdeep et al. \cite{Singh2022} explored the potential of using Hue sensors, termed HueSense, to distinguish unmodified \acp{LED} in indoor environments illuminated by four white \acp{LED} mounted on the ceiling, along with ambient light\cite{Singh2022}. Specifically, they developed an algorithm and conducted experiments using a robot equipped with three Hue sensors on top\cite{Singh2022}. These experiments were carried out in both static scenarios (e.g., when the robot was positioned directly under an \ac{LED}) and dynamic scenarios, where the robot moved to collect light features based on the ratio of spectral power in the dominant \ac{RGB} channels\cite{Singh2022}. Their findings showed 100\% accuracy in identifying the \acp{LED} in static scenarios, while in dynamic scenarios, the accuracy dropped to 80.1\%, mainly due to the impact of ambient light introducing noise\cite{Singh2022}. 

\subsubsubsection{Spectral-Fused Techniques}
Building on the successful \ac{LED} identification using Hue sensors as reported in \cite{Singh2022}, the same authors also developed a system called BLELight, a hybrid system combining \ac{uVLP} with unmodified \acp{LED} and \ac{BLE} (see Table~\ref{tbl:intensity-baseduVLPFingerprinitng}). In particular, they implemented this hybrid indoor localization system by training a \ac{DNN} model using \ac{IL} with 10 features extracted from the \ac{BLE} signals and the power ratios at dominant wavelengths ($\lambda_R$ (red), $\lambda_G$ (green), and $\lambda_B$ (blue)). In this setup, a network of nine white \acp{LED}, along with three Hue light sensors mounted on a moving target, was used to collect light features, while four \ac{BLE} anchors were employed to gather \ac{RSS} measurements. The results demonstrated that the BLELight hybrid approach achieved better accuracy compared to each technology used individually, with improvements of 64\% over \ac{BLE} alone and 47\% over \ac{uVLP}. Moreover, the system attained a \ac{MLA} of 0.20~m, outperforming joint training methods by up to 58\%. Finally, despite the promising results, further investigation is needed to assess the full potential of BLELight, particularly in scenarios where joint training of both signals of light and \ac{BLE} in the \ac{DNN} is implemented, and the features are not equally weighted\cite{Singh2023}.

Following HueSense \cite{Singh2022}, the study in \cite{Singh20244} introduced HueLoc, a localization system utilizing unmodulated light sources (e.g., nine \acp{LED}). The system leveraged the observation that each \ac{LED} emits a slightly different color spectrum, imperceptible to the human eye but detectable by color sensors. This unique spectral signature enables the identification of light sources solely based on their emission characteristics. HueLoc employed a regression-based learning approach to model how light intensity (e.g., \ac{RGB} power) at certain wavelengths changes with distance and angle for different \acp{LED}\cite{Singh20244}. Three off-the-shelf, power-efficient single-pixel color sensors (e.g., TCS34725) were integrated with a \ac{WiFi}-enabled Arduino board to simultaneously collect \ac{RGB} channel power as receivers. The system was evaluated in both static and dynamic scenarios. In the static setup, data were collected from each \ac{LED} by measuring power variations at dominant wavelengths at various distances, gathering more than 500 samples by placing the sensor module at fixed locations within the \acp{LED}' \ac{FoV}. In the dynamic scenario, the system was tested with a moving target (e.g., a robot) in a corridor, achieving 100\% \ac{LED} detection (i.e., one-to-one \ac{LED} mapping) and decimeter-level location accuracy, with a mean localization error of 7 cm. To enhance localization accuracy, the study further explored a fusion approach that combined light signature features with four \ac{BLE} receivers spanning an area of 25~$m^2$ using a three-stage incremental learning approach. Experimental results demonstrated a 75\% improvement over \ac{BLE}-only localization accuracy, achieving 90\% accuracy within 20 cm, with a \ac{2D} mean localization error of 12 cm and a \ac{3D} mean localization error of 18.4 cm\cite{Singh20244}. However, the study also identified key limitations. Since the system was trained using data from only one \ac{LED}, variations in power levels across \acp{LED} could impact overall performance. Additionally, relocating \acp{LED} would require retraining the system, making it sensitive to setup changes. Finally, temperature fluctuations could influence the extracted \ac{RGB} power values, affecting system reliability.

Finally, Liu et al.\cite{Liu2014} introduced a positioning method using a fingerprinting approach and \ac{MLE}, called Ambilight, which utilizes ambient light based on the principle that spectral variations between different locations are distinguishable. Their study employed an iRobot-based multi-sensor positioning platform equipped with light sensors and other positioning tools like \ac{BLE}, \ac{IMU}, and \ac{LiDAR}. A spectrometer sensor was used as the receiver in experiments conducted in two linear corridors within a building. The first corridor was a closed space without natural light, while the second had both natural light and artificial illumination from four \ac{FL} lamps fixed on the ceiling. The impact of natural light in the second experiment was addressed by normalizing the intensity measurements of the \ac{FL} lamps. Specifically, the background light (defined as the light measured when the lamps were turned off) was subtracted from the composite ambient light to isolate the calibrated lamp signal. Through real-world experiments, they reported a positioning \ac{RMS} error of 0.79~m. Furthermore, they demonstrated that using light intensity measurements at a 1~Hz sampling rate improved the accuracy of \ac{SLAM} with a \ac{LiDAR} sensor, reducing positioning errors by over 70\%\cite{Liu2014}. Despite the promising results, ambient light positioning remains challenging, particularly when the background light cannot be accurately measured and separated from the total ambient light to derive calibrated intensity measurements from the fluorescent lamps\cite{Liu2014}.
\begin{table*}[]
    \centering
    \setlength{\tabcolsep}{2.0pt} 
    \renewcommand{\arraystretch}{1.2} 
    \caption{Overview of Intensity-Based (IB) \ac{uVLP} Systems Employing Fingerprinting Methods.}
    \label{tbl:intensity-baseduVLPFingerprinitng}
    \begin{adjustbox}{width=\linewidth}
    \begin{tabular}{|c|c|c|c|c|c|c|c|}
    \hline
    \rowcolor{gray!40}

    \multicolumn{2}{|c|}{\rule{0pt}{10ex} \textbf{\fontsize{30pt}{36pt}\selectfont Scheme}} & 
    \textbf{\rule{0pt}{10ex} \fontsize{30pt}{36pt}\selectfont \ac{uVLP} Transmitter} & 
    \textbf{\rule{0pt}{10ex} \fontsize{30pt}{36pt}\selectfont \ac{uVLP} Receiver} & 
    \textbf{\rule{0pt}{10ex} \fontsize{30pt}{36pt}\selectfont Research Goal(s)} & 
    \textbf{\rule{0pt}{10ex} \fontsize{30pt}{36pt}\selectfont \MC{Implementation\\Techniques}} & 
    \textbf{\rule{0pt}{10ex} \fontsize{30pt}{36pt}\selectfont Main Achievements} & 
    \textbf{\rule{0pt}{10ex} \fontsize{30pt}{36pt}\selectfont Challenges} 
    
    \\ \hline
    \noalign{\vskip 5pt} 
    \hline

    \renewcommand{\arraystretch}{1.2} 
    \Huge \multirow{20}{*}{\rotatebox[origin=c]{90}{\textbf{\textit{PD-based}}}} 
    
    & \Huge \textbf{\MC{FiatLux \\ \cite{RaviFiatLuxF}}}
    & \Huge \MC{Static lighting conditions\\ (i.e, no windows)}
    & \Huge \MC{A light sensor worn on\\ a hat or as a pendant}
    & \Huge \MC{Determining the user’s\\ position and room}
    & \Huge \MC{Three fingerprinting \\algorithms: Bayesian, \\Range-Max, and Spatial}
    & \Huge \MC{Achieved user's position and room identification accuracy\\ of 91.2\% with Bayesian and 71.6\% with Range-Max\\ (light sensor on hat); 80.3\% with Bayesian and 68.1\%\\ with Range-Max (light sensor as pendant)}
    & \Huge \MC{Vulnerable to ambient light interference\\ (e.g., light from windows) and lower accuracy\\ in pendant condition due to body blocking light}
    \\ \cline{2-8}

    & \Huge \textbf{\MC{Golding et \\ al. \cite{Golding1999}}}
    & \Huge \MC{Arrangement of the \acp{FL}}
    & \Huge \MC{Developed board with \ac{3D}\\accelerometer, magnetometer,\\ \ac{FL} detector, and temperature\\ sensor, attached to a utility\\ belt worn by the user}
    & \Huge \MC{Determining a \\person's location}
    & \Huge \MC{\ac{ML} with fingerprinting}
    & \Huge \MC{Attained 98\% room-level accuracy with a 4.8~s latency\\ for navigation tasks (i.e., user's location)}
    & \Huge \MC{Performance dropped to 55\% error rate\\ (52 errors/visit) without (fused) sensor data}
    \\ \hline
    \noalign{\vskip 5pt} 
    \hline

    \Huge\multirow{-5.5}{*}{\rotatebox[origin=c]{90}{\textbf{\textit{Solar-based}}}} 
    &  \Huge\textbf{\MC{LuxTrace\\ \cite{Randall2007}}}
    &  \Huge\MC{\ac{FL} sources on the ceiling}
    &  \Huge\MC{Small solar panel embedded\\ in the user’s clothing}
    &  \Huge\MC{Estimating relative\\ displacement (distance \\from light sources)}
    &  \Huge\MC{Fingerprinting approach \\using light intensity\\ variations and \\\ac{RFID} for coarse \\location estimates}
    &  \Huge\MC{Achieved a localization error of about 0.2 meters over an \\8-meter path (0.25~m-0.33~m distance)}
    &  \Huge\MC{The system can not work well in dynamic\\ lighting conditions; accuracy is affected by receiver\\ orientation, natural light interference, light power \\variations, aging of light fixtures, dust accumulation,\\ and reflections from surrounding objects}
    \\ \hline

    \Huge\multirow{45}{*}{\rotatebox[origin=c]{90}{\textbf{\textit{\ac{ALS}-based}}}} 
    & \Huge \textbf{\MC{NaviLight \\ \cite{Zhao2017}}}
    & \Huge \MC{130 \ac{LFL} and \ac{LED} (office)\\ 38 \ac{LFL}, \ac{LED} and \ac{CFL} (mall)\\ 20 \ac{LFL} (parking)}
    & \Huge \MC{Various Android\\ smartphones}
    & \Huge \MC{Localizing a walking\\ user's position}
    & \Huge \MC{\ac{DTW} and coarse-grained\\ localization via \ac{KNN}}
    & \Huge \MC{Obtained 85\% accuracy within 0.5~m (office), 0.35~m\\ (mall), 3.9~m (parking), and 85\% of navigation errors\\ within 1.6~m, 2.2~m, 43~m, respectively}
    & \Huge \MC{Limited to indoor environments without sunlight;\\ sensitive to user diversity (e.g., phone height and\\ angle); \ac{LIF} maps are vulnerable to environmental\\ changes (e.g., room temperature)}
    \\ \cline{2-8}

     & \Huge \textbf{\MC{PILOT \\ \cite{Liu2019Graph}}}
    & \Huge \MC{125 rectangular-shaped \\\acp{LED} (supermarket)\\2448 circular-shaped\\ \acp{LED} (shopping mall)\\80 rectangular-shaped\\ \acp{LED} (office) }
    & \Huge \MC{Four Android smartphones \\ (Huawei Mate 8, Huawei P9, \\Samsung Galaxy S5, \\and Google Nexus 9)}
    & \Huge \MC{Localizing four walking\\ participant's position\\ and light peak detection}
    & \Huge \MC{Virtual graph \\representation and \ac{DTW}}
    & \Huge \MC{Achieved light peak detection accuracy of 100\%, 98.3\%,\\ and 100\% in a supermarket, shopping mall, and office.\\ Overlap detection accuracy improved from ~0.4 to 1\\ as edges increased from 1 to 10}
    & \Huge \MC{High computational cost and power consumption}
    \\ \cline{2-8}

    & \Huge \textbf{\MC{DeepML \\ \cite{Wang2018DeepML}}}
    & \Huge \MC{Ambient light}
    & \Huge \MC{Samsung Galaxy S7 \\Edge smartphone}
    & \Huge \MC{Estimating the location \\of the mobile device}
    & \Huge \MC{Deep \ac{LSTM} network}
    & \Huge \MC{Obtained maximum localization error of 3.7 m (laboratory),\\ 6.5 m (corridor); Improvement over magnetic-only \\localization (maximum errors: 5 m and 8.2 m); reduced \\data storage requirements and accelerated location\\ recognition compared to traditional fingerprinting methods}
    & \Huge \MC{Further investigation is needed when dealing with\\ magnetic distortions (e.g., elevators) and light fluctuations \\(e.g., switching lamps), which can cause \\inconsistent localization}
    \\ \cline{2-8}

    & \Huge \textbf{\MC{SurroundSense \\ \cite{Azizyan2009}}}
    & \Huge \MC{Ambient environmental \\features (light, sound)}
    & \Huge \MC{Smartphone (sensing \ac{WiFi}, \\sound, light, etc.)}
    &  \Huge \MC{Develop a multi-fusion \\indoor localization\\ system}
    &  \Huge \MC{Fingerprinting-based\\ localization}
    &  \Huge \MC{Tested in 51 stores with an average accuracy of\\ 87\% when using all sensing modalities}
    &  \Huge \MC{High energy consumption due to multiple sensing\\ modalities; Accelerometer requires time to converge, \\limiting real-time localization performance}
    \\ \cline{2-8}

    &  \Huge \textbf{\MC{LiLo\\ \cite{Wu2022}}}
    &  \Huge \MC{Ambient light}
    &  \Huge \MC{Smartphone (e.g., Google\\ Nexus 5, Android OS)}
    &  \Huge \MC{Recognizing \ac{ADL} (e.g., \\localization on key \\active area)}
    &  \Huge \MC{\ac{ML} (Bayesian classifier) \\ \& Fingerprinting-based \\\ac{AoA}}
    &  \Huge \MC{Obtained over 94\% recognition rates for \ac{ADL} and location\\ classification, with an average location error of 0.4~m}
    & \Huge \MC{The study needs to further investigate the efficient\\ management of an illuminance field map at different\\ periods of the day}
    \\ \cline{2-8}

     &  \Huge \textbf{\MC{GraphSLAM\\ \cite{Liang2020}}}
    &  \Huge \MC{21 \acp{FL}}
    &  \Huge \MC{Smartphone (e.g., Samsung \\ Galaxy S5, Android OS)}
    &  \Huge \MC{Smartphone localization}
    &  \Huge \MC{Fingerprinting and \ac{KF}}
    &  \Huge \MC{Achieved real-time localization on a smartphone in the\\
office area, with 50th percentile accuracy of 2.30 m \\and 90th percentile accuracy of 3.41 m}
    & \Huge \MC{Localization was tested with one user and device;\\ generalizability to others remains unverified.}
    \\ \hline

    \Huge \multirow{70}{*}{\rotatebox[origin=c]{90}{\textbf{\textit{Spectral Sensor-based}}}} 
    & \Huge\textbf{\MC{Salem et al. \\ \cite{Salem2021}}} 
    & \Huge \MC{10 raster \acp{FL}\\ (OSRAM LUMILUX\\ bulbs)}
    & \Huge  \MC{Helmet with \ac{IMU} \\sensor and\\ \ac{RGB} \ac{PD}}
    & \Huge \MC{Person positioning and\\ activity recognition}
    & \Huge \MC{\ac{ML} with fingerprinting}
    & \Huge \MC{Obtained 91.81\% accuracy in correctly mapping \\the person's position}
    & \Huge \MC{Minimal ambient light impact in controlled environment;\\ \ac{IMU} data integration for activity recognition\\ planned for future work}
    \\ \cline{2-8}

    & \Huge\textbf{\MC{Spectral-Loc\\\cite{Wang2023,Wang2022}}} 
    & \Huge\MC{\ac{ICL}, \ac{CFL} and \acp{LED} \\mounted on the ceiling}
    & \Huge\MC{Eight spectra sensors\\(e.g., AS7265x\cite{SparkFun2022}) \\ worn by the user}
    & \Huge\MC{Localizing users in \\indoor environments}
    & \Huge \MC{Fingerprinting based on\\ \ac{LSI} \& using (\ac{CNN})}
    & \Huge\MC{Sub-meter accuracy at the 90$^{th}$ percentile in meeting\\ rooms and open offices; error reduced from 1.05 m\\ to 0.98 m with more spectral sensors.}
    & \Huge\MC{Performance in crowded areas (e.g., malls, etc.) requires\\ further investigation; struggled in larger, more complex\\ office environments when compared to light \ac{IB} methods\\ (e.g., EHASS\cite{Umetsu2019})} \\
    \cline{2-8}

    & \Huge\textbf{\MC{HueSense\\\cite{Singh2022}}} 
    & \Huge\MC{Four \acp{LED} mounted on \\the ceiling \&\\ ambient light}
    & \Huge\MC{Three Hue sensors}
    & \Huge\MC{Identifying unmodified \\\acp{LED} using the\\ spectral power ratios \\of \ac{RGB} channels}
    & \Huge\MC{Developed algorithm \\based on hidden colour \\features of \acp{LED}}
    & \Huge\MC{Obtained 100\% accuracy in identifying \acp{LED} in static\\ scenarios; achieved 80.1\% accuracy in dynamic scenarios}
    & \Huge\MC{Ambient light introduced noise, reducing\\ identification accuracy in dynamic scenarios} \\
    \cline{2-8}

    &\Huge\textbf{\MC{BLELight\\\cite{Singh2023}}} 
    & \Huge\MC{Nine white \acp{LED} \&\\ Four \ac{BLE} anchors}
    & \Huge\MC{Three Hue Sensors}
    & \Huge\MC{Enhancing localization\\ accuracy by combining\\ spectral-based \ac{uVLP}\\ with \ac{BLE}}
    & \Huge\MC{\ac{DNN} with  \ac{IL} trained on\\ 10 \ac{BLE} features and \\spectral power ratios\\ of \ac{RGB}}
    & \Huge\MC{Obtained 64\% improvement over \ac{BLE} alone, and 47\%\\ improvement over \ac{uVLP} alone; outperformed\\ joint training approaches by up to 58\%,\\ achieving a \ac{MLA} of 0.20~m}
    & \Huge\MC{Further research needed on joint training of light\\ and \ac{BLE} signals. The system’s performance could \\be impacted if features are not equally weighted in\\ the \ac{DNN}} \\
    \cline{2-8}

    & \Huge \textbf{\MC{HueLoc\\\cite{Singh20244}}} 
    & \Huge \MC{Nine unmodulated\\ \acp{LED}}
    & \Huge \MC{Three off-the-shelf \\single-pixel color\\ sensors (e.g., \\TCS34725)\\ integrated with a \\\ac{WiFi}-assisted\\ Arduino board}
    & \Huge \MC{Identifying \acp{LED} and\\ performing localization}
    & \Huge \MC{Regression-based learning \\method; Three-stage \ac{IL} \\fusing light signature \\features with \ac{BLE}-\ac{RSS}\\ data (Hybrid HueLoc + \\\ac{BLE})}
    & \Huge \MC{Achieved 100\% \ac{LED} detection with decimeter-level\\ precision and a mean localization error of 7 cm; improved \\localization by 75\% over \ac{BLE}-only, with 90\% accuracy \\within 20 cm; \ac{2D} mean error: 12 cm, \ac{3D} mean error:\\ 18.4 cm.}
    & \Huge \MC{Moving or relocating \acp{LED} requires retraining the \\system; temperature fluctuations affect \ac{RGB} power values,\\ impacting system reliability; system was trained with data \\from only one \ac{LED}, making it sensitive to \ac{LED}\\ power variations} \\
    \cline{2-8}

     &\Huge \textbf{\MC{Ambilight\\\cite{Liu2014}}} 
    & \Huge \MC{Four \acp{FL} mounted\\ on the ceiling \\ \& Natural light \\(sunlight in\\ open areas)}
    & \Huge \MC{Spectrometer sensor}
    & \Huge \MC{Robot positioning}
    & \Huge \MC{Fingerprinting and\\ \ac{MLE}}
    & \Huge \MC{Achieved 0.79 m \ac{RMS} error; using light intensity at\\ 1 Hz sampling rate reduced positioning errors by over \\70\% in \ac{SLAM}}
    & \Huge \MC{Difficult to extract calibrated intensity values from mixed\\ ambient light sources; sunlight variations can affect\\ accuracy despite normalization techniques} \\
    \hline
    \end{tabular}
    \end{adjustbox}
    \end{table*} 
\subsection{Probabilistic approaches}
\label{subsec:probabllightintensity}
\subsubsection{\ac{PD}-based}
Amsters et al. \cite{Amsters2019,Amsters2018} proposed using an \ac{IEKF} to estimate the pose (position and orientation) of a robot (e.g., Kobuki robot), moving with a constant forward velocity (e.g., 0.2 $m/s$). The robot was equipped with five \acp{PD} placed on the topmost platform. The reduced computational load of this approach enables its implementation in smaller devices, while its performance in highly non-linear \ac{uVLP} measurement models is improved through local iterations during the update step \cite{Amsters2019}. The study concluded that the positioning accuracy was below 0.5 m on average in most cases \cite{Amsters2019}. In the best-case scenario, the average accuracy during experiments was approximately 0.4 m, which is comparable to IDyLL \cite{Xu2015}, reporting accuracy between 0.38 m and 0.74 m, and to Lightitude (1.93 m-2.24 m) \cite{Hu2015}, which utilizes a smartphone assisted by an \ac{IMU}. However, despite the promise of their approach in estimating a mobile robot's pose, the experimental results neither provide real-time position estimation nor extend beyond a small test area within the environment \cite{Amsters2019} (see Table~\ref{tbl:uVLPProbablisticMethods}).

\subsubsection{\ac{ALS}-based \ac{uVLP}-\ac{PDR} Fusion}
Literature review of \ac{ALS}-based probabilistic methods shows they all fuse with \ac{PDR} to improve localization accuracy and robustness. Xu et al. \cite{Xu2015} proposed IDyLL, an \ac{IB}-\ac{uVLP} system leveraging smartphone light sensors (e.g., \ac{ALS}) and \ac{PDR}.
The system estimates user displacement via step counts, stride length, velocity, and heading from \ac{IMU} data, while detecting peaks in ambient light intensity measured by the phone’s \ac{PD} as the user approaches or moves away from luminaires. These light peaks correspond to luminaire locations and serve as spatial landmarks for localization. IDyLL processes the light signal by low-pass filtering, squaring, and smoothing to enhance peak detection, which helps mitigate ambiguity caused by unmodulated luminaires and sensor noise. By fusing detected light peaks with \ac{PDR}, the system refines position estimates, achieving mean localization errors between 0.38 m and 0.74 m across multiple building floors\cite{Xu2015}. However, IDyLL assumes spatially varying lighting patterns, making it vulnerable to homogeneous lighting, high ceiling-mounted luminaires, sunlight interference, shadows, and requires the phone to be face-up. Its computational cost also limits applicability on resource-constrained devices. To address degraded performance from missing or burned-out luminaires, Xu et al. \cite{QiangXu2015} integrated a burned-out luminaire detection method using simulation-based illumination modeling and dynamic time warping to match measured and predicted light patterns. This enhancement improved location accuracy by 15–60\%, reducing average error from 0.4 m to 0.34 m in tests with up to five burned-out luminaires along a 140 m indoor trajectory.

Lightitude is a \ac{uVLP} system proposed in\cite{Hu2015} based on the observation that \ac{RLS} varies depending on the light sources' locations. The study developed a light strength model to associate each position in a six-dimensional variable space (roll, pitch, yaw, and the \ac{3D} coordinates of the receiving device) with \ac{RLS}. Using a \ac{PF} as the localization module, the system collected data from inertial sensors and \ac{RLS} on an Android OS. Testing was conducted using Google Nexus 4 and Nexus 7 smartphones in an office space of approximately 720~$m^2$ with 39 \acp{CFL} and a school library floor of about 960~$m^2$ with 85 \acp{CFL}. The system achieved a \ac{MLA} of 1.93~m and 2.24~m, respectively\cite{Hu2015}. Furthermore, under shading conditions caused by obstacles and human presence, Lightitude could locate users between target shelves in the library with an accuracy of 85\%.

Jimenez presented a light-matching-based indoor localization method in \cite{Jimenez2013} that utilizes lamp location, orientation, and shape information, along with inverse-square law modeling of illumination intensity, to estimate user positions. The system's ability to distinguish between different light sources is limited as it relies on asymmetries and irregularities in luminaire placement. Localization performance was primarily evaluated through simulation (with 15 unmodulated lamps), with experimental validation conducted in a complementary study\cite{Jimnez2014}. The study was tested on walking trajectories inside a building, where the user carried a foot-mounted \ac{IMU} (e.g., Xsens MTi), an \ac{RFID} reader, and a Samsung Galaxy S3 smartphone\cite{Jimnez2014}. The smartphone, running an Android application, collected data from external inertial sensors, the \ac{RFID} reader, and internal sensors (e.g., \ac{WiFi}, sound, etc.). By integrating an \ac{PF} with \ac{PDR} data and updating it using measurement models that incorporate data from \ac{RFID}, \ac{WiFi}, light, \ac{GPS}, and magnetometers, light-matching achieved a location error of less than 1 meter in most cases. The proposed system depends on a map of the building, the coordinates and characteristics of the lamps, and the locations of \ac{RFID} and \ac{WiFi} access points\cite{Jimnez2014}.

In a study by Zhuang et al. \cite{Zhuang2024}, the system collects data from the smartphone's internal \ac{IMU} and magnetometer, using the Mahony complementary filtering algorithm to compute the smartphone's attitude. By calculating the dot product between the normal vector of the smartphone plane and the vector connecting the smartphone to the \ac{LED}, they determined the cosine of the incident angle. The smartphone is rotated to different angles using the \ac{ALS} to capture optical power at various attitudes. They then convert the positioning equations into a multi-objective optimization problem, determining the smartphone's \ac{3D} coordinates using the ordinary least squares (OLS) method. The proposed method is evaluated in a $5\,\mathrm{m} \times 5\,\mathrm{m} \times 2.2\,\mathrm{m}$ area, showing an average \ac{3D} positioning error of 0.2~m, with the 90$^{\text{th}}$ percentile error within 0.46~m\cite{Zhuang2024}. However, the method supports only static positioning and uses a single \ac{LED}, limiting its applicability to dynamic and full \ac{uVLP} scenarios.

The study in \cite{Wang20188} presents LiMag, which combines \ac{uVLP} with a magnetic field. While \ac{VL} offers strong location differentiation but is sensitive to environmental changes, the magnetic field provides stability but weaker differentiation in large areas\cite{Wang20188,Wang2018DeepML}. LiMag integrates lighting infrastructure and the magnetic field with a cloud server and a smartphone (light sensor, magnetometer, gyroscope, and accelerometer) for localization. It employs a \ac{PF} and a long-trajectory calibration scheme employing \ac{DTW} to improve tracking via a pedestrian motion model and hybrid fingerprinting. The pedestrian model detects steps, estimates stride length, and determines walking direction, while trajectory calibration improves \ac{PF} convergence. LiMag was tested in three indoor scenarios: an office (520~$m^2$, 362 \ac{CFL}/\acp{LED}), a shopping mall (1,800~$m^2$, 660 \ac{CFL}/\ac{LFL}/\acp{LED}), and a parking area (640~$m^2$, 120 \ac{CFL}/\acp{LED}). Employing Android smartphones (Huawei Mate 9, Samsung S6) with five participants, it achieved a 75$^{th}$ percentile localization accuracy of 1.8 m, 2.2 m, and 3.3 m, respectively. Under sunlight interference, localization accuracy was 2.5 m (hybrid/LiMag), 3 m (light fingerprints), and 5.6 m (magnetic-only)\cite{Wang20188}. 

Finally, in \cite{Yang2025}, uLiDR was proposed as a smartphone-based pedestrian localization system by integrating \ac{uVLP} and \ac{PDR}. It consists of four modules: \ac{AoA}-based initialization for estimating a precise initial position for \ac{PDR}, \ac{uVLP}, \ac{PDR}, and an optimization-based integration framework (e.g., factor graph). Specifically, given the initial \ac{LED} position, the pedestrian rotates the smartphone while keeping it at the same location to collect \ac{IMU} readings and \ac{RSS} values. The \ac{AoA}-based scheme leverages the \ac{PD}'s attitude, obtained from the Attitude and Heading Reference System (AHRS) algorithm, and \ac{RSS} values to estimate the smartphone's \ac{3D} position. During walking, after acquiring \ac{RSS} values from the \ac{uVLP} system and verifying their reliability, uLiDR uses the attitude from AHRS to compensate for the smartphone's tilt and then integrates the \ac{RSS} values with step length and attitude into the optimization-based positioning framework. This integration framework accounts for \ac{PD} tilt and refines the smartphone's position estimation. To assess the static positioning performance of uLiDR, the study was conducted in a $5\,\mathrm{m} \times 5\,\mathrm{m}$ room with one unmodulated \ac{LED} installed on the ceiling at a height of 2.2~m and a distance of 1.44~m from the smartphone, covering 16 test points. The study achieved an average positioning error of 24.77~cm and 28.85~cm for \ac{2D} and \ac{3D}, respectively\cite{Yang2025}. Moreover, uLiDR was evaluated in two dynamic scenarios: an indoor corridor (total path length: 138.49~m) with twelve unmodulated \acp{LED} and an outdoor area with four unmodulated road lamps serving as beacons spaced approximately 10–20~m apart. For evaluation, the study used \ac{LiDAR} mounted on a helmet worn by the walking participant to provide a highly accurate reference trajectory as ground truth. In the indoor corridor, uLiDR achieved an average positioning error of 0.49~m, outperforming IDyLL\cite{Xu2015} (1.08~m) and \ac{PDR} (2.09~m)\cite{Yang2025}. In the outdoor experiment, the average positioning errors were 1.90~m for uLiDR, 4.39~m for IDyLL\cite{Xu2015}, and 5.38~m for \ac{PDR}\cite{Yang2025}. While promising, uLiDR has several drawbacks. First, it can be influenced by high walking speeds (e.g., above 1.3~m/s) and different participants, as variations in walking patterns impact the \ac{PDR} system's accuracy in step detection, heading estimation, and step length estimation\cite{Yang2025}. Moreover, the outdoor experiment was conducted at night, and further evaluation is needed under daylight conditions with ambient light interference.

\begin{table*}
    \centering
    \setlength{\tabcolsep}{2.0pt} 
    \caption{Overview of Intensity-Based (IB) \ac{uVLP} Systems Employing Probabilistic Methods.}
    \label{tbl:uVLPProbablisticMethods}
    \begin{adjustbox}{width=\linewidth}
    \renewcommand{\arraystretch}{1.3} 
    \begin{tabular}{|c|c|c|c|c|c|c|c|}
    \hline
    \rowcolor{gray!40}
    \multicolumn{2}{|c|}{\rule{0pt}{10ex} \textbf{\fontsize{30pt}{36pt}\selectfont Scheme}} & 
    \textbf{\rule{0pt}{10ex} \fontsize{30pt}{36pt}\selectfont \ac{uVLP} Transmitter} & 
    \textbf{\rule{0pt}{10ex} \fontsize{30pt}{36pt}\selectfont \ac{uVLP} Receiver} & 
    \textbf{\rule{0pt}{10ex} \fontsize{30pt}{36pt}\selectfont Research Goal(s)} & 
    \textbf{\rule{0pt}{10ex} \fontsize{30pt}{36pt}\selectfont \MC{Implementation\\ Techniques}} & 
    \textbf{\rule{0pt}{10ex} \fontsize{30pt}{36pt}\selectfont Main Achievements} & 
    \textbf{\rule{0pt}{10ex} \fontsize{30pt}{36pt}\selectfont Challenges} 
    \\ \hline
    \noalign{\vskip 5pt} 
    \hline

    \renewcommand{\arraystretch}{1.2} 

    \multirow{-4.8}{*}{\rotatebox[origin=c]{90}{\textbf{\textit{\Huge PD-based}}}} 
    & \Huge  \textbf{\MC{Amsters et \\ al.\cite{Amsters2019,Amsters2018}}}
    & \Huge  \MC{Four \acp{LED} mounted\\ on the ceiling}
    & \Huge  \MC{Five \acp{PD} are placed on \\the topmost platform of a robot}
    & \Huge  \MC{Estimate the pose \\(position\and\\ orientation)\\ of a mobile robot}
    & \Huge  \MC{\ac{IEKF} using initial pose \\and light intensity \\measurements from \\five \acp{PD}}
    & \Huge  \MC{Achieved 0.4~m accuracy in estimating the mobile robot’s\\ pose, comparable to IDyLL (0.38–0.74 m) \cite{Xu2015} and\\ Lightitude (1.93-2.24~m) \cite{Hu2015}}
    & \Huge  \MC{Lacks real-time estimation, limited to a\\ small test area, and requires an initial pose estimate}
    \\ \hline

    \multirow{65}{*}{\rotatebox[origin=c]{90}{\textbf{\textit{\Huge \ac{ALS}-based}}}} 

    & \Huge \textbf{\MC{IDyLL \\ \cite{Xu2015}}}
    & \Huge \MC{57, 42 and 21 \acp{LED} \\in three different areas}
    & \Huge \MC{An Android OS\\ smartphone camera}
    & \Huge \MC{Localizing a walking\\ user's position \\and luminary detection}
    & \Huge \MC{\ac{PF} with sensor\\ fusion (e.g., \ac{PDR} \\and others)}
    & \Huge \MC{Achieved mean localization errors of 0.38~m, 0.42~m,\\ and 0.74~m across three different floors of buildings;\\ Obtained average precision of 95\% and recall\\ of 92\% for luminary detection}
    & \Huge \MC{Assumes detectable lighting variations (e.g.,\\ homogeneous lighting, high luminaries),\\ sensitive to environmental interference\\ (e.g., shadowing), orientation changes cause\\ false alarms, requires switching to base\\ \ac{PDR} in certain scenarios}
    \\ \cline{2-8}

    & \Huge \textbf{\MC{Lightitude \\ \cite{Hu2015}}}
    & \Huge \MC{39 \ac{CFL} (Office space) \\ and 85 \acp{CFL} (school library)}
    & \Huge \MC{An Android OS\\ smartphone camera}
    & \Huge \MC{Localizing a walking\\ user's position}
    & \Huge \MC{\ac{PF} based on \ac{RLS}}
    & \Huge \MC{Achieved mean accuracy of 1.93~m in the office\\ and 2.24~m in the library}
    & \Huge \MC{The proposed system needs to be evaluated in the\\ presence of ambient light interference}
    \\ \cline{2-8}
    
    & \Huge \textbf{\MC{Jimenez et\\ al. \cite{Jimenez2013}}}
    & \Huge \MC{15 \acp{LED} (simulation) \\ 98 \acp{LED} (experimental)}
    & \Huge \MC{Smartphone (Samsung Galaxy S3) \\with external inertial sensors\\ (e.g., foot-mounted \ac{IMU}, etc.)}
    & \Huge \MC{Estimating the location \\of the moving participant}
    & \Huge \MC{\ac{PF} fusion with \ac{PDR} \\and measurement models \\(light, \ac{RFID}, etc.)}
    & \Huge \MC{Yielded location errors of less than 1 m in most\\ cases during experimental validation}
    & \Huge \MC{The system relies on a building map, lamp \\coordinates and characteristics, and \ac{RFID}\\/\ac{WiFi} access point locations}
    \\ \cline{2-8}

    &  \Huge \textbf{\MC{Zhuang et \\al. \cite{Zhuang2024}}}
    &  \Huge \MC{An unmodulated \ac{LED}}
    &  \Huge \MC{Smartphone’s built-in \ac{ALS}}
    &  \Huge \MC{Determining the \\smartphone's \\\ac{3D} coordinates}
    &  \Huge \MC{Mahony complementary \\filtering algorithm \\ + Ordinary Least\\ Squares}
    &  \Huge \MC{Achieved an average \ac{3D} positioning error of 0.2 m,\\ with the 90$^{th}$ percentile error within 0.46 m}
    &  \Huge \MC{It currently supports only static positioning,\\ limiting its use in dynamic environments}
    \\ \cline{2-8}

    & \Huge \textbf{\MC{LiMag \\ \cite{Wang20188}}}
    & \Huge \MC{362 \acp{CFL}/\acp{LED} (office) \\ 660 \acp{CFL}/\acp{LFL}/\acp{LED} (mall)\\ 120 \acp{CFL}/\acp{LED} (parking)}
    & \Huge \MC{Smartphone (\ac{ALS}, magnetometer,\\ gyroscope, accelerometer)}
    & \Huge \MC{Localizing five walking\\ participant's position}   
    & \Huge \MC{\ac{PF}, \ac{DTW}, and \\hybrid fingerprinting \\using magnetic field\\ and light intensity}
    & \Huge \MC{Achieved a 75$^{th}$ percentile localization accuracy of\\ 1.8 m and 2.2 m, respectively, in the office and mall}
    & \Huge \MC{Reliance on walking detection and step counting,\\ which can limit accuracy in some scenarios}
    \\ \cline{2-8}

    & \Huge \textbf{\MC{uLiDR \\ \cite{Yang2025}}}
    & \Huge \MC{12 \acp{LED} (indoor test) \\ 4 road lamps (outdoor test)}    
    & \Huge \MC{An Android smartphone \\ (e.g., Samsung Galaxy S7)}    
    & \Huge \MC{Smartphone's position\\ estimating}
    & \Huge \MC{\ac{AoA}-based fusion \\with \ac{PDR}}    
    & \Huge \MC{Obtianed 24.77 cm (\ac{2D}) and 28.85 cm (\ac{3D}) static\\ positioning errors in a 5~m~$\times$~5~m room; For dynamic\\ positioning, it achieved 0.49~m error in an indoor corridor\\ (138.49~m path), outperforming IDyLL (1.08~m)\cite{Xu2015} and\\ \ac{PDR} (2.09 m), and 1.90 m error in an outdoor area,\\ outperforming IDyLL (4.39 m)\cite{Xu2015} and \ac{PDR} (5.38 m)}    
    & \Huge \MC{Struggles with adjacent \ac{LED} interference;\\ Limited to underground/night (Sunlight Sensitivity);\\ Affected by atypical walking (complex trajectories)}    
    \\ \hline
    \end{tabular}
\end{adjustbox}
\end{table*}
\subsection{\ac{CF}-based Techniques}
\label{sec:CFbaseduVLP}
To date, \ac{CF}-based \ac{uVLP} systems using light intensity receivers (see Fig.~\ref{fig:uVLPTechniquesClassification}) have mainly relied on \acp{PD}. This section reviews these \ac{PD}-based approaches in detail. Note that \ac{CF}-based \ac{uVLP} techniques leveraging imaging receivers are discussed separately in Section~\ref{subsec:CF-baseduVLPImage}. 

Following this line of research (see Table~\ref{tbl:cf-baseduVLP}), Hamidi-Rad et al. \cite{HamidiRad2017} investigated the use of \ac{ML} classification methods (e.g., \ac{KNN} and \ac{CNN}) for identifying light bulbs (e.g., determining their associated locations), including \acp{LED} and \acp{CFL}. To end this, they designed a light fingerprinting system collecting light signals at a sampling frequency of 1 MHz. By employing signal overlapping segmentation, the segmented data was normalized and then processed using \ac{FFT} for feature extraction. The extracted features were subsequently used in \ac{KNN} and \ac{CNN} models in a real-world experiment. In the first experiment, which involved classifying five \acp{CFL} and one \ac{LED}, \ac{KNN} achieved an accuracy of 93.11\%, while \ac{CNN} achieved 97.10\%. In a second experiment with eight \acp{LED}, \ac{KNN} achieved 93.04\%, and \ac{CNN} achieved 94.48\%. However, when the training and testing data were collected two days apart, the classification accuracy dropped to 76.11\% for \ac{CNN} and 76.41\% for \ac{KNN}. Moreover, the study concluded that \ac{CFL} bulbs exhibit greater variability in their high-frequency switching patterns compared to \ac{LED} bulbs\cite{HamidiRad2017}.

A study reported in \cite{Bastiaens2020} found that the observed light power at the \acp{CF} (e.g., above 80~kHz) can provide adequate \ac{SNR} for accurate decimeter-level positioning when applying the \ac{MUSIC} algorithm to the \ac{RSS} of light in real-world scenarios\cite{Bastiaens2020,Bastiaens2020(2)}. This research thoroughly examined the stability of the \ac{CF} with light dimming and across five different types of \acp{LED} over time. After successfully evaluating the \ac{CF}'s stability, the authors employed a demodulation approach that utilized \ac{FFT} and zero padding. The paper also explored various \ac{2D} localization techniques, including trilateration-based localization, \ac{CMD} localization, \ac{MBF} localization, and simultaneous positioning and orientation (SPAO). The results demonstrate that \ac{CF}-based \ac{uVLP} incurs only a moderate accuracy loss compared to conventional \ac{VLP}. For point source-like \acp{LED} in a $4\,\mathrm{m} \times 4\,\mathrm{m}$ room, trilateration-based median $p_{50}$ and $p_{90}$ errors increase from 5.3~cm to 7.9~cm and from 9.6~cm to 15.6~cm, respectively. However, applying a robust \ac{MBF} method improves \ac{uVLP} accuracy to 5.0~cm ($p_{50}$) and 10.6~cm ($p_{90}$). Compared to \ac{VLP}'s 3.5~cm ($p_{50}$) and 6.8~cm ($p_{90}$), \ac{uVLP} achieves comparable performance while offering lower cost and higher maintained illuminance, highlighting its strong practical potential. However, the measurements and subsequent simulations also revealed \ac{uVLP}'s limitations for low-latency industrial tracking applications. Additionally, the study did not address all potential demodulation and filtering techniques, suggesting that further investigation into alternative methods may be necessary to enhance the positioning accuracy of \ac{uVLP}\cite{Bastiaens2020,Bastiaens2020(2)}. 

Zhang et al.\cite{Zhang2016LiTell2} proposed LiTell2, a \ac{uVLP} system that combines \ac{AoA} information with \acp{PD} through a specially designed light sensor. LiTell2 extracts unique features (e.g., \ac{CF}) from existing \ac{FL} or \ac{LED} lights and uses two co-located \acp{PD} with varying angular responses to derive \ac{AoA}, achieving precise localization with minimal fingerprinting costs. LiTell2’s capabilities were validated through real-world experiments, later named Pulsar, which demonstrated its performance in \ac{2D} localization\cite{Zhang2017}. Tests were conducted in environments with 64 to 157 light sources, involving various \acp{CF}. Results showed a median error of 5~cm for horizontal $(x,y)$ and 20~cm for height $(h)$ when Pulsar's sensor was placed 2~m below a 1.5~m-spaced light array. At higher ceilings (e.g., 4~m), the errors increased slightly to 6~cm $(x,y)$ and 31~cm $(h)$, while extreme tests at 6~m heights showed systematic errors exceeding 1~m due to weaker \ac{RSS}. The lateral light array separation also impacted accuracy, with errors increasing from 7~cm to 1.27~m as spacing grew from 1.5~m to 2.5~m. Comparisons between 4 compact \acp{CFL} and 4 T8 tube \acp{FL} with diffusive covers (1.2~m apart) revealed that Pulsar’s localization degraded by only 10~cm when switching from \acp{CFL} to tube \acp{FL}. This highlights the need for further investigation into Pulsar's robustness and precision with varied light configurations\cite{Zhang2017}.

Finally, Bastiaens et al. \cite{Bastiaens2022} benchmarked the \ac{2D} positioning performance of aSDS-TWR (asymmetric double-sided two-way ranging) \ac{UWB} and \ac{RSS}-based \ac{uVLP} in \ac{LoS} and various \ac{NLoS} conditions. Using comprehensive positioning data gathered over an $8\,\mathrm{m} \times 6\,\mathrm{m}$ evaluation area with 15 \ac{VLP}-enabled point-source \acp{LED} and eight \ac{UWB} anchors, they compared accuracy against a high-precision ground truth system (e.g., eight MoCap cameras). The \ac{uVLP} \ac{PD} receiver and \ac{UWB} tag were mounted on a height-adjustable cart for stateless (multilateration) and stateful (Kalman filter) localization. In \ac{LoS}-dominated conditions, both \ac{UWB} and \ac{VLP} provided accurate positioning, with median ($p_{50}$) and 90$^{th}$ percentile ($p_{90}$) errors of 5~cm and 10~cm, respectively. Near obstacles, \ac{VLP} maintained decimeter-level $p_{90}$ accuracy, while \ac{UWB} exhibited errors in the 10–20 cm range due to multipath effects. Achieving a $p_{50}$ of 21.3 cm at a 25 Hz update rate demonstrated \ac{uVLP}’s viability beyond lab settings\cite{Bastiaens2022}. While promising, the study was conducted in a controlled laboratory environment and needs further evaluation in challenging conditions, such as human blockage, outages, storage racks, and ambient light interference\cite{Bastiaens2022}.
\begin{table*} 
    \centering
    \setlength{\tabcolsep}{2.0pt} 
    \renewcommand{\arraystretch}{1.2} 
    \caption{Summary of \ac{CF}-Based \ac{uVLP} Systems Using Light Intensity and Imaging Receivers.}
    \label{tbl:cf-baseduVLP}
    \begin{adjustbox}{width=\linewidth}
        \begin{tabular}{|c|c|c|c|c|c|c|c|}
        \hline
        \rowcolor{gray!50}
        \multicolumn{2}{|c|}{\rule{0pt}{10ex} \textbf{\fontsize{30pt}{36pt}\selectfont Scheme}} & 
        \textbf{\rule{0pt}{10ex} \fontsize{30pt}{36pt}\selectfont \ac{uVLP} Transmitter} & 
        \textbf{\rule{0pt}{10ex} \fontsize{30pt}{36pt}\selectfont \ac{uVLP} Receiver} & 
        \textbf{\rule{0pt}{10ex} \fontsize{30pt}{36pt}\selectfont Research Goal(s)} & 
        \textbf{\rule{0pt}{10ex} \fontsize{30pt}{36pt}\selectfont \MC{Implementation\\ Techniques}} & 
        \textbf{\rule{0pt}{10ex} \fontsize{30pt}{36pt}\selectfont Main Achievements} & 
        \textbf{\rule{0pt}{10ex} \fontsize{30pt}{36pt}\selectfont Challenges} 
        \\ \hline
        \noalign{\vskip 5pt} 
        \hline
    
    \renewcommand{\arraystretch}{1.2} 
    \Huge \multirow{45}{*}{\rotatebox[origin=c]{90}{\textbf{\textit{PD-based}}}} 
    & \Huge\textbf{\MC{Hamidi-Rad et\\ al.\cite{HamidiRad2017}}}
    & \Huge\MC{5 \acp{CFL} and 9 \acp{LED}\\ bulbs}
    & \Huge\MC{High-frequency light sensor \\(TSL14S), 8-bit/sample \ac{ADC} \\oscilloscope (PicoScope 2000) \\linked with Raspberry Pi}
    & \Huge\MC{Identifying light bulbs and\\ determining their associated \\locations using \ac{ML}\\ classification}
    &\Huge \MC{Feature extraction \\using \ac{FFT} and\\ classification using\\ \ac{KNN} and \ac{CNN}}
    & \Huge\MC{Achieved light source classification accuracy of 93.11\%\\ (\ac{CFL} + \ac{LED}) and 93.04\% (\acp{LED}) with \ac{KNN}, and 97.10\%\\ (\ac{CFL} + \ac{LED}) and 94.48\% (\acp{LED}) with \ac{CNN}. Demonstrated\\ that \ac{CFL} bulbs exhibit greater variability in switching \\patterns compared to (\acp{LED})}
    & \Huge\MC{Classification accuracy dropped to 76\% when\\ training and testing data were collected two\\ days apart}    
    \\ \cline{2-8}

    & \Huge \textbf{\MC{Bastiaens et \\ al.\cite{Bastiaens2020,Bastiaens2020(2)}}}
    & \Huge \MC{Five types of \acp{LED} \\(circular, rectangular, \\etc.)}
    & \Huge \MC{A Thorlabs PDA36A2}
    & \Huge \MC{Estimating the \\receiver position}
    & \Huge \MC{\ac{MUSIC} and \ac{FFT} along \\with \ac{MBF}, trilateration, \\and the \ac{CMD} localization\\ algorithm}
    & \Huge \MC{Achieved decimeter-level accuracy in real-world scenarios; $p_{50}$\\ \ac{RMS} error: 3.5 cm → 5.0 cm (with four point-source-like \acp{LED}\\ in a 4~m $\times$ 4~m room); $p_{90}$ error improvement: 10.6 cm → 6.8 cm \\when \ac{VLP} enhanced \ac{uVLP}; positioning feasible within a 2.25 m \\range without modifying lighting infrastructure}
    & \Huge \MC{Low-latency industrial tracking remains \\challenging; limited scale (2.25 m) affects \\broader applicability; unexplored demodulation\\ and filtering techniques might further\\ improve accuracy}
    \\ \cline{2-8}

    & \Huge\textbf{\MC{LiTell2/Pulsar \\ \cite{Zhang2016LiTell2}/ \cite{Zhang2017}}}
    & \Huge\MC{\ac{FL} and \ac{LED} \\lights (64 to 157 \\light sources)}
    & \Huge\MC{Two co-located \acp{PD} with\\ varying angular responses}
    & \Huge\MC{Enhancing \ac{uVLP} accuracy \\by combining \ac{AoA} \\information with \ac{PD}-based\\ sensing while minimizing\\ fingerprinting costs}
    & \Huge\MC{Extracting \acp{CF} from \ac{FL} \\or \ac{LED} lights, \\fingerprinting algorithm,\\ \ac{AoA}-based localization}
    & \Huge\MC{Median localization error: 5 cm $(x,y)$, 20 cm (h) at 2~m\\ height with 1.5~m-spaced lights; Errors: 6 cm $(x,y)$, 31 cm (h) \\at 4~m ceiling height; accuracy degradation: 7 cm → 1.27 m as \\light spacing increases 1.5~m → 2.5~m; localization error +10 cm\\ when switching from\\ \acp{CFL} to tube \acp{FL}}
    & \Huge\MC{Systematic errors $>$ 1~m at 6~m height due to\\ weaker \ac{RSS}; increased localization error with \\greater light array spacing; further investigation\\ needed for robustness across different light\\ configurations}
    \\ \cline{2-8}

    & \Huge\textbf{\MC{Bastiaens et \\al. \cite{Bastiaens2022}}}
    & \Huge\MC{15 \ac{VLP}-enabled \\point-source \acp{LED}}
    & \Huge\MC{A \ac{PD} receiver for \ac{uVLP}\\ (and \ac{UWB} tag for\\ \ac{UWB} positioning)}
    & \Huge\MC{Benchmarking \ac{2D} \\positioning of aSDS-TWR \\\ac{UWB} vs. \ac{RSS}-based \ac{uVLP} \\in \ac{LoS} and \ac{NLoS}}
    & \Huge\MC{Stateless (multilateration) \\and stateful (Kalman \\filter) localization}
    & \Huge\MC{In \ac{LoS} conditions, both \ac{UWB} and \ac{VLP} achieved 5 cm ($p_{50}$) and\\ 10 cm ($p_{90}$) accuracy; near obstacles, \ac{VLP} maintained decimeter\\level $p_{90}$ accuracy, while \ac{UWB} had errors between 10–20 cm due to\\ path loss and multipath effects; \ac{uVLP} achieved a $p_{50}$ of 21.3 cm at a\\ 25 Hz update rate, demonstrating its viability \\beyond lab environments}
    & \Huge\MC{The study was conducted in a controlled lab\\ environment; further evaluation is needed in \\real-world conditions, including human\\ blockage, outages, storage racks, and ambient \\light interference}    
    \\ \hline
    \noalign{\vskip 5pt} 
    \hline
  
    \Huge\multirow{15}{*}{\rotatebox[origin=c]{90}{\textbf{\textit{Camera-based}}}} 
    & \Huge\textbf{\MC{LiTell\\\cite{Zhang2016,Zhang2019}}}
    & \Huge\MC{162 \acp{FL} spaced 1.5 to \\3 meters apart}
    & \Huge\MC{Four types of smartphone \\cameras (e.g., \\Nexus 5, etc.)}    
    & \Huge\MC{\ac{FL} identification \& \\Indoor positioning}
    & \Huge\MC{Fingerprinting algorithm \\using location-related \\\acp{CF} from light sources}   
    & \Huge\MC{Achieved meter-level granularity sufficient for navigation; mean\\ accuracy of 90.3\%, with a variation of 11\% from best to worst case; \\successfully distinguished \acp{FL} along the path in indoor environments \\with densely placed lights}
    & \Huge\MC{It is unsuitable for environments with\\ significant temperature variations, which\\ affect the \ac{CF} feature. Moreover, it has a high \\misidentification rate (40\%) and requires\\ high-resolution rear cameras\\ }
    \\\cline{2-8}
   
    & \Huge\textbf{\MC{Carver et \\ al.\cite{Carver2017}}}
    & \Huge\MC{Two types of \ac{FL} \\shapes, looped\\ and tube}
    & \Huge\MC{Three types of smartphones\\ camera (e.g., Google Nexus 5)}    
    & \Huge\MC{Light identification\\ using smartphone \\front-facing cameras}
    & \Huge\MC{Image processing\\ and \ac{FFT} algorithm}
    & \Huge\MC{Demonstrated the potential of using smartphone front-facing cameras\\ for \ac{CF} from \acp{FL}}
    & \Huge\MC{Requires further research to address the\\ limitations of ambient light and enhance the \\robustness of the \ac{CF} extraction for \\front-facing camera-based uVLP}
    \\  
    \hline
    \end{tabular}
    \end{adjustbox}
    \end{table*}

\subsection{Lessons Learned and Takeaways}
This section summarizes the lessons learned exclusively for undemultiplexed light-intensity-based localization methods, particularly fingerprinting and probabilistic approaches. For \ac{CF}-based \ac{uVLP} methods employing intensity receivers or image sensors, we refer to Section~\ref{subsec:DemuxuVLP}.

A comprehensive analysis of the studies in Table~\ref{tbl:intensity-baseduVLPFingerprinitng} reveals that a significant portion of \ac{uVLP} research relies on total received light intensity measurements combined with fingerprinting techniques \cite{Randall2007, RaviFiatLuxF, Golding1999, Wang2018DeepML, Zhao2017, Liu2019Graph, Wu2022, Azizyan2009}. This trend originates from the absence of modulation in undemultiplexed \ac{uVLP}, preventing differentiation of individual contributions from multiple \acp{LED}. Consequently, traditional localization methods such as triangulation or trilateration are inapplicable. Recent advances, however, have enabled demultiplexing in \ac{uVLP} through exploitation of the \ac{CF} of \acp{LED}, facilitating more advanced positioning techniques (see Table~\ref{tbl:cf-baseduVLP}). Despite fingerprinting's popularity, it has several limitations: it typically requires labor-intensive offline training, is sensitive to environmental dynamics (e.g., lighting fluctuations, furniture rearrangements), and encounters scalability issues in large or dynamic settings. 

Studies such as NaviLight \cite{Zhao2017} and others \cite{Amsters2018, Amsters2019} have demonstrated that static single-point intensity measurements alone are inadequate for precise positioning. Instead, capturing intensity sequences along user trajectories introduces beneficial spatial and temporal variability, enhancing distinguishability and robustness \cite{Zhao2017}. Nevertheless, fingerprinting remains practical and effective for \ac{uVLP}, offering simplicity and achieving meter-level accuracy under typical indoor conditions. Four primary sensor categories have been employed for fingerprinting (see Fig.~\ref{fig:uVLPTechniquesClassification}): \acp{PD} \cite{RaviFiatLuxF, Golding1999}, \acp{ALS} (e.g., LiLo \cite{Wu2022}, NaviLight \cite{Zhao2017}, PILOT \cite{Liu2019Graph}), spectral sensors (e.g., HueLoc \cite{Singh20244}, HueSense \cite{Singh2022}), and solar-based receivers (e.g., LuxTrace \cite{Randall2007}).  \ac{ALS}-based \ac{uVLP} has gained interest due to its widespread integration in smartphones. Methods employing \acp{ALS}, either alone~\cite{Zhao2017, Liu2019Graph} or as part of multi-sensor fusion frameworks (e.g., DeepML~\cite{Wang2018DeepML}, which improved localization accuracy from 5~m with magnetic fields alone to 3.7~m when combined with \ac{uVLP}; SurroundSense~\cite{Azizyan2009}, which increased accuracy from 70\% using \ac{WiFi} alone to 87\% by incorporating environmental data such as sound, light, and color), have significantly enhanced positioning accuracy. Further improvements are achieved by integrating raw intensity data probabilistically (e.g., via \ac{EKF}). Studies such as uLiDR\cite{Yang2025} (fusion with \ac{AoA} and \ac{PDR}), IDyLL\cite{Xu2015} (fusion with \ac{PDR}), and LiMag\cite{Wang20188} (fusion with magnetic fields) have reported centimeter- to meter-level accuracy. Under sunlight interference, LiMag\cite{Wang20188} reported 2.5~m accuracy when combining magnetic and light intensity data, compared to 3~m with fingerprinting alone, and 5.6~m using only magnetic data.

Moreover, the studies summarized in Table~\ref{tbl:intensity-baseduVLPFingerprinitng} highlight \ac{SB}-\ac{uVLP} as a promising research direction. Unlike \ac{IB}-\ac{uVLP}, which relies on total received intensity (a low-cost but accuracy-limited approach due to external influences\cite{Alijani2025}) \ac{SB}-\ac{uVLP} enhances discrimination by analyzing intensity per wavelength bin (e.g., 18 bins\cite{Wang2023,Wang2022}). This improves performance but increases sensor cost and complexity\cite{Alijani2025}. In \ac{SB}-\ac{uVLP}, capturing accurate \ac{LSI} is challenging due to spatial variability (attenuation, occlusions) and temporal dynamics (ambient and artificial lighting variations) \cite{Hu2024}. Thus, creating detailed \ac{LSI} maps becomes a labor-intensive process in indoor \ac{LSI}-based \ac{IoT} applications (e.g., material estimation\cite{Hu2024}). HueSense \cite{Singh2022}, for instance, achieved 100\% static accuracy but dropped to 80.1\% under dynamic ambient conditions. Fusion methods combining \ac{uVLP} with \ac{BLE} (e.g., BLELight \cite{Singh2023}, HueLoc \cite{Singh20244}) or techniques like Mixture of Gaussians (MoG) for ambient light classification (Iris \cite{Hu2023}) enhance robustness. In this regard, Spectral-Loc \cite{Wang2022,Wang2023} showed additional spectral sensors reduce localization error from 1.05~m to 0.98~m (90th percentile), maintaining sub-meter accuracy at the 75th percentile despite additional indoor light interference. Finally, solar-based \ac{uVLP} offers great potential for green \ac{IoT} and battery-less \ac{6G} applications but remains underexplored. LuxTrace \cite{Randall2007} highlights promising results, with diverse spectral responses from various solar cell materials potentially enhancing localization accuracy and efficiency \cite{Wang2024}. Accurate characterization of both intensity and spectral variations is critical for optimizing power conversion and positioning performance \cite{Ma2020}. Therefore, future research should focus on detailed characterization and broader exploration of solar-cell-based \ac{uVLP} systems.

\section{Imaging-Based \ac{uVLP} Studies}
\label{sec:imagbaseduVLP}
This section reviews \ac{uVLP} investigations that utilize image sensors (\ac{CCD} and \ac{CMOS}), as classified in in Fig.~\ref{fig:uVLPTechniquesClassification}.
\subsection{Undemultiplexed Methods}
\subsubsection{Light Source Localization via Visual Landmarks}
\label{subsec:ImageprocessingUVLP}
In \cite{Facchinetti1995SelfPositioningRN}, an early \ac{uVLP} approach was proposed for robot self-positioning using images of non-modulated ceiling lights (e.g., \acp{FL}) and their structural features, which typically consist of homogeneously oriented perpendicular patterns. A top-mounted camera captured ceiling images during movement, which were processed to estimate the dominant orientation. To resolve the 90° ambiguity, four predefined orientation templates (one per quadrant) were compared with the current image, and the best match was used to determine the robot’s \ac{2D} position and orientation. By limiting the template orientation range, the algorithm achieved faster processing and millimeter-level accuracy in a 70 cm triangular path test\cite{Facchinetti1995SelfPositioningRN}.

The study in \cite{Launay} presents an indoor navigation system for a mobile robot using a ceiling-facing camera and unmodified lights (e.g., 241 \acp{FL}) for pose correction. The robot estimates its pose via odometry and periodically corrects it when ceiling lights are detected. To map light positions to the robot’s frame, images are taken from two known positions under each light. Light sources are identified through histogram-based thresholding and pixel brightness checks, and pose is estimated using moment-based features after distortion correction. A consistent map is built offline for route planning. The system enabled 90~m navigation in corridors under 10~m wide by correcting odometry errors. However, it assumes one light per image, requires a known initial position and lamp database, lacks multi-hypothesis support, and fails under strong sunlight without computationally intensive processing.

Chen et al.\cite{Chen2014} proposed a vision-based indoor localization method that uses the corners of rectangular \ac{FL} lampshades on the ceiling as visual features. An upward-facing camera on a robot captures grayscale images (640$\times$480), which are segmented using intensity histogram thresholding and morphological operations to extract the lamp corners. In a \(5.6\,\text{m} \times 4.5\,\text{m}\) area with a ceiling height of 2.5\text{m} and nine installed \ac{FL} units, 36 corners were used as reference features. An \ac{EKF} was employed to fuse visual observations with odometry. To evaluate performance, both positioning accuracy and robustness tests were conducted\cite{Chen2014}. The method outperformed Launay’s\cite{launay2001fluorescent} and Wang’s~\cite{Wang2005} approaches, achieving maximum position errors of 0.021m (x-axis) and 0.017m (y-axis), and a peak orientation error of 2.44°. In a robustness test with some lamps turned off, the method sustained low localization errors, with a maximum orientation error of 3.17° and position errors of 0.078m (x-axis) and 0.032m (y-axis)~\cite{Chen2014}.

Finally, in \cite{WANG1997}, an autonomous electric wheelchair system with self-localization capability was proposed, employing a camera linked to two image processing boards integrated with an \ac{FPGA} for high-speed image processing. Real-world tests were conducted in a room with 18 ceiling-mounted \acp{FL}, taking into account environmental obstacles. The system successfully identified 15 landmarks across 20 wheelchair movements, achieving a maximum position error of 0.35m and an orientation error of 17° at the final destination. However, the need for an \ac{FPGA} and a high-speed microprocessor may reduce the system’s suitability for low-power, battery-operated applications.

\subsubsection{Image-Based Fingerprinting}
\label{subsec:imagebasedfingerprint}
The iLAMP system, introduced by Zhu et al.~\cite{Zhu2017}, is a smartphone-based \ac{uVLP} solution that utilizes spatial radiance patterns (i.e., unique radiance intensity distributions across the surfaces of light fixtures) captured by a smartphone camera for indoor positioning. It employs an image-based fingerprinting technique, extracting and matching distinctive radiance textures. Additional intrinsic visual features, including color patterns derived from \ac{RGB} values and infrared-to-visible intensity ratios, further enhance the positioning accuracy. These supplementary features are captured via the smartphone's \ac{ALS} and camera \ac{RGB} outputs, without requiring modifications to existing lighting systems such as \acp{LED} and \acp{FL}\cite{Zhu2017}. Field tests conducted using a Nexus 5X smartphone at approximately 1.2 meters above the floor revealed varying accuracy depending on ceiling height and environmental complexity. Tests covered diverse scenarios including offices (588 \acp{FL} at a 2.5-meter ceiling, mixed 190 \acp{LED} and \acp{FL} at a 3-meter ceiling), a semi-open parking ramp (232 \acp{FL} at a 2.5-meter ceiling), and a retail store (330 \acp{FL} at a 6-meter ceiling). Using spatial radiance patterns alone, accuracy exceeded 96\% in typical buildings with ceilings up to 3 meters but dropped to about 82\% in environments with a 6-meter ceiling. Incorporating intrinsic visual features improved overall matching accuracy to nearly 100\%.

LiTell \cite{Zhang2016,Zhang2019} primarily suffers from low reliability due to its reliance on flickering frequencies as identification features, restricting it to \acp{FL} \cite{Zhang20199}. To address this, Auto-Litell \cite{Zhang20199} was proposed, featuring offline and online phases. In the offline phase, a \ac{CNN} extracts stable, subtle features from images of unmodulated light signatures. The online phase then utilizes softmax-based fingerprint matching against a pre-established location database to estimate user positions. Experiments validated Auto-Litell’s robustness and accuracy across varied environments: an office (50 \ac{FL}, 3~m ceiling), shopping mall (512~\acp{LED} and 57~\ac{FL}, 3.4~m ceiling), basement parking (60~\ac{FL}, 2 m ceiling), and supermarket (120~
\ac{FL}, 2.4 m ceiling). The system consistently maintained nearly 100\% accuracy in identifying lights, even under random smartphone tilting conditions. Compared to iLAMP \cite{Zhu2017}, which dropped to 30\% accuracy under tilt and averaged 28.1\% across diverse light sources, Auto-Litell achieved an average accuracy of 96.36\%. In summary, Auto-Litell demonstrates superior robustness and accuracy across multiple lighting conditions and orientations but still requires full deployment on mobile devices for practical end-user applications.

The LiLoc system \cite{Wang2018} identifies unmodified light sources based on \ac{HVF}. Instead of analyzing frequency components, LiLoc extracts distinct radiance patterns emitted by light fixtures from images captured by a smartphone’s front-facing camera. LiLoc operates by utilizing \ac{ALS} and applying an augmented \ac{PF} algorithm to map light intensity fingerprints from various sources, including \ac{FL}, \ac{ICL}, and \acp{LED}, for localization purposes. To further improve positioning accuracy, LiLoc integrates a beacon identification algorithm with \ac{PDR} and \acp{HVF} extraction. Experimental evaluations in diverse indoor environments demonstrated that LiLoc achieved median localization errors of 2.2 m in office spaces, 3.5 m in shopping malls, and 3 m in parking areas, even in the presence of different light sources and sunlight interference. However, this method depends on access to RAW pixel data, a feature not universally available across smartphones, and the extracted patterns may exhibit inconsistencies across different devices \cite{Wang2018}.

Finally, Liang et al. \cite{Liang2022} proposed an inertial-aided \ac{VLP} system that combines modulated and unmodulated ceiling-mounted \acp{LED} to reduce retrofitting costs, leveraging fingerprinting-based localization. The system uses a rolling shutter camera and an unsynchronized \ac{IMU} to detect mapped landmarks (fingerprints) from modulated \acp{LED} and opportunistic features from unmodulated \acp{LED} or undecodable ones through blob detection and optional \ac{VLC} decoding. These features are fed into a stochastic cloning sliding-window \ac{EKF}. Mapped landmarks, acting as predefined fingerprints, provide absolute position constraints to correct drift, while opportunistic features help improve relative pose estimation during landmark outages. In extensive tests conducted in a room with 25 \acp{LED}, the system achieved centimeter-level accuracy and 1-2° orientation error across 14 datasets. The system outperformed baseline methods under challenging lighting conditions, but its performance is limited by the reliance on circular \acp{LED}, fixed ceiling lighting, and constrained roll/pitch variations.
\begin{table*} 
    \centering
    \setlength{\tabcolsep}{2.0pt} 
    \renewcommand{\arraystretch}{1.2} 
    \caption{Overview of Imaging-Based \ac{uVLP} Systems Employing Undemultiplexed Techniques}
    \label{tbl:imagebaseduVLP}
    \begin{adjustbox}{width=\linewidth}
        \begin{tabular}{|c|c|c|c|c|c|c|c|}
        \hline
        \rowcolor{gray!50}
        \multicolumn{2}{|c|}{\rule{0pt}{10ex} \textbf{\fontsize{30pt}{36pt}\selectfont Scheme}} & 
        \textbf{\rule{0pt}{10ex} \fontsize{30pt}{36pt}\selectfont \ac{uVLP} Transmitter} & 
        \textbf{\rule{0pt}{10ex} \fontsize{30pt}{36pt}\selectfont \ac{uVLP} Receiver} & 
        \textbf{\rule{0pt}{10ex} \fontsize{30pt}{36pt}\selectfont Research Goal(s)} & 
        \textbf{\rule{0pt}{10ex} \fontsize{30pt}{36pt}\selectfont \MC{Implementation\\ Techniques}} & 
        \textbf{\rule{0pt}{10ex} \fontsize{30pt}{36pt}\selectfont Main Achievements} & 
        \textbf{\rule{0pt}{10ex} \fontsize{30pt}{36pt}\selectfont Challenges} 
        \\ \hline
        \noalign{\vskip 5pt} 
        \hline
    \renewcommand{\arraystretch}{1.2} 
    \Huge \multirow{35}{*}{\rotatebox[origin=c]{90}{\textbf{\textit{Image-based Processing}}}} 
    & \Huge \textbf{\MC{Facchinetti et \\ al. \cite{Facchinetti1995SelfPositioningRN}}}
    & \Huge \MC{Ceiling-mounted \\lighting landmarks \\(e.g., \acp{FL})}
    & \Huge \MC{Vertically mounted\\ camera on the robot}
    & \Huge \MC{Self-positioning \\for a mobile robot}
    & \Huge \MC{Computer vision based\\ algorithm (image\\ processing)}
    & \Huge \MC{Reported millimeter-level positioning accuracy\\ with a 70 cm mobile robot in a 54~m triangular race}
    & \Huge \MC{Requires comprehensive image recording \\and processing}
    \\ \cline{2-8}
    & \Huge \textbf{\MC{Launay et\\ al. \cite{Launay}}}
    & \Huge \MC{241 \acp{FL}}
    & \Huge \MC{One installed upward-facing \\camera on the robot}
    & \Huge \MC{Enhance indoor robot \\navigation by reducing\\ odometry drift}
    & \Huge \MC{Fusing odometry with \\light-based pose\\ correction}
    & \Huge \MC{Effective landmark-based navigation within 10~m\\ spacing in a 2.5~m-wide corridor; Error compensation \\via distorted environmental maps; 90~m navigation\\ using only odometry and light-based corrections}
    & \Huge \MC{Requires a known initial position and lacks\\ multiple hypotheses, reducing reliability in \\complex environments. Strong sunlight near \\windows hinders landmark detection, causing \\navigation failures}
    \\ \cline{2-8}
    & \Huge \textbf{\MC{Chen et\\ al. \cite{Chen2014}}}
    & \Huge \MC{9 \acp{FL}}
    & \Huge \MC{One installed upward-facing \\camera on the robot}
    & \Huge \MC{Indoor robot \\navigation}
    & \Huge \MC{Fusing odometry with \\light-based pose\\ correction (\ac{EKF})}
    & \Huge \MC{The method outperformed those of Launay~\cite{launay2001fluorescent}\\ and Wang~\cite{Wang2005}, with maximum position\\ errors of 0.021m (x) and 0.017m (y), and \\a peak orientation error of 2.44°.}
    & \Huge \MC{Evaluation needed under ambient light interference\\ and with other light sources (e.g., \acp{LED})}
    \\ \cline{2-8}
    & \Huge \textbf{\MC{Wang et\\ al. \cite{WANG1997}}}
    & \Huge \MC{18 \acp{FL}}
    & \Huge \MC{One installed upward-facing \\camera on the wheelchair}
    & \Huge \MC{Indoor wheelchair \\navigation}
    & \Huge \MC{Image processing}
    & \Huge \MC{Achieved a maximum position error of 0.35m\\ and a peak orientation error of 17°}
    & \Huge \MC{Evaluation needed under ambient light interference\\ and with other light sources (e.g., \acp{LED})}
  \\ \hline
    \Huge\multirow{60}{*}{\rotatebox[origin=c]{90}{\textbf{\textit{Image-based Fingerprinting}}}} 
    & \Huge\textbf{\MC{iLAMP\\\cite{Zhu2017}}}
    & \Huge\MC{Office 1 (588 \acp{FL}, \\2.5 m height), \\\&  Office 2 (190 \acp{LED}\\ + \acp{FL}, 3 m ceiling), \\\& parking ramp (232 \\\acp{FL}, 2.5 m ceiling), \\\& retail store (330 \acp{FL},\\ 6 m ceiling)}
    & \Huge\MC{Smartphone camera (e.g., \\Nexus 5X) + Smartphone \\\ac{ALS}}
    & \Huge\MC{Smartphone's \ac{3D} \\positioning (heading \\direction)}
    & \Huge\MC{Sensor assisted \\photogrammetry \\technique and \ac{DTW}}
    & \Huge\MC{Obtained over 96\% accuracy in recognizing light\\ landmarks in buildings with up to 3m ceiling height, and \\over 82\% accuracy in buildings with 6m ceiling height. \\Combining main and assisted features improved matching \\accuracy to nearly 100\%, with a p90 \ac{3D} location error of 3.5 \\cm (2.8°) for smartphone positioning.}
    & \Huge\MC{High computational cost for feature extraction\\ and matching, resulting in significant \\latency (e.g., hundreds of milliseconds).\\ Tested with a single \ac{FL} and 25 random\\ target positions, limiting robustness\\ across diverse conditions} 
    \\\cline{2-8}
    & \Huge\textbf{\MC{Auto-Litell \\ \cite{Zhang20199}}}
    & \Huge\MC{Office with 50 \acp{FL} \\(3~m ceiling height) \\\&  Mall with 512 \acp{LED} \\and 57 \acp{FL} (3.4~m \\ceiling height) \& \\Basement parking lot \\with 60 \acp{FL} (2~m\\ ceiling height) \&\\ Supermarket with \\120 \acp{FL} (2.4~m\\ ceiling height)}
    & \Huge\MC{Smartphone camera (e.g., \\LG G4, OPPO R17,\\ and Lenovo PHAB2 Pro)}
    & \Huge\MC{Light identification \& \\user localization}
    & \Huge\MC{Deep learning model \\(e.g., \ac{CNN}) and image \\processing based \\on features map}
    & \Huge\MC{Achieved nearly 100\% accuracy under both horizontal and \\tilted conditions, while iLAMP's accuracy dropped significantly \\to 30\% under tilted conditions; attained a mean accuracy of \\96.36\% across all four types of light sources (\ac{LED}, \ac{FL}, etc.)}
    & \Huge\MC{Full-fledged mobile device deployment is \\required to make Auto-LiTell accessible and \\usable for end-users despite its high precision}
    \\\cline{2-8}
    & \Huge\textbf{\MC{LiLoc \\  \cite{Wang2018}}}
    & \Huge \MC{\acp{FL}, \acp{ICL}, and \acp{LED}}
    & \Huge \MC{Smartphone’s front-facing \\camera  (\ac{ALS} sensor)}
    & \Huge \MC{Estimating target\\ location}
    & \Huge \MC{\ac{PF} algorithm\\ fusion with \ac{PDR}  and\\ fingerprinting}
    & \Huge \MC{Achieved median localization errors of 2.2 m (office), \\3.5 m (mall), 3 m (parking); Handled sunlight\\ interference and varied light sources}
    & \Huge \MC{Dependence on RAW pixel data, inconsistent across\\ devices; Radiance pattern extraction may vary\\ between devices}
    \\\cline{2-8}
    & \Huge\textbf{\MC{Liang et \\ al. \cite{Liang2022}}}
    & \Huge\MC{Both modulate and \\unmodulated \acp{LED}}    
    & \Huge\MC{Smartphone camera \\(i.e., rolling shutter camera)}    
    & \Huge\MC{Feasibility of using\\ both modulated and \\unmodulated light \\for \ac{VLP}}    
    & \Huge\MC{Imgae processing and \\\ac{EKF} estimator}    
    & \Huge\MC{Achieved a global positioning accuracy of a few \\centimeters and orientation accuracy of up to 1–2 degrees}    
    & \Huge\MC{The system is limited to circular-shaped \acp{LED}\\ and needs broader testing; \ac{LED} power \\variations affect performance; requires ceiling\\ lights and an upward-facing camera; limited \\handling of roll and pitch orientation changes} \\ 
    \hline
    \end{tabular}
    \end{adjustbox}
    \end{table*}
    
\subsection{Demultiplexed Techniques (Camera-based \ac{CF} Studies)}
\label{subsec:CF-baseduVLPImage}
Smartphone cameras are rarely used for \ac{CF}-based \ac{uVLP} due to their limited sampling rate (see Table~\ref{tbl:cf-baseduVLP}). However, LiTell~\cite{Zhang2016,Zhang2019,Zhang2016LiTell2} demonstrates that smartphone cameras (e.g., Nexus 5) can capture unique \ac{CF} signatures ($>$80\,kHz) from \acp{FL} despite low frame rates, by exploiting the camera’s analog bandwidth, exposure optimization, and \ac{SNR} boosting via  sequential imaging combination. LiTell collected 2-second samples from 162 \acp{FL} over 3 hours (5 measurements per lamp), using a linear search algorithm to match extracted \acp{CF} with a fingerprint database (e.g., a map of locations and \ac{CF}). In real indoor environments with densely spaced \acp{FL} (1.5–3\,m apart), it achieved meter-level granularity sufficient for navigation, even without a sub-light localization module and could distinguish the \acp{FL} along the path, with a mean accuracy of 90.3\% and a variation of 11\% from the best to worst case. With a 1.2\,m \ac{FL} mounted at a height of 1.8\,m above the phone, LiTell achieved 10\,cm accuracy (90\% of cases) when the phone was level and stationary, and a 15\,cm median accuracy during user motion, with 90\% of errors within 25\,cm. Accuracy degrades with distance due to pixel footprint expansion but remains within 0.5\,m under typical indoor conditions. The \ac{SNR} also drops with distance but stays above 3\,dB at 2\,m, the typical operational range. LiTell relies on rear cameras (more capable than front cameras), which require the phone to be held horizontally. When using front cameras, more images are needed to maintain adequate \ac{SNR}. However, its performance is limited by camera dynamic range (max ~2.5 m), resulting in blind spots, and environmental temperature variations can affect \ac{CF} stability. Power consumption is another concern for practical deployment\cite{Zhang2016,Zhang2019,Wang2018}.

The primary limitation of LiTell \cite{Zhang2019,Zhang2016} lies in its dependence on a smartphone’s rear camera, which results in a poor user experience due to the constant need to switch between the camera and the screen\cite{Carver2017}. With advancements in front-facing camera technology, there is growing potential for a \ac{CF}-based \ac{uVLP} system that leverages front-facing cameras, enabling a more seamless and user-friendly experience \cite{Carver2017}. Carver et al. \cite{Carver2017} explored this direction by using the front-facing camera of a conventional smartphone (e.g., Google Nexus 5) to distinguish between unmodified \acp{FL}. Their study involved 40-minute MP4 recordings of two common \ac{FL} shapes (looped and tube) positioned 131~cm above the device in a controlled environment. An image processing algorithm extracted aliased \acp{CF} by applying a \ac{FFT} to the temporal light intensity vector, which was obtained by summing pixel intensities within the isolated light region across video frames\cite{Carver2017}. Although this method shows promise, it was limited to classifying light sources based on extracted \acp{CF} and did not evaluate performance for positioning. Additionally, the impact of ambient lighting was not considered, as each light was tested individually. This aspect warrants further investigation for practical deployment.

\subsection{Lessons Learned and Takeaways}
\label{sec:LessonsCF-baseduVLP}
\subsubsection{Undemultiplexed-based Methods}
A review of \ac{uVLP} studies reveals that image sensors (\ac{CCD} and \ac{CMOS}), as summarized in Table~\ref{tbl:imagebaseduVLP}, are used far less frequently than light-intensity-based receivers due to high power consumption, latency, and processing demands. Nonetheless, they offer high spatial resolution for precise localization. Ceiling-pattern-based methods, primarily explored using \acp{FL} within \ac{SLAM} frameworks, require camera calibration and intensive image processing~\cite{Chen2014,Facchinetti1995SelfPositioningRN,Launay,WANG1997}. Systems like iLAMP \cite{Zhu2017}, Auto-Litell \cite{Zhang20199}, and LiLoc\cite{Wang2018}, which use image-based fingerprinting, highlight the value of combining multiple visual features to improve robustness under varying lighting and orientation conditions. However, reliance on fixed layouts, ceiling height, and limited tilt, along with high power, computation, and low sampling rates, limits real-time applicability. For example, while iLAMP\cite{Zhu2017} offers lower power consumption (0.93 W) compared to similar systems like Litell (2.7 W), its computationally intensive image-processing tasks result in relatively high latencies, typically in the range of hundreds of milliseconds. In detailed tests, iLAMP demonstrated a 90$^{\text{th}}$ percentile \ac{3D} positioning error of 3.5 cm and a heading error of 2.8° when tested using \acp{FL}. 

Finally, the tilt of the smartphone (and also \ac{PD}) during \ac{uVLP} operation must be accounted for\cite{Yang2025,Zhu2017,Zhu2017}. Experimental results show that Auto-LiTell \cite{Zhang20199} and iLAMP \cite{Zhu2017} demonstrate nearly 100\% accuracy in horizontal alignment \cite{Zhang20199}. However, when the smartphone is tilted, iLAMP’s accuracy drops dramatically to 30\%, with an average accuracy of 28.1\% across different lighting conditions (e.g., \acp{LED}, \acp{FL}) \cite{Zhang20199}. One solution explored in Auto-LiTell uses a \ac{CNN} to extract \acp{HVF} from the \acp{LED}, maintaining near-100\% accuracy even with random smartphone tilt. This approach highlights the potential of \acp{CNN} to improve system robustness to orientation changes\cite{Zhang20199}. Another potential solution could be fusing \ac{uVLP} with data from an \ac{IMU} to track the device’s attitude, which could further enhance the system's accuracy and stability by compensating for the tilt and providing more reliable orientation information\cite{Yang2025}. 

\subsubsection{Demultiplexed-based Methods}
\label{subsec:DemuxuVLP}
The \ac{CF}-\ac{uVLP} (or demultiplexed \ac{uVLP}) approach, which demultiplexes signals based on the \ac{LED}'s \ac{CF}, has been investigated using both \acp{PD}~\cite{Bastiaens2020, Bastiaens2022,HamidiRad2017,Zhang2017} and cameras~\cite{Zhang2016,Carver2017} (see Table~\ref{tbl:cf-baseduVLP}). Although it offers lower \ac{SNR} and requires more complex signal processing than modulated \ac{VLP}, it retains many of its benefits~\cite{Alijani2025}. Bastiaens et al.~\cite{Bastiaens2022} experimentally compared three \ac{IPS} technologies: \ac{UWB}, modulated \ac{VLP}, and \ac{CF}-based \ac{uVLP}. In an $8\,\mathrm{m} \times 6\,\mathrm{m}$ testbed, \ac{VLP} and \ac{UWB} achieved median \ac{2D} positioning errors of 5~cm and 10~cm (90th percentile) under \ac{LoS} conditions, respectively. \ac{CF}-based \ac{uVLP}, while infrastructure-free and low-cost, provided a median accuracy of approximately 20~cm. Under \ac{NLoS} conditions, \ac{VLP} outperformed \ac{UWB}, with minimal degradation observed in \ac{CF}-based \ac{uVLP} when object tilt was limited. Additionally, Pulsar~\cite{Zhang2017} highlighted the impact of receiver placement on accuracy. Moving the receiver from 2~m beneath a 1.5~m-spaced ceiling light array to a 4~m ceiling slightly increased the median error from 5~cm to 6~cm, while at 6~m height, errors exceeded 1~m due to reduced \ac{RSS}. These results emphasize the sensitivity of \ac{CF}-based \ac{uVLP} to receiver height and environmental variations in signal strength. 

Overall, \ac{CF}-based \ac{uVLP} can achieve centimeter- to meter-level localization accuracy, as demonstrated in systems such as LiTell~\cite{Zhang2016,Zhang2019} using smartphone cameras, or in \ac{PD}-based \ac{CF} approaches~\cite{Bastiaens2020, Bastiaens2022,Zhang2017,Zhang2016LiTell2}. However, when using camera-based \ac{CF} techniques, one must consider the power consumption of smartphone cameras, which are inherently power-intensive \cite{Zhu2017,Zhang20199}. Rear cameras provide better quality at higher power consumption, while front cameras are more energy-efficient but need more images to compensate for lower quality \cite{Zhang2016,Zhang2019}. Carver et al. \cite{Carver2017} showed that JPEG compression on front cameras removes high-frequency details (not applicable for \ac{CF}-based \ac{uVLP}), making MP4 a preferable alternative. Smartphone cameras typically suffer from low sampling rates and limited sensitivity to high-frequency light signals due to noise and hardware constraints \cite{Zhang20199,Zhang2016LiTell2,Zhang2016}. Their dynamic range is dominated by low-frequency, high-intensity signals, while hardware-induced noise (e.g., pixel heterogeneity, salt-and-pepper noise) and physical obstructions (e.g., latticed covers) further degrade \ac{CF} signals\cite{Zhang20199,Zhang2016LiTell2,Zhang2016}. LiTell  addresses these issues by optimizing camera sampling, leveraging the rolling shutter effect, and amplifying features to boost \ac{SNR}, enabling \ac{CF} detection above 80kHz. Its simple location-matching approach stores \ac{CF}-location pairs in a database, allowing smartphones to identify nearby \acp{FL} and achieve meter-level accuracy, even in dense environments—similar to \ac{PD}-based \ac{uVLP} systems in \cite{Bastiaens2020}. However, the effective working range of LiTell\cite{Zhang2016} is limited to 2.5 meters due to the low dynamic range of commercial smartphones' cameras, which could result in several blind spots\cite{Wang20188}. Finally, it is worth mentioning that LiTell’s method is limited to \acp{FL}, which restricts its usage in other indoor navigation systems\cite{Singh20244}. It can only discriminate lights with a low accuracy of 60\% even when using high-resolution rear cameras\cite{Zhang20199}.

\section{Open Challenges and Future Roadmap}
\label{Sec:FutureDirections}
\subsection{Key Challenges Ahead}
\label{sec:challenges}
\subsubsection{Challenges with built-in \acp{PD} of smartphones}
Smartphone-integrated \acp{PD} (e.g., \ac{ALS} sensors) have very low sampling rates due to system and power constraints, often falling below 10~Hz. For instance, the Huawei Mate 30 utilizes the ALS-STK3638 \ac{PD} with a sampling rate of only 2.85~Hz; the Samsung Galaxy S7 uses the YMD49XX \ac{PD} with a rate of 5.55~Hz; and the Xiaomi Redmi K40 features the TCS3701 \ac{PD} with a rate of 3.7~Hz~\cite{Sato2022,Otsuka2025,Wang2019,Yang2025}. Such low sampling rates are inadequate to satisfy the Nyquist–Shannon sampling theorem: the received optical power from each \ac{LED} at its \ac{CF} cannot be accurately identified or extracted using demultiplexing techniques (e.g., \ac{FFT}), as \acp{CF} are typically in the frequency range of 30–160~kHz~\cite{Zhang2016,Zhang2019,Zhang2017,Zhang20199,lirias3591348,Bastiaens2020}. When sampled at sub-Nyquist rates, aliasing occurs—high-frequency components are misrepresented as lower-frequency artifacts in the sampled signal (e.g., in \ac{ALS}-based \ac{uVLP} systems)~\cite{Sato2022,Otsuka2025,Wang2019,Yang2025}. This distorts the received signal and significantly degrades the accuracy of identifying the \acp{CF} of \acp{LED}, which are essential for \ac{CF}-based \ac{uVLP} systems. However, using higher sampling rates or exploiting aliased \ac{CF} components in a controlled manner could enable smartphones to distinguish between different \ac{LED} sources, thereby supporting practical light-based positioning.

\subsubsection{Impact of Tilt and Alignment in \ac{uVLP} Systems}
\ac{Rx} tilt, such as a smartphone or \ac{PD} rotating during user movement (e.g., walking), can degrade the performance of \ac{uVLP} systems by altering the angle of incidence between incoming light and the receiver, which affects the \ac{RSS} \cite{Yang2025,Zhuang2024}. This variation becomes critical when using amplitude-based methods, such as trilateration, in \ac{CF}-based \ac{uVLP} systems \cite{Zhuang2024,Bastiaens2020}. Smartphones, with their built-in \acp{IMU}, provide the necessary information to implement compensation algorithms, as explained in Section~\ref{sec:imusolution}. For proximity-based approaches, the effect is minimal, as the \ac{CF} itself remains unaffected by tilt. In \ac{SB}-\ac{uVLP}, tilt can distort the spectral response by shifting the relative contributions of multiple light sources. In smartphone-based systems, tilt also affects the sensor’s \ac{FoV}, potentially occluding some \acp{LED} or altering their apparent positions \cite{Rahman20200}. However, in image-based systems that rely only on the detected positions of light sources (e.g., features as shown in Fig.~\ref{fig:imagebaseduVLP}) rather than their orientation, the effect of tilt is generally limited. 

\subsubsection{Power, latency, and privacy issues of camera-based \ac{uVLP}}
Camera-based \ac{uVLP} systems leverage ubiquitous, high-resolution image sensors to extract rich environmental information. However, camera-based \ac{uVLP} systems typically exhibit significantly higher power consumption and latency compared to \ac{PD}-based systems \cite{lirias3591348}. For instance, LiTell consumes approximately 2.7 W \cite{Zhang2016, Zhang2019}, while iLAMP reduces this to 0.93 W \cite{Zhu2017}, which, although lower, remains energy-intensive. Moreover, image sensors are substantially more costly than \acp{PD} \cite{lirias3591348}. These limitations—elevated power consumption, computational overhead, cost, and latency—pose challenges for deployment in battery-constrained platforms such as smartphones, drones, and mobile service robots, particularly in green \ac{IoT} and real-time indoor localization scenarios. For instance, the LiTell system \cite{Zhang2016, Zhang2019} demonstrates camera-based \ac{uVLP} with latencies exceeding 2 seconds due to image capture and processing delays. Such sluggish updates are detrimental in dynamic scenarios (i.e., high-speed applications), where timely localization is critical—for example, in mobile robots performing real-time navigation, obstacle avoidance, or path planning. Finally, camera-based systems raise potential privacy concerns due to the nature of image capture \cite{Liu2019Graph}.

\subsubsection{Impact of ambient light interference on \ac{uVLP} systems}
Ambient light from natural sources (e.g., sunlight) and artificial sources (e.g., \acp{FL}, \acp{LED}, incandescent fixtures\cite{Moreira1997}) not used for \ac{uVLP} can significantly interfere with \ac{uVLP} systems\cite{Xu2015,Zhao2017,RaviFiatLuxF,Yang2025,Faulkner2019,Faulkner2020}. This interference may arise directly from these sources or indirectly through reflections (e.g., walls, glass) and scattering (e.g., dust, smoke or mist) \cite{Alijani2025}. Depending on the system design, ambient light may act as interference or be exploited as a transmitter, as in Ambilight \cite{Liu2014}. When ambient light is not intended for use in a \ac{uVLP} system, it can degrade system performance\cite{Xu2015,Zhao2017,RaviFiatLuxF,Yang2025,Faulkner2019,Faulkner2020}. Since \ac{uVLP} relies on subtle variations in \ac{RSS} from unmodulated \acp{LED}, additional ambient illumination introduces a \ac{DC} offset, which reduces the \ac{SNR} and masks \ac{LED}-specific features (e.g., \ac{CF}). Temporal variations in ambient light (e.g., changes in sunlight intensity) can alter the \ac{RSS} even when the receiver does not move \cite{Yang2024}. This temporal drift causes a mismatch between the calibration and operational conditions, degrading the positioning accuracy. Moreover, ambient light intensity may vary spatially. Even with symmetrically mounted luminaires, ambient light coverage across a space is uneven\cite{Yang2024}. As a result, \ac{RSS} values collected at different locations may differ despite equal distances to the \ac{LED}, or conversely, may appear similar despite different distances. This spatial inconsistency causes the \ac{RSS}–distance relationship to deviate from the Lambertian emission model, hindering accurate calibration and reducing \ac{uVLP} precision\cite{Yang2024}.

\subsection{Shaping the Future of \ac{uVLP} Technologies}
\label{subsec:futureroadmap}
This section outlines several emerging and promising research directions within the domain of \ac{uVLP}, some of which directly address one or more of the previously mentioned challenges.
\subsubsection{Hybrid \ac{VLP} with Modulated and Unmodulated Sources}
There is strong potential in leveraging both \ac{VLP} and \ac{uVLP} to balance accuracy and cost\cite{Liang2022}. Hybrid positioning systems that integrate modulated and unmodulated light sources can address this trade-off effectively\cite{Liang2022}. In such systems, only a subset of \acp{LED} may be modulated to provide precise location fixes when detected\cite{lirias3591348}. These modulated lights can be strategically installed at key locations, such as the start and end of corridors, to maximize coverage\cite{lirias3591348}. Between these anchors, unmodulated sources can support coarse trajectory estimation\cite{lirias3591348}. This hybrid approach offers system designers greater flexibility in managing the balance between performance and hardware cost. Additionally, combining both source types can improve the scalability of (u)VLP systems\cite{lirias3591348}.

\subsubsection{Integration of \ac{uVLP} with \ac{IMU} Sensors}
\label{sec:imusolution}
Despite the potential of \ac{LBS}, the cost of dedicated localization infrastructure remains a significant barrier, leading to increased interest in infrastructure-less positioning using signals of opportunity~\cite{Bastiaens2020,Liu2014}. One widely adopted approach is \ac{PDR} using \acp{IMU}, which relies solely on smartphone sensors and eliminates the need for additional infrastructure~\cite{Yang2025,deCock2022,DeCock2023,Cock2021}. Recent advancements in \ac{IMU} hardware and sensor fusion algorithms have enabled best-case positioning accuracies on the order of 1 meter~\cite{deCock2022}. In \ac{uVLP}, minor tilting of the \ac{PD} or camera during pedestrian movement can significantly degrade positioning accuracy. This issue can be effectively addressed by integrating smartphone-derived attitude information into the positioning framework~\cite{Yang2025}. A combined positioning system that fuses \ac{PDR} and \ac{uVLP} can exploit their complementary characteristics: \ac{IMU}-based attitude estimation mitigates tilt-induced errors in \ac{uVLP}, while \ac{uVLP} corrects the cumulative step-length and heading errors inherent in \ac{PDR} by providing relatively absolute position fixes \cite{Yang2025}. This integrated approach offers enhanced robustness, particularly in dynamic scenarios involving variable walking patterns and sensor uncertainties. Furthermore, incorporating \ac{AoA} estimation methods that explicitly account for smartphone orientation could further improve positional accuracy, enabling precise localization even in complex indoor environments\cite{Yang2025}.

\subsubsection{\ac{ML} and Signal Processing in \ac{uVLP}}
\ac{ML} and advanced signal processing techniques present promising solutions to the challenges identified in Section~\ref{sec:challenges}. Unlike conventional \ac{VLP}, \ac{uVLP} does not employ modulators, leading to low \ac{SNR} conditions, particularly in \ac{CF}-based methods (less so in image-based methods), thereby reducing positioning accuracy~\cite{lirias3591348}. Thus, advancing \ac{uVLP} performance largely depends on sophisticated signal processing and the integration of \ac{ML} approaches~\cite{HamidiRad2017}.
One promising strategy involves \ac{ML}-based classification algorithms to automatically extract and rank features from \ac{LED} signals, facilitating effective demultiplexing of individual \ac{LED} contributions~\cite{HamidiRad2017}. Such approaches are particularly beneficial in scenarios where \acp{LED} undergo intermittent control for energy efficiency purposes, such as constant current reduction (CCR) dimming, rather than traditional modulation methods like \ac{PWM}~\cite{boqi_pwm_dimming}. Energy-efficient dimming techniques often modify the light spectrum or reduce signal levels, potentially rendering \ac{CF} detection ineffective and compromising positioning accuracy. \ac{ML} techniques, such as transformers\cite{Kamath2022}, can enable utilization of the entire spectral response, possibly enhancing robustness compared to conventional \ac{CF}-based demultiplexing methods. Moreover, \ac{ML} can facilitate efficient classification in fingerprint-based approaches (both light intensity and image-based) and improves performance in fused \ac{uVLP} systems which integrates data from multiple sensors, thus effectively compensating for ambient light interference and reducing calibration costs.

In addition to \ac{ML}\cite{Rekkas2023,Zhang2025}, advanced signal processing methods can further enhance \ac{uVLP} systems. Techniques such as windowing and zero-padding mitigate spectral leakage and enhance frequency resolution in demultiplexing algorithms\cite{Bastiaens2020}. Selecting appropriate window types preserves critical signal amplitude, essential for accurate positioning. Overall, the integration of \ac{ML} and advanced signal processing represents a promising research direction towards achieving more robust and accurate \ac{uVLP}-based indoor positioning systems.

\subsubsection{Reconfigurable Intelligent Surfaces (RIS) for uVLP}
\acp{RIS} have emerged as a key enabler for \ac{6G} and have been adapted for \ac{VLC} and \ac{VLP} as optical \acp{RIS}, using passive elements such as metasurfaces, mirror arrays, or liquid crystals (LCs) to steer light based on Snell’s law~\cite{Aboagye2023,Abdelhady2021,Sun20222,Aboagye2022}. These systems primarily address \ac{LoS} blockages and receiver orientation issues. By employing low-loss reflective materials, optical \acp{RIS} enable efficient light steering and intensity amplification~\cite{Aboagye2023,Abdelhady2021,Sun20222,Aboagye2022}, achieving higher \ac{SNR} at the optical receiver than standard wall reflections~\cite{Aboagye2023,Abdelhady2021,Sun20222,Aboagye2022}. This makes them a promising solution for improving signal quality in \ac{uVLP}, which inherently suffers from low \ac{SNR} due to the absence of modulation, thereby aiding \ac{CF} detection. Moreover, in the context of \ac{uVLP}, most optical \acp{RIS} are not frequency-selective and function by passively redirecting or diffusing light to enhance coverage or mitigate shadowing. Achieving frequency-selective behavior—such as boosting specific \acp{CF} for \ac{CF}-based \ac{uVLP} would require specialized metasurface designs, which remains an open and largely unexplored research direction.

\section{Conclusion}
\label{Sec:conclusion}
Visible Light Positioning (VLP) has attracted growing interest as a key enabler of next-generation indoor positioning systems (IPS), particularly within the framework of sixth-generation (6G) wireless networks. However, conventional VLP systems typically rely on modulation drivers to distinguish contributions from individual light-emitting diodes (LEDs) at the receiver, which increases deployment costs, degrades illumination quality, and encounters limitations due to the low sampling rates of camera-based systems—factors that have collectively hindered large-scale adoption. Recent advances have introduced unmodulated VLP (uVLP), using unmodulated/unmodified LEDs to address these challenges. In this paper, we presented a comprehensive survey of \ac{uVLP} systems, beginning with an overview of \ac{VLP} fundamentals and highlighting the key distinctions between modulated and unmodulated approaches. We categorized and thoroughly reviewed \ac{uVLP} systems based on the types of receivers employed—light intensity receivers (including photodiodes (PDs), solar cells, ambient light sensors (ALS), and spectral sensors) and image-based receivers—and further classified them into demultiplexing and non-demultiplexing positioning techniques.

Despite considerable progress, uVLP remains in its infancy, facing several open challenges. These include aliasing effects caused by the low sampling rates of smartphone PDs, sensitivity to tilt and misalignment between transmitters and receivers, latency and privacy issues in camera-based systems, and vulnerability to ambient light interference. Promising future research directions include the development of hybrid VLP systems that integrate modulated and unmodulated light sources, multi-sensor fusion strategies, advanced signal processing and machine learning techniques, and the application of Reconfigurable Intelligent Surfaces (RIS) to enhance system robustness and adaptability.

\section*{Acknowledgment}
This work was supported by the imec.icon project PULSAR, a  research project bringing together academic researchers and industry partners. The PULSAR project was cofinanced by imec and received project support from Flanders Innovation \& Entrepreneurship (project nr. HBC.2023.0606).

\section*{Acronyms}
\begin{acronym}[mmWAVE]
  \input{acronyms.acro}
\end{acronym}

\bibliographystyle{IEEEtran} 
\bibliography{references} 

\begin{thebibliography}{100}
\providecommand{\url}[1]{#1}
\csname url@samestyle\endcsname
\providecommand{\newblock}{\relax}
\providecommand{\bibinfo}[2]{#2}
\providecommand{\BIBentrySTDinterwordspacing}{\spaceskip=0pt\relax}
\providecommand{\BIBentryALTinterwordstretchfactor}{4}
\providecommand{\BIBentryALTinterwordspacing}{\spaceskip=\fontdimen2\font plus
\BIBentryALTinterwordstretchfactor\fontdimen3\font minus \fontdimen4\font\relax}
\providecommand{\BIBforeignlanguage}[2]{{%
\expandafter\ifx\csname l@#1\endcsname\relax
\typeout{** WARNING: IEEEtran.bst: No hyphenation pattern has been}%
\typeout{** loaded for the language `#1'. Using the pattern for}%
\typeout{** the default language instead.}%
\else
\language=\csname l@#1\endcsname
\fi
#2}}
\providecommand{\BIBdecl}{\relax}
\BIBdecl

\bibitem{Zafari2019}
\BIBentryALTinterwordspacing
F.~Zafari, A.~Gkelias, and K.~K. Leung, ``{A Survey of Indoor Localization Systems and Technologies},'' \emph{IEEE Communications Surveys \& Tutorials}, vol.~21, no.~3, p. 2568–2599, 2019. [Online]. Available: \url{http://dx.doi.org/10.1109/COMST.2019.2911558}
\BIBentrySTDinterwordspacing

\bibitem{Farahsari2022}
\BIBentryALTinterwordspacing
P.~S. Farahsari, A.~Farahzadi, J.~Rezazadeh, and A.~Bagheri, ``{A Survey on Indoor Positioning Systems for IoT-Based Applications},'' \emph{IEEE Internet of Things Journal}, vol.~9, no.~10, p. 7680–7699, May 2022. [Online]. Available: \url{http://dx.doi.org/10.1109/JIOT.2022.3149048}
\BIBentrySTDinterwordspacing

\bibitem{Yassin2017}
\BIBentryALTinterwordspacing
A.~Yassin, Y.~Nasser, M.~Awad, A.~Al-Dubai, R.~Liu, C.~Yuen, R.~Raulefs, and E.~Aboutanios, ``{Recent Advances in Indoor Localization: A Survey on Theoretical Approaches and Applications},'' \emph{IEEE Communications Surveys \& Tutorials}, vol.~19, no.~2, p. 1327–1346, 2017. [Online]. Available: \url{http://dx.doi.org/10.1109/COMST.2016.2632427}
\BIBentrySTDinterwordspacing

\bibitem{Horikawa2014}
S.~Horikawa, ``{Pervasive Visible Light Positioning System using White LED Lighting},'' in \emph{Technical Report of IEICE DSP}, vol. 103, no. 719, 2014.

\bibitem{Hassan2015}
\BIBentryALTinterwordspacing
N.~U. Hassan, A.~Naeem, M.~A. Pasha, T.~Jadoon, and C.~Yuen, ``{Indoor Positioning Using Visible LED Lights: A Survey},'' \emph{ACM Computing Surveys}, vol.~48, no.~2, p. 1–32, Nov. 2015. [Online]. Available: \url{http://dx.doi.org/10.1145/2835376}
\BIBentrySTDinterwordspacing

\bibitem{Do2016}
\BIBentryALTinterwordspacing
T.-H. Do and M.~Yoo, ``{An in-Depth Survey of Visible Light Communication Based Positioning Systems},'' \emph{Sensors}, vol.~16, no.~5, p. 678, May 2016. [Online]. Available: \url{http://dx.doi.org/10.3390/s16050678}
\BIBentrySTDinterwordspacing

\bibitem{Luo2017}
\BIBentryALTinterwordspacing
J.~Luo, L.~Fan, and H.~Li, ``{Indoor Positioning Systems Based on Visible Light Communication: State of the Art},'' \emph{IEEE Communications Surveys \& Tutorials}, vol.~19, no.~4, p. 2871–2893, 2017. [Online]. Available: \url{http://dx.doi.org/10.1109/COMST.2017.2743228}
\BIBentrySTDinterwordspacing

\bibitem{Zhuang2018}
\BIBentryALTinterwordspacing
Y.~Zhuang, L.~Hua, L.~Qi, J.~Yang, P.~Cao, Y.~Cao, Y.~Wu, J.~Thompson, and H.~Haas, ``{A Survey of Positioning Systems Using Visible LED Lights},'' \emph{IEEE Communications Surveys \& Tutorials}, vol.~20, no.~3, p. 1963–1988, 2018. [Online]. Available: \url{http://dx.doi.org/10.1109/COMST.2018.2806558}
\BIBentrySTDinterwordspacing

\bibitem{Pathak2015}
\BIBentryALTinterwordspacing
P.~H. Pathak, X.~Feng, P.~Hu, and P.~Mohapatra, ``{Visible Light Communication, Networking, and Sensing: A Survey, Potential and Challenges},'' \emph{IEEE Communications Surveys \& Tutorials}, vol.~17, no.~4, p. 2047–2077, 2015. [Online]. Available: \url{http://dx.doi.org/10.1109/COMST.2015.2476474}
\BIBentrySTDinterwordspacing

\bibitem{Wang2024VLP}
\BIBentryALTinterwordspacing
R.~Wang, G.~Niu, Q.~Cao, C.~S. Chen, and S.-W. Ho, ``{A Survey of Visible-Light-Communication-Based Indoor Positioning Systems},'' \emph{Sensors}, vol.~24, no.~16, p. 5197, Aug. 2024. [Online]. Available: \url{http://dx.doi.org/10.3390/s24165197}
\BIBentrySTDinterwordspacing

\bibitem{Afzalan2019}
\BIBentryALTinterwordspacing
M.~Afzalan and F.~Jazizadeh, ``{Indoor Positioning Based on Visible Light Communication: A Performance-based Survey of Real-world Prototypes},'' \emph{ACM Computing Surveys}, vol.~52, no.~2, p. 1–36, May 2019. [Online]. Available: \url{http://dx.doi.org/10.1145/3299769}
\BIBentrySTDinterwordspacing

\bibitem{Kouhini2021}
\BIBentryALTinterwordspacing
S.~M. Kouhini, C.~Kottke, Z.~Ma, R.~Freund, V.~Jungnickel, M.~Muller, D.~Behnke, M.~M. Vazquez, and J.-P. M.~G. Linnartz, ``{LiFi Positioning for Industry 4.0},'' \emph{IEEE Journal of Selected Topics in Quantum Electronics}, vol.~27, no.~6, p. 1–15, Nov. 2021. [Online]. Available: \url{http://dx.doi.org/10.1109/JSTQE.2021.3095364}
\BIBentrySTDinterwordspacing

\bibitem{Mapunda2020}
\BIBentryALTinterwordspacing
G.~A. Mapunda, R.~Ramogomana, L.~Marata, B.~Basutli, A.~S. Khan, and J.~M. Chuma, ``{Indoor Visible Light Communication: A Tutorial and Survey},'' \emph{Wireless Communications and Mobile Computing}, vol. 2020, pp. 1--46, Dec. 2020. [Online]. Available: \url{http://dx.doi.org/10.1155/2020/8881305}
\BIBentrySTDinterwordspacing

\bibitem{Bastiaens2023AnEA}
\BIBentryALTinterwordspacing
S.~Bastiaens, M.~Alijani, W.~Joseph, and D.~Plets, ``{An Experimental Analysis of Visible Light Positioning in NLoS Environments},'' in \emph{IPIN-WiP}, 2023. [Online]. Available: \url{https://api.semanticscholar.org/CorpusID:266562382}
\BIBentrySTDinterwordspacing

\bibitem{Maheepala2020}
\BIBentryALTinterwordspacing
M.~Maheepala, A.~Z. Kouzani, and M.~A. Joordens, ``{Light-Based Indoor Positioning Systems: A Review},'' \emph{IEEE Sensors Journal}, vol.~20, no.~8, p. 3971–3995, Apr. 2020. [Online]. Available: \url{http://dx.doi.org/10.1109/JSEN.2020.2964380}
\BIBentrySTDinterwordspacing

\bibitem{Bastiaens2024}
\BIBentryALTinterwordspacing
S.~Bastiaens, M.~Alijani, W.~Joseph, and D.~Plets, ``{Visible Light Positioning as a Next-Generation Indoor Positioning Technology: A Tutorial},'' \emph{IEEE Communications Surveys \& Tutorials}, p. 1–1, 2024. [Online]. Available: \url{http://dx.doi.org/10.1109/COMST.2024.3372153}
\BIBentrySTDinterwordspacing

\bibitem{Bastiaens2022}
\BIBentryALTinterwordspacing
S.~Bastiaens, J.~V.-V. Gerwen, N.~Macoir, K.~Deprez, C.~De~Cock, W.~Joseph, E.~De~Poorter, and D.~Plets, ``{Experimental Benchmarking of Next-Gen Indoor Positioning Technologies (Unmodulated) Visible Light Positioning and Ultra-Wideband},'' \emph{IEEE Internet of Things Journal}, vol.~9, no.~18, p. 17858–17870, Sep. 2022. [Online]. Available: \url{http://dx.doi.org/10.1109/JIOT.2022.3161791}
\BIBentrySTDinterwordspacing

\bibitem{Du2019}
\BIBentryALTinterwordspacing
P.~Du, S.~Zhang, C.~Chen, H.~Yang, W.-D. Zhong, R.~Zhang, A.~Alphones, and Y.~Yang, ``{Experimental Demonstration of 3D Visible Light Positioning Using Received Signal Strength With Low-Complexity Trilateration Assisted by Deep Learning Technique},'' \emph{IEEE Access}, vol.~7, p. 93986–93997, 2019. [Online]. Available: \url{http://dx.doi.org/10.1109/ACCESS.2019.2928014}
\BIBentrySTDinterwordspacing

\bibitem{Bastiaens2020}
\BIBentryALTinterwordspacing
S.~Bastiaens, K.~Deprez, L.~Martens, W.~Joseph, and D.~Plets, ``{A Comprehensive Study on Light Signals of Opportunity for Subdecimetre Unmodulated Visible Light Positioning},'' \emph{Sensors}, vol.~20, no.~19, p. 5596, Sep. 2020. [Online]. Available: \url{http://dx.doi.org/10.3390/s20195596}
\BIBentrySTDinterwordspacing

\bibitem{DeLausnay2015}
\BIBentryALTinterwordspacing
S.~De~Lausnay, L.~De~Strycker, J.-P. Goemaere, N.~Stevens, and B.~Nauwelaers, ``{A Visible Light Positioning system using Frequency Division Multiple Access with square waves},'' in \emph{2015 9th International Conference on Signal Processing and Communication Systems (ICSPCS)}.\hskip 1em plus 0.5em minus 0.4em\relax IEEE, Dec. 2015. [Online]. Available: \url{http://dx.doi.org/10.1109/ICSPCS.2015.7391787}
\BIBentrySTDinterwordspacing

\bibitem{Berton2020}
\BIBentryALTinterwordspacing
G.~Berton, ``{Minimising the modulation index in Visible Light Positioning},'' 2020, master's thesis, Ghent University. [Online]. Available: \url{http://lib.ugent.be/catalog/rug01:002945959}
\BIBentrySTDinterwordspacing

\bibitem{Zhang2014}
\BIBentryALTinterwordspacing
W.~Zhang, M.~I.~S. Chowdhury, and M.~Kavehrad, ``{Asynchronous indoor positioning system based on visible light communications},'' \emph{Optical Engineering}, vol.~53, no.~4, p. 045105, Apr. 2014. [Online]. Available: \url{http://dx.doi.org/10.1117/1.OE.53.4.045105}
\BIBentrySTDinterwordspacing

\bibitem{BastiaensPhdBook}
{S. Bastiaens}, ``\BIBforeignlanguage{{eng}}{{Towards centimetre-order indoor localisation with RSS-based visible light positioning}},'' Ph.D. dissertation, {Ghent University}, {2022}.

\bibitem{Bastiaens2018}
\BIBentryALTinterwordspacing
S.~Bastiaens, D.~Plets, L.~Martens, and W.~Joseph, ``{Impact of Nonideal LED Modulation on RSS-based VLP Performance},'' in \emph{2018 IEEE 29th Annual International Symposium on Personal, Indoor and Mobile Radio Communications (PIMRC)}.\hskip 1em plus 0.5em minus 0.4em\relax IEEE, Sep. 2018, p. 1–5. [Online]. Available: \url{http://dx.doi.org/10.1109/PIMRC.2018.8581009}
\BIBentrySTDinterwordspacing

\bibitem{Xu2015}
\BIBentryALTinterwordspacing
Q.~Xu, R.~Zheng, and S.~Hranilovic, ``{IDyLL: indoor localization using inertial and light sensors on smartphones},'' in \emph{Proceedings of the 2015 ACM International Joint Conference on Pervasive and Ubiquitous Computing}, ser. UbiComp ’15.\hskip 1em plus 0.5em minus 0.4em\relax ACM, Sep. 2015. [Online]. Available: \url{http://dx.doi.org/10.1145/2750858.2807540}
\BIBentrySTDinterwordspacing

\bibitem{Liu2019}
\BIBentryALTinterwordspacing
X.~Liu, X.~Wei, and L.~Guo, ``{DIMLOC: Enabling High-Precision Visible Light Localization Under Dimmable LEDs in Smart Buildings},'' \emph{IEEE Internet of Things Journal}, vol.~6, no.~2, p. 3912–3924, Apr. 2019. [Online]. Available: \url{http://dx.doi.org/10.1109/JIOT.2019.2893251}
\BIBentrySTDinterwordspacing

\bibitem{GPinto2012}
\BIBentryALTinterwordspacing
A.~M. G.~Pinto, A.~P. Moreira, and P.~G.~Costa, ``{Indoor Localization System based on Artificial Landmarks and Monocular Vision},'' \emph{TELKOMNIKA (Telecommunication, Computing, Electronics and Control)}, vol.~10, no.~4, Dec. 2012. [Online]. Available: \url{http://dx.doi.org/10.12928/TELKOMNIKA.v10i4.640}
\BIBentrySTDinterwordspacing

\bibitem{Amsters2019}
\BIBentryALTinterwordspacing
R.~Amsters, E.~Demeester, N.~Stevens, and P.~Slaets, ``{In-Depth Analysis of Unmodulated Visible Light Positioning Using the Iterated Extended Kalman Filter},'' \emph{Sensors}, vol.~19, no.~23, p. 5198, Nov. 2019. [Online]. Available: \url{http://dx.doi.org/10.3390/s19235198}
\BIBentrySTDinterwordspacing

\bibitem{Amsters2018}
\BIBentryALTinterwordspacing
R.~Amsters, E.~Demeester, P.~Slaets, and N.~Stevens, ``{Unmodulated Visible Light Positioning Using the Iterated Extended Kalman Filter},'' in \emph{2018 International Conference on Indoor Positioning and Indoor Navigation (IPIN)}.\hskip 1em plus 0.5em minus 0.4em\relax IEEE, Sep. 2018, p. 1–8. [Online]. Available: \url{http://dx.doi.org/10.1109/IPIN.2018.8533849}
\BIBentrySTDinterwordspacing

\bibitem{RaviFiatLuxF}
\BIBentryALTinterwordspacing
N.~Ravi and L.~Iftode, ``{FiatLux: Fingerprinting Rooms Using Light Intensity}.'' [Online]. Available: \url{https://api.semanticscholar.org/CorpusID:8539806}
\BIBentrySTDinterwordspacing

\bibitem{Faulkner2020}
\BIBentryALTinterwordspacing
{N. Faulkner, F. Alam, M. Legg, and S. Demidenko}, ``{Watchers on the Wall: Passive Visible Light-Based Positioning and Tracking With Embedded Light-Sensors on the Wall},'' \emph{IEEE Transactions on Instrumentation and Measurement}, vol.~69, no.~5, p. 2522–2532, May 2020. [Online]. Available: \url{http://dx.doi.org/10.1109/TIM.2019.2953373}
\BIBentrySTDinterwordspacing

\bibitem{Konings2020}
\BIBentryALTinterwordspacing
D.~Konings, N.~Faulkner, F.~Alam, E.~M.-K. Lai, and S.~Demidenko, ``{FieldLight: Device-Free Indoor Human Localization Using Passive Visible Light Positioning and Artificial Potential Fields},'' \emph{IEEE Sensors Journal}, vol.~20, no.~2, p. 1054–1066, Jan. 2020. [Online]. Available: \url{http://dx.doi.org/10.1109/JSEN.2019.2944178}
\BIBentrySTDinterwordspacing

\bibitem{Faulkner2019}
\BIBentryALTinterwordspacing
N.~Faulkner, F.~Alam, M.~Legg, and S.~Demidenko, ``{Smart Wall: Passive Visible Light Positioning with Ambient Light Only},'' in \emph{2019 IEEE International Instrumentation and Measurement Technology Conference (I2MTC)}.\hskip 1em plus 0.5em minus 0.4em\relax IEEE, May 2019, p. 1–6. [Online]. Available: \url{http://dx.doi.org/10.1109/I2MTC.2019.8826875}
\BIBentrySTDinterwordspacing

\bibitem{Singh2023}
\BIBentryALTinterwordspacing
J.~Singh, T.~Farnham, and Q.~Wang, ``{When BLE Meets Light: Multi-modal Fusion for Enhanced Indoor Localization},'' in \emph{Proceedings of the 29th Annual International Conference on Mobile Computing and Networking}, ser. ACM MobiCom ’23, vol.~17.\hskip 1em plus 0.5em minus 0.4em\relax ACM, Oct. 2023, p. 1–3. [Online]. Available: \url{http://dx.doi.org/10.1145/3570361.3615746}
\BIBentrySTDinterwordspacing

\bibitem{Singh2022}
\BIBentryALTinterwordspacing
J.~Singh, Q.~Wang, M.~Zuniga, and T.~Farnham, ``{HueSense: Featuring LED Lights Through Hue Sensing},'' in \emph{Proceedings of the 1st ACM Workshop on AI Empowered Mobile and Wireless Sensing}, ser. MORSE ’22.\hskip 1em plus 0.5em minus 0.4em\relax ACM, Oct. 2022, p. 19–24. [Online]. Available: \url{http://dx.doi.org/10.1145/3556558.3558582}
\BIBentrySTDinterwordspacing

\bibitem{Armstrong2013}
\BIBentryALTinterwordspacing
J.~Armstrong, Y.~Sekercioglu, and A.~Neild, ``{Visible light positioning: a roadmap for international standardization},'' \emph{IEEE Communications Magazine}, vol.~51, no.~12, p. 68–73, Dec. 2013. [Online]. Available: \url{http://dx.doi.org/10.1109/MCOM.2013.6685759}
\BIBentrySTDinterwordspacing

\bibitem{Jiao2017}
\BIBentryALTinterwordspacing
Z.~Jiao, B.~Zhang, M.~Liu, and C.~Li, ``{Visible Light Communication Based Indoor Positioning Techniques},'' \emph{IEEE Network}, vol.~31, no.~5, pp. ?--?, 2017. [Online]. Available: \url{http://dx.doi.org/10.1109/MNET.2017.1600264}
\BIBentrySTDinterwordspacing

\bibitem{Chen2019}
\BIBentryALTinterwordspacing
P.~Chen, D.~Che, and Y.~Yin, ``{A survey on visible light positioning from the hardware perspective},'' in \emph{Proceedings of the ACM Turing Celebration Conference - China}, ser. ACM TURC 2019.\hskip 1em plus 0.5em minus 0.4em\relax ACM, May 2019. [Online]. Available: \url{http://dx.doi.org/10.1145/3321408.3321416}
\BIBentrySTDinterwordspacing

\bibitem{Rahman2020}
\BIBentryALTinterwordspacing
A.~B. M.~M. Rahman, T.~Li, and Y.~Wang, ``{Recent Advances in Indoor Localization via Visible Lights: A Survey},'' \emph{Sensors}, vol.~20, no.~5, p. 1382, Mar. 2020. [Online]. Available: \url{http://dx.doi.org/10.3390/s20051382}
\BIBentrySTDinterwordspacing

\bibitem{Liu2021}
\BIBentryALTinterwordspacing
X.~Liu, L.~Guo, and X.~Wei, ``{Indoor Visible Light Applications for Communication, Positioning, and Security},'' \emph{Wireless Communications and Mobile Computing}, vol. 2021, no.~1, Jan. 2021. [Online]. Available: \url{http://dx.doi.org/10.1155/2021/1730655}
\BIBentrySTDinterwordspacing

\bibitem{Tran2022}
\BIBentryALTinterwordspacing
H.~Q. Tran and C.~Ha, ``{Machine learning in indoor visible light positioning systems: A review},'' \emph{Neurocomputing}, vol. 491, p. 117–131, Jun. 2022. [Online]. Available: \url{http://dx.doi.org/10.1016/j.neucom.2021.10.123}
\BIBentrySTDinterwordspacing

\bibitem{ZHANG2023}
\BIBentryALTinterwordspacing
S.~Zhang, P.~Du, H.~Yang, R.~Zhang, C.~Chen, and A.~Alphones, ``{Recent Progress in Visible Light Positioning and Communication Systems},'' \emph{IEICE Transactions on Communications}, vol. E106.B, no.~2, p. 84–100, Feb. 2023. [Online]. Available: \url{http://dx.doi.org/10.1587/transcom.2022CEI0001}
\BIBentrySTDinterwordspacing

\bibitem{Zhu2024}
\BIBentryALTinterwordspacing
Z.~Zhu, Y.~Yang, M.~Chen, C.~Guo, J.~Cheng, and S.~Cui, ``{A Survey on Indoor Visible Light Positioning Systems: Fundamentals, Applications, and Challenges},'' \emph{IEEE Communications Surveys \& Tutorials}, p. 1–1, 2024. [Online]. Available: \url{http://dx.doi.org/10.1109/COMST.2024.3471950}
\BIBentrySTDinterwordspacing

\bibitem{Alijani2025}
\BIBentryALTinterwordspacing
M.~Alijani, C.~D. Cock, W.~Joseph, and D.~Plets, ``{Device-Free Visible Light Sensing: A Survey},'' \emph{IEEE Communications Surveys \& Tutorials}, p. 1–1, 2025. [Online]. Available: \url{http://dx.doi.org/10.1109/COMST.2025.3546166}
\BIBentrySTDinterwordspacing

\bibitem{Facchinetti1995SelfPositioningRN}
\BIBentryALTinterwordspacing
C.~Facchinetti, F.~Ti{\`e}che, and H.~Hugli, ``{Self-Positioning Robot Navigation Using Ceiling Image Sequences},'' 1995. [Online]. Available: \url{https://api.semanticscholar.org/CorpusID:29304450}
\BIBentrySTDinterwordspacing

\bibitem{Liang2022}
\BIBentryALTinterwordspacing
Q.~Liang, Y.~Sun, L.~Wang, and M.~Liu, ``{A Novel Inertial-Aided Visible Light Positioning System Using Modulated LEDs and Unmodulated Lights as Landmarks},'' \emph{IEEE Transactions on Automation Science and Engineering}, vol.~19, no.~4, p. 3049–3067, Oct. 2022. [Online]. Available: \url{http://dx.doi.org/10.1109/TASE.2021.3105700}
\BIBentrySTDinterwordspacing

\bibitem{Yang2025}
\BIBentryALTinterwordspacing
X.~Yang, H.~Zhang, Y.~Zhuang, Y.~Wang, M.~Shi, and Y.~Xu, ``{uLiDR: An inertial-assisted unmodulated visible light positioning system for smartphone-based pedestrian navigation},'' \emph{Information Fusion}, vol. 113, p. 102579, Jan. 2025. [Online]. Available: \url{http://dx.doi.org/10.1016/j.inffus.2024.102579}
\BIBentrySTDinterwordspacing

\bibitem{Wang2013}
\BIBentryALTinterwordspacing
T.~Q. Wang, Y.~A. Sekercioglu, A.~Neild, and J.~Armstrong, ``{Position Accuracy of Time-of-Arrival Based Ranging Using Visible Light With Application in Indoor Localization Systems},'' \emph{Journal of Lightwave Technology}, vol.~31, no.~20, p. 3302–3308, Oct. 2013. [Online]. Available: \url{http://dx.doi.org/10.1109/JLT.2013.2281592}
\BIBentrySTDinterwordspacing

\bibitem{Jung2011}
\BIBentryALTinterwordspacing
S.-Y. Jung, S.~Hann, and C.-S. Park, ``{TDOA-based optical wireless indoor localization using LED ceiling lamps},'' \emph{IEEE Transactions on Consumer Electronics}, vol.~57, no.~4, p. 1592–1597, Nov. 2011. [Online]. Available: \url{http://dx.doi.org/10.1109/TCE.2011.6131130}
\BIBentrySTDinterwordspacing

\bibitem{Zhang2018}
\BIBentryALTinterwordspacing
S.~Zhang, W.-D. Zhong, P.~Du, and C.~Chen, ``{Experimental Demonstration of Indoor Sub-Decimeter Accuracy VLP System Using Differential PDOA},'' \emph{IEEE Photonics Technology Letters}, vol.~30, no.~19, p. 1703–1706, Oct. 2018. [Online]. Available: \url{http://dx.doi.org/10.1109/LPT.2018.2866402}
\BIBentrySTDinterwordspacing

\bibitem{Xu2023}
\BIBentryALTinterwordspacing
{Y. Xu et al.}, ``{A Novel Differential Phase of Arrival-based Experimental Visible Light Positioning System},'' in \emph{2023 13th International Conference on Indoor Positioning and Indoor Navigation (IPIN)}.\hskip 1em plus 0.5em minus 0.4em\relax IEEE, Sep. 2023, p. 1–6. [Online]. Available: \url{http://dx.doi.org/10.1109/IPIN57070.2023.10332503}
\BIBentrySTDinterwordspacing

\bibitem{Steendam2018}
\BIBentryALTinterwordspacing
H.~Steendam, ``{A 3-D Positioning Algorithm for AOA-Based VLP With an Aperture-Based Receiver},'' \emph{IEEE Journal on Selected Areas in Communications}, vol.~36, no.~1, p. 23–33, Jan. 2018. [Online]. Available: \url{http://dx.doi.org/10.1109/JSAC.2017.2774478}
\BIBentrySTDinterwordspacing

\bibitem{Kuo2014}
\BIBentryALTinterwordspacing
Y.-S. Kuo, P.~Pannuto, K.-J. Hsiao, and P.~Dutta, ``{Luxapose: indoor positioning with mobile phones and visible light},'' in \emph{Proceedings of the 20th annual international conference on Mobile computing and networking}, ser. MobiCom’14.\hskip 1em plus 0.5em minus 0.4em\relax ACM, Sep. 2014. [Online]. Available: \url{http://dx.doi.org/10.1145/2639108.2639109}
\BIBentrySTDinterwordspacing

\bibitem{EWLam2019}
\BIBentryALTinterwordspacing
{E. W. Lam and T. D. C. Little}, ``{Visible light positioning: moving from 2D planes to 3D spaces [Invited]},'' \emph{Chinese Optics Letters}, vol.~17, no.~3, p. 030604, 2019. [Online]. Available: \url{http://dx.doi.org/10.3788/COL201917.030604}
\BIBentrySTDinterwordspacing

\bibitem{Keskin20188}
\BIBentryALTinterwordspacing
M.~F. Keskin, A.~D. Sezer, and S.~Gezici, ``{Localization via Visible Light Systems},'' \emph{Proceedings of the IEEE}, vol. 106, no.~6, p. 1063–1088, Jun. 2018. [Online]. Available: \url{http://dx.doi.org/10.1109/JPROC.2018.2823500}
\BIBentrySTDinterwordspacing

\bibitem{ElGamal2005}
\BIBentryALTinterwordspacing
A.~El~Gamal and H.~Eltoukhy, ``{CMOS image sensors},'' \emph{IEEE Circuits and Devices Magazine}, vol.~21, no.~3, p. 6–20, May 2005. [Online]. Available: \url{http://dx.doi.org/10.1109/MCD.2005.1438751}
\BIBentrySTDinterwordspacing

\bibitem{Yamazato2017}
\BIBentryALTinterwordspacing
T.~Yamazato, ``{Overview of visible light communications with emphasis on image sensor communications},'' in \emph{2017 23rd Asia-Pacific Conference on Communications (APCC)}.\hskip 1em plus 0.5em minus 0.4em\relax IEEE, Dec. 2017, p. 1–6. [Online]. Available: \url{http://dx.doi.org/10.23919/APCC.2017.8304093}
\BIBentrySTDinterwordspacing

\bibitem{KAMAKURA2017}
\BIBentryALTinterwordspacing
K.~KAMAKURA, ``{Image Sensors Meet LEDs},'' \emph{IEICE Transactions on Communications}, vol. E100.B, no.~6, p. 917–925, 2017. [Online]. Available: \url{http://dx.doi.org/10.1587/transcom.2016LCI0001}
\BIBentrySTDinterwordspacing

\bibitem{Bastiaens2021}
\BIBentryALTinterwordspacing
S.~Bastiaens, J.~Mommerency, K.~Deprez, W.~Joseph, and D.~Plets, ``{Received Signal Strength Visible Light Positioning-based Precision Drone Landing System},'' in \emph{2021 International Conference on Indoor Positioning and Indoor Navigation (IPIN)}.\hskip 1em plus 0.5em minus 0.4em\relax IEEE, Nov. 2021, p. 1–8. [Online]. Available: \url{http://dx.doi.org/10.1109/IPIN51156.2021.9662584}
\BIBentrySTDinterwordspacing

\bibitem{Plets2017}
\BIBentryALTinterwordspacing
D.~Plets, A.~Eryildirim, S.~Bastiaens, N.~Stevens, L.~Martens, and W.~Joseph, ``{A Performance Comparison of Different Cost Functions for RSS-Based Visible Light Positioning Under the Presence of Reflections},'' in \emph{Proceedings of the 4th ACM Workshop on Visible Light Communication Systems}, ser. MobiCom ’17.\hskip 1em plus 0.5em minus 0.4em\relax ACM, Oct. 2017, p. 37–41. [Online]. Available: \url{http://dx.doi.org/10.1145/3129881.3129888}
\BIBentrySTDinterwordspacing

\bibitem{Jin2025}
\BIBentryALTinterwordspacing
L.~Jin, Q.~Jiang, S.~Yu, H.~Xu, W.~E, M.~Xie, X.~Li, and D.~Ma, ``{Lightweight RSS-based Visible Light Positioning Considering Dust Accumulation},'' \emph{Journal of Lightwave Technology}, p. 1–10, 2025. [Online]. Available: \url{http://dx.doi.org/10.1109/JLT.2025.3542181}
\BIBentrySTDinterwordspacing

\bibitem{oppenheim1999discrete}
A.~V. Oppenheim, R.~W. Schafer, and J.~R. Buck, \emph{{Discrete-Time Signal Processing}}, 2nd~ed.\hskip 1em plus 0.5em minus 0.4em\relax Upper Saddle River, NJ: Prentice Hall, 1999.

\bibitem{BarcoAlvrez2024}
\BIBentryALTinterwordspacing
J.~Barco~Alvárez, J.~C. Torres~Zafra, J.~S. Betancourt, M.~Morales~Cespedes, and C.~I. del Valle~Morales, ``{Navigating in Light: Precise Indoor Positioning Using Trilateration and Angular Diversity in a Semi-Spherical Photodiode Array with Visible Light Communication},'' \emph{Electronics}, vol.~13, no.~18, p. 3597, Sep. 2024. [Online]. Available: \url{http://dx.doi.org/10.3390/electronics13183597}
\BIBentrySTDinterwordspacing

\bibitem{Plets2019}
\BIBentryALTinterwordspacing
D.~Plets, Y.~Almadani, S.~Bastiaens, M.~Ijaz, L.~Martens, and W.~Joseph, ``{Efficient 3D trilateration algorithm for visible light positioning},'' \emph{Journal of Optics}, vol.~21, no.~5, p. 05LT01, Apr. 2019. [Online]. Available: \url{http://dx.doi.org/10.1088/2040-8986/ab1389}
\BIBentrySTDinterwordspacing

\bibitem{Zhang2016}
\BIBentryALTinterwordspacing
{C. Zhang and X. Zhang}, ``{LiTell: robust indoor localization using unmodified light fixtures},'' in \emph{Proceedings of the 22nd Annual International Conference on Mobile Computing and Networking}, ser. MobiCom'16.\hskip 1em plus 0.5em minus 0.4em\relax ACM, Oct. 2016. [Online]. Available: \url{http://dx.doi.org/10.1145/2973750.2973767}
\BIBentrySTDinterwordspacing

\bibitem{Zhang2016LiTell2}
\BIBentryALTinterwordspacing
C.~Zhang, S.~Zhou, and X.~Zhang, ``{Visible Light Localization Using Incumbent Light Fixtures: Demo Abstract},'' in \emph{Proceedings of the 14th ACM Conference on Embedded Network Sensor Systems CD-ROM}, ser. SenSys ’16.\hskip 1em plus 0.5em minus 0.4em\relax ACM, Nov. 2016, p. 304–305. [Online]. Available: \url{http://dx.doi.org/10.1145/2994551.2996534}
\BIBentrySTDinterwordspacing

\bibitem{Zhang2017}
\BIBentryALTinterwordspacing
{C. Zhang, and X. Zhang}, ``{Pulsar: Towards Ubiquitous Visible Light Localization},'' in \emph{Proceedings of the 23rd Annual International Conference on Mobile Computing and Networking}, ser. MobiCom ’17.\hskip 1em plus 0.5em minus 0.4em\relax ACM, Oct. 2017. [Online]. Available: \url{http://dx.doi.org/10.1145/3117811.3117821}
\BIBentrySTDinterwordspacing

\bibitem{ni_usb6212}
{National Instruments}, ``Usb-6212 specifications,'' \url{https://www.ni.com/docs/en-U.S./bundle/usb-6212-specs/page/specs.html}, accessed: 2024-11-21.

\bibitem{Wang2023}
\BIBentryALTinterwordspacing
Y.~Wang, J.~Hu, H.~Jia, W.~Hu, M.~Hassan, A.~Uddin, B.~Kusy, and M.~Youssef, ``{Spectral-Loc: Indoor Localization Using Light Spectral Information},'' \emph{Proceedings of the ACM on Interactive, Mobile, Wearable and Ubiquitous Technologies}, vol.~7, no.~1, pp. 1--26, Mar. 2023. [Online]. Available: \url{http://dx.doi.org/10.1145/3580878}
\BIBentrySTDinterwordspacing

\bibitem{Wang2022}
\BIBentryALTinterwordspacing
{Y. Wang, J. Hu, H. Jia, W. Hu, M. Hassan, A. Uddin, B. Kusy, and M. Youssef}, ``Indoor localization using light spectral information,'' in \emph{Proceedings of the 28th Annual International Conference on Mobile Computing And Networking}, ser. MobiCom '22.\hskip 1em plus 0.5em minus 0.4em\relax New York, NY, USA: Association for Computing Machinery, 2022, pp. 820--822. [Online]. Available: \url{https://doi.org/10.1145/3495243.3558249}
\BIBentrySTDinterwordspacing

\bibitem{Carver2017}
\BIBentryALTinterwordspacing
C.~Carver, S.~Wu, A.~Rogers, M.~Stafford, N.~S. Artan, and Z.~Dong, ``{Indoor Localization Through Visible Light Characterization Using Front-Facing Smartphone Camera},'' in \emph{2017 IEEE 14th International Conference on Mobile Ad Hoc and Sensor Systems (MASS)}.\hskip 1em plus 0.5em minus 0.4em\relax IEEE, Oct. 2017, p. 575–579. [Online]. Available: \url{http://dx.doi.org/10.1109/MASS.2017.102}
\BIBentrySTDinterwordspacing

\bibitem{Bellini2002}
\BIBentryALTinterwordspacing
C.~Bellini, S.~Panzieri, and F.~Pascucci, ``{A real-time architecture for low-cost vision based robots navigation},'' \emph{IFAC Proceedings Volumes}, vol.~35, no.~1, p. 479–484, 2002. [Online]. Available: \url{http://dx.doi.org/10.3182/20020721-6-ES-1901.00892}
\BIBentrySTDinterwordspacing

\bibitem{Chen2014}
\BIBentryALTinterwordspacing
X.~Chen and Y.~Jia, ``{Indoor localization for mobile robots using lampshade corners as landmarks: Visual system calibration, feature extraction and experiments},'' \emph{International Journal of Control, Automation and Systems}, vol.~12, no.~6, p. 1313–1322, Oct. 2014. [Online]. Available: \url{http://dx.doi.org/10.1007/s12555-013-0076-y}
\BIBentrySTDinterwordspacing

\bibitem{Alves2013}
\BIBentryALTinterwordspacing
P.~Alves, H.~Costelha, and C.~Neves, ``{Localization and navigation of a mobile robot in an office-like environment},'' in \emph{2013 13th International Conference on Autonomous Robot Systems}.\hskip 1em plus 0.5em minus 0.4em\relax IEEE, Apr. 2013, p. 1–6. [Online]. Available: \url{http://dx.doi.org/10.1109/Robotica.2013.6623536}
\BIBentrySTDinterwordspacing

\bibitem{launay2001fluorescent}
\BIBentryALTinterwordspacing
F.~Launay, A.~Ohya, and S.~Yuta, ``Autonomous indoor mobile robot navigation by detecting fluorescent tubes,'' 2001. [Online]. Available: \url{https://www.cs.tsukuba.ac.jp/~ohya/pdf/ICAR2001-FAB.pdf}
\BIBentrySTDinterwordspacing

\bibitem{Wang20188}
\BIBentryALTinterwordspacing
Q.~Wang, H.~Luo, A.~Men, F.~Zhao, and Y.~Huang, ``{An Infrastructure-Free Indoor Localization Algorithm for Smartphones},'' \emph{Sensors}, vol.~18, no.~10, p. 3317, Oct. 2018. [Online]. Available: \url{http://dx.doi.org/10.3390/s18103317}
\BIBentrySTDinterwordspacing

\bibitem{Wang2018DeepML}
\BIBentryALTinterwordspacing
X.~Wang, Z.~Yu, and S.~Mao, ``{DeepML: Deep LSTM for Indoor Localization with Smartphone Magnetic and Light Sensors},'' in \emph{2018 IEEE International Conference on Communications (ICC)}.\hskip 1em plus 0.5em minus 0.4em\relax IEEE, May 2018, p. 1–6. [Online]. Available: \url{http://dx.doi.org/10.1109/ICC.2018.8422562}
\BIBentrySTDinterwordspacing

\bibitem{Bridgelux_BXRE-35E2000-C-73}
\BIBentryALTinterwordspacing
{Bridgelux}, ``{BXRE-35E2000-C-73 LED Array},'' 2019. [Online]. Available: \url{https://www.bridgelux.com/sites/default/files/resource_media/Bridgelux%20DS101%20Gen%207%20V13%20Array%20Data%20Sheet%2020190930%20Rev%20N.pdf}
\BIBentrySTDinterwordspacing

\bibitem{NI_USB6212_Specs}
\BIBentryALTinterwordspacing
N.~Instruments, \emph{{USB-6212 Specifications}}, National Instruments, 2021. [Online]. Available: \url{https://www.ni.com/docs/en-US/bundle/usb-6212-specs/page/specs.html}
\BIBentrySTDinterwordspacing

\bibitem{SparkFun2022}
\BIBentryALTinterwordspacing
SparkFun, ``{SparkFun Triad Spectroscopy Sensor - AS7265x (Qwiic)},'' 2022, accessed: 2024-08-19. [Online]. Available: \url{https://www.sparkfun.com/products/15050}
\BIBentrySTDinterwordspacing

\bibitem{ams_AS7265x_2018}
\BIBentryALTinterwordspacing
{{AMS AG}}, ``{AS7265x Smart 18‑Channel VIS to NIR Spectral‑ID 3‑Sensor Chipset with Electronic Shutter Datasheet},'' {ams AG}, {Premstätten, Austria}, {Datasheet} {v1‑04}, July 2018, revision 1‑04, published July 9, 2018. [Online]. Available: \url{https://cdn.sparkfun.com/assets/c/2/9/0/a/AS7265x_Datasheet.pdf}
\BIBentrySTDinterwordspacing

\bibitem{PDA100A2_Manual}
\BIBentryALTinterwordspacing
{Thorlabs}, \emph{PDA100A2 Silicon Photodiode Power Sensor Manual}, Thorlabs, accessed: 2025-05-15. [Online]. Available: \url{https://www.montana.edu/ddickensheets/courses/eele482/handouts/PDA100A2-Manual.pdf}
\BIBentrySTDinterwordspacing

\bibitem{Thorlabs_PDA36A2}
\BIBentryALTinterwordspacing
Thorlabs, \emph{{PDA36A2 Si Switchable Gain Detector, 350 - 1100 nm, 12 MHz BW, 13 mm², Universal 8-32 / M4 Taps}}, Thorlabs, 2018, accessed: 2025-03-16. [Online]. Available: \url{https://www.thorlabs.com/thorproduct.cfm?partnumber=PDA36A2}
\BIBentrySTDinterwordspacing

\bibitem{Zhu2017}
\BIBentryALTinterwordspacing
S.~Zhu and X.~Zhang, ``{Enabling High-Precision Visible Light Localization in Today’s Buildings},'' in \emph{Proceedings of the 15th Annual International Conference on Mobile Systems, Applications, and Services}, ser. MobiSys’17.\hskip 1em plus 0.5em minus 0.4em\relax ACM, Jun. 2017. [Online]. Available: \url{http://dx.doi.org/10.1145/3081333.3081335}
\BIBentrySTDinterwordspacing

\bibitem{Barwar2023}
\BIBentryALTinterwordspacing
M.~K. Barwar, L.~K. Sahu, P.~R. Tripathi, P.~Bhatnagar, K.~K. Gupta, A.~H. Chander, and J.~M. Guerrero, ``{Demystifying the Devices Behind the LED Light: LED Driver Circuits},'' \emph{IEEE Industrial Electronics Magazine}, vol.~17, no.~1, p. 55–66, Mar. 2023. [Online]. Available: \url{http://dx.doi.org/10.1109/MIE.2022.3164526}
\BIBentrySTDinterwordspacing

\bibitem{Zhang2016(2)}
\BIBentryALTinterwordspacing
C.~Zhang, S.~Zhou, and X.~Zhang, ``{Visible Light Localization Using Incumbent Light Fixtures: Demo Abstract},'' in \emph{Proceedings of the 14th ACM Conference on Embedded Network Sensor Systems CD-ROM}, ser. SenSys ’16, vol.~51.\hskip 1em plus 0.5em minus 0.4em\relax ACM, Nov. 2016, p. 304–305. [Online]. Available: \url{http://dx.doi.org/10.1145/2994551.2996534}
\BIBentrySTDinterwordspacing

\bibitem{Wang2024}
\BIBentryALTinterwordspacing
Y.~Wang, ``{PhD Forum Abstract: Advancing Solar Cells: Beyond Energy Harvesting to Positioning and Communication},'' in \emph{2024 23rd ACM/IEEE International Conference on Information Processing in Sensor Networks (IPSN)}.\hskip 1em plus 0.5em minus 0.4em\relax IEEE, May 2024. [Online]. Available: \url{http://dx.doi.org/10.1109/IPSN61024.2024.00060}
\BIBentrySTDinterwordspacing

\bibitem{Lang2018}
\BIBentryALTinterwordspacing
{T. Lang et al.}, ``{LED-Based Visible Light Communication and Positioning Technology and SoCs (Invited)},'' in \emph{2018 14th IEEE International Conference on Solid-State and Integrated Circuit Technology (ICSICT)}.\hskip 1em plus 0.5em minus 0.4em\relax IEEE, Oct. 2018, p. 1–4. [Online]. Available: \url{http://dx.doi.org/10.1109/ICSICT.2018.8565725}
\BIBentrySTDinterwordspacing

\bibitem{statista_led_lighting}
Statista, ``Led lighting in the u.s.'' \url{https://www.statista.com/topics/1144/led-lighting-in-the-us/#topicOverview}, 2024, accessed: 2024-10-10.

\bibitem{GrandViewResearch2025}
\BIBentryALTinterwordspacing
G.~V. Research, ``{LED Lighting Market Size, Share \& Trends Analysis Report},'' 2025, accessed: 2025-02-09. [Online]. Available: \url{https://www.grandviewresearch.com/industry-analysis/led-lighting-market}
\BIBentrySTDinterwordspacing

\bibitem{Yan2021}
\BIBentryALTinterwordspacing
S.~Yan, Z.~Yin, and G.~Tan, ``{CurveLight: An Accurate and Practical Indoor Positioning System},'' in \emph{Proceedings of the 19th ACM Conference on Embedded Networked Sensor Systems}, ser. SenSys ’21.\hskip 1em plus 0.5em minus 0.4em\relax ACM, Nov. 2021, p. 152–164. [Online]. Available: \url{http://dx.doi.org/10.1145/3485730.3485934}
\BIBentrySTDinterwordspacing

\bibitem{Munir2019}
\BIBentryALTinterwordspacing
B.~Munir and V.~Dyo, ``{Passive Localization Through Light Flicker Fingerprinting},'' \emph{IEEE Sensors Journal}, vol.~19, no.~24, p. 12137–12144, Dec. 2019. [Online]. Available: \url{http://dx.doi.org/10.1109/JSEN.2019.2936899}
\BIBentrySTDinterwordspacing

\bibitem{Ma2020}
\BIBentryALTinterwordspacing
X.~Ma, S.~Bader, and B.~Oelmann, ``{Estimating Harvestable Energy in Time-Varying Indoor Light Conditions},'' in \emph{Proceedings of the 8th International Workshop on Energy Harvesting and Energy-Neutral Sensing Systems}, ser. SenSys ’20.\hskip 1em plus 0.5em minus 0.4em\relax ACM, Nov. 2020, p. 71–76. [Online]. Available: \url{http://dx.doi.org/10.1145/3417308.3430270}
\BIBentrySTDinterwordspacing

\bibitem{Liu2014}
\BIBentryALTinterwordspacing
J.~Liu, Y.~Chen, A.~Jaakkola, T.~Hakala, J.~Hyyppa, L.~Chen, J.~Tang, R.~Chen, and H.~Hyyppa, ``{The uses of ambient light for ubiquitous positioning},'' in \emph{{2014 IEEE/ION Position, Location and Navigation Symposium - PLANS 2014}}.\hskip 1em plus 0.5em minus 0.4em\relax IEEE, May 2014. [Online]. Available: \url{http://dx.doi.org/10.1109/PLANS.2014.6851363}
\BIBentrySTDinterwordspacing

\bibitem{Zhang2013}
\BIBentryALTinterwordspacing
Y.~Zhang, G.~Song, G.~Qiao, Y.~Wang, and W.~Wang, ``{Ambient light intensity based topology switching control for multi-robot system},'' in \emph{2013 IEEE International Conference on Information and Automation (ICIA)}, vol.~10.\hskip 1em plus 0.5em minus 0.4em\relax IEEE, Aug. 2013, p. 793–798. [Online]. Available: \url{http://dx.doi.org/10.1109/ICInfA.2013.6720402}
\BIBentrySTDinterwordspacing

\bibitem{Sprute2017}
\BIBentryALTinterwordspacing
D.~Sprute, A.~P\"{o}rtner, R.~Rasch, S.~Battermann, and M.~K\"{o}nig, \emph{{Ambient Assisted Robot Object Search}}.\hskip 1em plus 0.5em minus 0.4em\relax Springer International Publishing, 2017, p. 112–123. [Online]. Available: \url{http://dx.doi.org/10.1007/978-3-319-66188-9_10}
\BIBentrySTDinterwordspacing

\bibitem{Wu2022}
\BIBentryALTinterwordspacing
J.~Wu, Y.~Feng, and C.~K. Chang, ``{LiLo: ADL Localization with Conventional Luminaries and Ambient Light Sensor},'' \emph{Electronics}, vol.~11, no.~16, p. 2503, Aug. 2022. [Online]. Available: \url{http://dx.doi.org/10.3390/electronics11162503}
\BIBentrySTDinterwordspacing

\bibitem{Zhang2019}
\BIBentryALTinterwordspacing
C.~Zhang and X.~Zhang, ``{Visible Light Localization Using Conventional Light Fixtures and Smartphones},'' \emph{IEEE Transactions on Mobile Computing}, vol.~18, no.~12, p. 2968–2983, Dec. 2019. [Online]. Available: \url{http://dx.doi.org/10.1109/TMC.2018.2888973}
\BIBentrySTDinterwordspacing

\bibitem{Teledyne2024}
\BIBentryALTinterwordspacing
T.~V. Solutions, ``Rolling vs global shutter,'' 2024, accessed: 2025-03-06. [Online]. Available: \url{https://www.teledynevisionsolutions.com/learn/learning-center/imaging-fundamentals/rolling-vs-global-shutter/}
\BIBentrySTDinterwordspacing

\bibitem{Lee2015}
\BIBentryALTinterwordspacing
H.-Y. Lee, H.-M. Lin, Y.-L. Wei, H.-I. Wu, H.-M. Tsai, and K.~C.-J. Lin, ``{RollingLight: Enabling Line-of-Sight Light-to-Camera Communications},'' in \emph{Proceedings of the 13th Annual International Conference on Mobile Systems, Applications, and Services}, ser. MobiSys’15.\hskip 1em plus 0.5em minus 0.4em\relax ACM, May 2015, p. 167–180. [Online]. Available: \url{http://dx.doi.org/10.1145/2742647.2742651}
\BIBentrySTDinterwordspacing

\bibitem{Rizk2022}
\BIBentryALTinterwordspacing
H.~Rizk, D.~Ma, M.~Hassan, and M.~Youssef, ``{Indoor Localization using Solar Cells},'' in \emph{2022 IEEE International Conference on Pervasive Computing and Communications Workshops and other Affiliated Events (PerCom Workshops)}.\hskip 1em plus 0.5em minus 0.4em\relax IEEE, Mar. 2022, p. 38–41. [Online]. Available: \url{http://dx.doi.org/10.1109/PerComWorkshops53856.2022.9767256}
\BIBentrySTDinterwordspacing

\bibitem{Hui2020}
\BIBentryALTinterwordspacing
R.~Hui, \emph{Photodetectors}.\hskip 1em plus 0.5em minus 0.4em\relax Elsevier, 2020, p. 125–154. [Online]. Available: \url{http://dx.doi.org/10.1016/B978-0-12-805345-4.00004-4}
\BIBentrySTDinterwordspacing

\bibitem{Paschotta2006}
\BIBentryALTinterwordspacing
R.~Paschotta, \emph{Photodiodes - an encyclopedia article}.\hskip 1em plus 0.5em minus 0.4em\relax RP Photonics AG, 2006. [Online]. Available: \url{http://dx.doi.org/10.61835/8jq}
\BIBentrySTDinterwordspacing

\bibitem{Alijani2024}
\BIBentryALTinterwordspacing
M.~Alijani, W.~Joseph, E.~De~Poorter, M.~Deruyck, and D.~Plets, ``{ChickSense: Toward Poultry Welfare With Machine Learning-Assisted Visible Light Sensing},'' \emph{IEEE Sensors Journal}, vol.~24, no.~21, p. 36178–36193, Nov. 2024. [Online]. Available: \url{http://dx.doi.org/10.1109/JSEN.2024.3463209}
\BIBentrySTDinterwordspacing

\bibitem{Deprez2020}
\BIBentryALTinterwordspacing
K.~Deprez, S.~Bastiaens, L.~Martens, W.~Joseph, and D.~Plets, ``{Passive Visible Light Detection of Humans},'' \emph{Sensors}, vol.~20, no.~7, p. 1902, Mar. 2020. [Online]. Available: \url{http://dx.doi.org/10.3390/s20071902}
\BIBentrySTDinterwordspacing

\bibitem{Zhuang2024}
\BIBentryALTinterwordspacing
Y.~Zhuang, Y.~Wang, X.~Yang, and T.~Ma, ``{Visible light positioning system using a smartphone’s built-in ambient light sensor and inertial measurement unit},'' \emph{Optics Letters}, vol.~49, no.~8, p. 2105, Apr. 2024. [Online]. Available: \url{http://dx.doi.org/10.1364/OL.519674}
\BIBentrySTDinterwordspacing

\bibitem{Wang2019}
\BIBentryALTinterwordspacing
Z.~Wang, Z.~Yang, Q.~Huang, L.~Yang, and Q.~Zhang, ``{ALS-P: Light Weight Visible Light Positioning via Ambient Light Sensor},'' in \emph{IEEE INFOCOM 2019 - IEEE Conference on Computer Communications}.\hskip 1em plus 0.5em minus 0.4em\relax IEEE, Apr. 2019, p. 1306–1314. [Online]. Available: \url{http://dx.doi.org/10.1109/INFOCOM.2019.8737575}
\BIBentrySTDinterwordspacing

\bibitem{Liu2024}
\BIBentryALTinterwordspacing
A.~Liu, W.~Shi, M.~Safari, W.~Liu, and J.~Cao, ``{Design Guidelines for Optical Camera Communication Systems: A Tutorial},'' \emph{IEEE Photonics Journal}, vol.~16, no.~4, p. 1–25, Aug. 2024. [Online]. Available: \url{http://dx.doi.org/10.1109/JPHOT.2024.3424885}
\BIBentrySTDinterwordspacing

\bibitem{Sato2022}
\BIBentryALTinterwordspacing
T.~Sato, S.~Shimada, H.~Murakami, H.~Watanabe, H.~Hashizume, and M.~Sugimoto, ``{ALiSA: A Visible-Light Positioning System Using the Ambient Light Sensor Assembly in a Smartphone},'' \emph{IEEE Sensors Journal}, vol.~22, no.~6, p. 4989–5000, Mar. 2022. [Online]. Available: \url{http://dx.doi.org/10.1109/JSEN.2021.3074580}
\BIBentrySTDinterwordspacing

\bibitem{Otsuka2025}
\BIBentryALTinterwordspacing
S.~Otsuka, Y.~Onishi, M.~Nakamura, H.~Hashizume, and M.~Sugimoto, ``{ALS+PDR: Indoor Pedestrian Dead Reckoning Using a Smartphone Ambient Light Sensor},'' \emph{IEEE Journal of Indoor and Seamless Positioning and Navigation}, vol.~3, p. 43–52, 2025. [Online]. Available: \url{http://dx.doi.org/10.1109/JISPIN.2025.3541991}
\BIBentrySTDinterwordspacing

\bibitem{Yang2018}
\BIBentryALTinterwordspacing
L.~Yang, Z.~Wang, W.~Wang, and Q.~Zhang, ``{NALoc: Nonlinear Ambient-Light-Sensor-based Localization System},'' \emph{Proceedings of the ACM on Interactive, Mobile, Wearable and Ubiquitous Technologies}, vol.~2, no.~4, p. 1–22, Dec. 2018. [Online]. Available: \url{http://dx.doi.org/10.1145/3287077}
\BIBentrySTDinterwordspacing

\bibitem{Elamrawy2025}
\BIBentryALTinterwordspacing
F.~M. Elamrawy, A.~A. El~Aziz, S.~Khamis, and H.~Kasem, ``{Performance analysis of an indoor visible light communication system using LED configurations and diverse photodetectors},'' \emph{Scientific Reports}, vol.~15, no.~1, May 2025. [Online]. Available: \url{http://dx.doi.org/10.1038/s41598-025-99643-z}
\BIBentrySTDinterwordspacing

\bibitem{Alam2019}
\BIBentryALTinterwordspacing
F.~Alam, N.~Faulkner, M.~Legg, and S.~Demidenko, ``{Indoor Visible Light Positioning Using Spring-Relaxation Technique in Real-World Setting},'' \emph{IEEE Access}, vol.~7, p. 91347–91359, 2019. [Online]. Available: \url{http://dx.doi.org/10.1109/ACCESS.2019.2927922}
\BIBentrySTDinterwordspacing

\bibitem{Li2017}
\BIBentryALTinterwordspacing
Z.~Li, A.~Yang, H.~Lv, L.~Feng, and W.~Song, ``{Fusion of Visible Light Indoor Positioning and Inertial Navigation Based on Particle Filter},'' \emph{IEEE Photonics Journal}, vol.~9, no.~5, p. 1–13, Oct. 2017. [Online]. Available: \url{http://dx.doi.org/10.1109/JPHOT.2017.2733556}
\BIBentrySTDinterwordspacing

\bibitem{Plets20199}
\BIBentryALTinterwordspacing
D.~Plets, S.~Bastiaens, M.~Ijaz, Y.~Almadani, L.~Martens, W.~Raes, N.~Stevens, and W.~Joseph, ``{Three-dimensional Visible Light Positioning: an Experimental Assessment of the Importance of the LEDs’ Locations},'' in \emph{2019 International Conference on Indoor Positioning and Indoor Navigation (IPIN)}.\hskip 1em plus 0.5em minus 0.4em\relax IEEE, Sep. 2019, p. 1–6. [Online]. Available: \url{http://dx.doi.org/10.1109/IPIN.2019.8911763}
\BIBentrySTDinterwordspacing

\bibitem{Almadani2021}
\BIBentryALTinterwordspacing
Y.~Almadani, M.~Ijaz, B.~Adebisi, S.~Rajbhandari, S.~Bastiaens, W.~Joseph, and D.~Plets, ``{An experimental evaluation of a 3D visible light positioning system in an industrial environment with receiver tilt and multipath reflections},'' \emph{Optics Communications}, vol. 483, p. 126654, Mar. 2021. [Online]. Available: \url{http://dx.doi.org/10.1016/j.optcom.2020.126654}
\BIBentrySTDinterwordspacing

\bibitem{Shao2018}
\BIBentryALTinterwordspacing
S.~Shao, A.~Khreishah, and I.~Khalil, ``{RETRO: Retroreflector Based Visible Light Indoor Localization for Real-time Tracking of IoT Devices},'' in \emph{IEEE INFOCOM 2018 - IEEE Conference on Computer Communications}.\hskip 1em plus 0.5em minus 0.4em\relax IEEE, Apr. 2018, p. 1025–1033. [Online]. Available: \url{http://dx.doi.org/10.1109/INFOCOM.2018.8485817}
\BIBentrySTDinterwordspacing

\bibitem{Sun2025}
\BIBentryALTinterwordspacing
X.~Sun, Y.~Zhuang, Z.~Zheng, H.~Zhang, B.~Wang, X.~Wang, and J.~Zhou, ``{Tightly coupled integration of Visible Light Positioning, GNSS, and INS for indoor/outdoor transition areas},'' \emph{Information Fusion}, vol. 117, p. 102781, May 2025. [Online]. Available: \url{http://dx.doi.org/10.1016/j.inffus.2024.102781}
\BIBentrySTDinterwordspacing

\bibitem{Specim2024}
\BIBentryALTinterwordspacing
Specim, ``{How Does Spectral Sensing Work?}'' 2024, accessed: 2024-08-20. [Online]. Available: \url{https://www.specim.com/technology/how-does-spectral-sensing-work/}
\BIBentrySTDinterwordspacing

\bibitem{Jiawei2024}
{J. Hu et al.}, ``{LiDARSpectra: Synthetic Indoor Spectral Mapping with Low-cost LiDARs},'' in \emph{2024 23rd ACM/IEEE International Conference on Information Processing in Sensor Networks (IPSN)}, 2024, pp. 75--87.

\bibitem{Kranjc2006}
\BIBentryALTinterwordspacing
T.~Kranjc, ``{Light intensity measurement},'' in \emph{Unconventional Imaging II}, V.~L. Gamiz, P.~S. Idell, and M.~S. Strojnik, Eds., vol. 6307.\hskip 1em plus 0.5em minus 0.4em\relax SPIE, Aug. 2006, p. 63070Q. [Online]. Available: \url{http://dx.doi.org/10.1117/12.681721}
\BIBentrySTDinterwordspacing

\bibitem{Xiaomi}
\BIBentryALTinterwordspacing
DXOMARK, ``{Xiaomi Mi 11 Ultra Camera Test: Large Sensor Power},'' 2021, accessed: 2024-08-19. [Online]. Available: \url{https://www.dxomark.com/xiaomi-mi-11-ultra-camera-review-large-sensor-power/}
\BIBentrySTDinterwordspacing

\bibitem{Danys2022}
\BIBentryALTinterwordspacing
L.~Danys, I.~Zolotova, D.~Romero, P.~Papcun, E.~Kajati, R.~Jaros, P.~Koudelka, J.~Koziorek, and R.~Martinek, ``{Visible Light Communication and localization: A study on tracking solutions for Industry 4.0 and the Operator 4.0},'' \emph{Journal of Manufacturing Systems}, vol.~64, p. 535–545, Jul. 2022. [Online]. Available: \url{http://dx.doi.org/10.1016/j.jmsy.2022.07.011}
\BIBentrySTDinterwordspacing

\bibitem{Sisinni2018}
\BIBentryALTinterwordspacing
E.~Sisinni, A.~Saifullah, S.~Han, U.~Jennehag, and M.~Gidlund, ``{Industrial Internet of Things: Challenges, Opportunities, and Directions},'' \emph{IEEE Transactions on Industrial Informatics}, vol.~14, no.~11, p. 4724–4734, Nov. 2018. [Online]. Available: \url{http://dx.doi.org/10.1109/TII.2018.2852491}
\BIBentrySTDinterwordspacing

\bibitem{Rekkas2023}
\BIBentryALTinterwordspacing
V.~P. Rekkas, L.~A. Iliadis, S.~P. Sotiroudis, A.~D. Boursianis, P.~Sarigiannidis, D.~Plets, W.~Joseph, S.~Wan, C.~G. Christodoulou, G.~K. Karagiannidis, and S.~K. Goudos, ``{Artificial Intelligence in Visible Light Positioning for Indoor IoT: A Methodological Review},'' \emph{IEEE Open Journal of the Communications Society}, vol.~4, p. 2838–2869, 2023. [Online]. Available: \url{http://dx.doi.org/10.1109/OJCOMS.2023.3327211}
\BIBentrySTDinterwordspacing

\bibitem{Salem2021}
\BIBentryALTinterwordspacing
Z.~Salem, A.~P. Weiss, and F.-P. Wenzl, ``{A low-complexity approach for visible light positioning and space-resolved human activity recognition},'' in \emph{Multimodal Sensing and Artificial Intelligence: Technologies and Applications II}, S.~Negahdaripour, E.~Stella, D.~Ceglarek, and C.~M\"{o}ller, Eds.\hskip 1em plus 0.5em minus 0.4em\relax SPIE, Jun. 2021, p.~14. [Online]. Available: \url{http://dx.doi.org/10.1117/12.2593291}
\BIBentrySTDinterwordspacing

\bibitem{Golding1999}
\BIBentryALTinterwordspacing
A.~Golding and N.~Lesh, ``{Indoor navigation using a diverse set of cheap, wearable sensors},'' in \emph{Digest of Papers. Third International Symposium on Wearable Computers}, ser. ISWC-99.\hskip 1em plus 0.5em minus 0.4em\relax IEEE Comput. Soc, p. 29–36. [Online]. Available: \url{http://dx.doi.org/10.1109/ISWC.1999.806640}
\BIBentrySTDinterwordspacing

\bibitem{Randall2007}
\BIBentryALTinterwordspacing
J.~Randall, O.~Amft, J.~Bohn, and M.~Burri, ``{LuxTrace: indoor positioning using building illumination},'' \emph{Personal and Ubiquitous Computing}, vol.~11, no.~6, p. 417–428, Mar. 2007. [Online]. Available: \url{http://dx.doi.org/10.1007/s00779-006-0097-0}
\BIBentrySTDinterwordspacing

\bibitem{Ullah2023}
\BIBentryALTinterwordspacing
M.~H. Ullah, G.~Gelli, and F.~Verde, ``{Visible light backscattering with applications to the Internet of Things: State-of-the-art, challenges, and opportunities},'' \emph{Internet of Things}, vol.~22, p. 100768, Jul. 2023. [Online]. Available: \url{http://dx.doi.org/10.1016/j.iot.2023.100768}
\BIBentrySTDinterwordspacing

\bibitem{Singh2020}
\BIBentryALTinterwordspacing
J.~Singh and U.~Raza, ``{Passive visible light positioning systems: an overview},'' in \emph{Proceedings of the Workshop on Light Up the IoT}, ser. MobiCom ’20.\hskip 1em plus 0.5em minus 0.4em\relax ACM, Sep. 2020. [Online]. Available: \url{http://dx.doi.org/10.1145/3412449.3412553}
\BIBentrySTDinterwordspacing

\bibitem{Wang2020}
\BIBentryALTinterwordspacing
Q.~Wang and M.~Zuniga, ``{Passive visible light networks: taxonomy and opportunities},'' in \emph{Proceedings of the Workshop on Light Up the IoT}, ser. MobiCom ’20.\hskip 1em plus 0.5em minus 0.4em\relax ACM, Sep. 2020, p. 42–47. [Online]. Available: \url{http://dx.doi.org/10.1145/3412449.3412551}
\BIBentrySTDinterwordspacing

\bibitem{Murdoch1981}
\BIBentryALTinterwordspacing
J.~B. Murdoch, ``{Inverse Square Law Approximation of Illuminance},'' \emph{Journal of the Illuminating Engineering Society}, vol.~10, no.~2, p. 96–106, Jan. 1981. [Online]. Available: \url{http://dx.doi.org/10.1080/00994480.1980.10748595}
\BIBentrySTDinterwordspacing

\bibitem{Taylor2000}
A.~E. Taylor \emph{et~al.}, \emph{{Illumination Fundamentals}}, 2000.

\bibitem{Kahn1997}
\BIBentryALTinterwordspacing
J.~Kahn and J.~Barry, ``{Wireless infrared communications},'' \emph{Proceedings of the IEEE}, vol.~85, no.~2, pp. 265--298, 1997. [Online]. Available: \url{http://dx.doi.org/10.1109/5.554222}
\BIBentrySTDinterwordspacing

\bibitem{QiangXu2015}
\BIBentryALTinterwordspacing
Q.~Xu and R.~Zheng, ``{Automated detection of burned-out luminaries using indoor positioning},'' in \emph{2015 International Conference on Indoor Positioning and Indoor Navigation (IPIN)}.\hskip 1em plus 0.5em minus 0.4em\relax IEEE, Oct. 2015, p. 1–9. [Online]. Available: \url{http://dx.doi.org/10.1109/IPIN.2015.7346947}
\BIBentrySTDinterwordspacing

\bibitem{Moreno2008}
\BIBentryALTinterwordspacing
I.~Moreno and C.-C. Sun, ``{Modeling the radiation pattern of LEDs},'' \emph{Optics Express}, vol.~16, no.~3, p. 1808, 2008. [Online]. Available: \url{http://dx.doi.org/10.1364/OE.16.001808}
\BIBentrySTDinterwordspacing

\bibitem{Bhalerao2016}
\BIBentryALTinterwordspacing
M.~Bhalerao, M.~Sumathi, and S.~Sonavane, ``{Line of sight model for visible light communication using Lambertian radiation pattern of LED},'' \emph{International Journal of Communication Systems}, vol.~30, no.~11, Dec. 2016. [Online]. Available: \url{http://dx.doi.org/10.1002/dac.3250}
\BIBentrySTDinterwordspacing

\bibitem{Vatansever2017}
\BIBentryALTinterwordspacing
Z.~Vatansever and M.~Brandt-Pearce, ``{Visible Light Positioning with Diffusing Lamps Using an Extended Kalman Filter},'' in \emph{2017 IEEE Wireless Communications and Networking Conference (WCNC)}.\hskip 1em plus 0.5em minus 0.4em\relax IEEE, Mar. 2017, p. 1–6. [Online]. Available: \url{http://dx.doi.org/10.1109/WCNC.2017.7925652}
\BIBentrySTDinterwordspacing

\bibitem{Verniers2017}
\BIBentryALTinterwordspacing
K.~Verniers, L.~Van~der Perre, and N.~Stevens, ``{Impact of wavelength dependency of LED and photodiode in visible light positioning},'' in \emph{2017 20th International Symposium on Wireless Personal Multimedia Communications (WPMC)}.\hskip 1em plus 0.5em minus 0.4em\relax IEEE, Dec. 2017, p. 217–222. [Online]. Available: \url{http://dx.doi.org/10.1109/WPMC.2017.8301811}
\BIBentrySTDinterwordspacing

\bibitem{Ghassemlooynew}
\BIBentryALTinterwordspacing
\emph{{Visible Light Communications: Theory and Applications}}.\hskip 1em plus 0.5em minus 0.4em\relax CRC Press, Jun. 2017. [Online]. Available: \url{http://dx.doi.org/10.1201/9781315367330}
\BIBentrySTDinterwordspacing

\bibitem{Sander2021}
\BIBentryALTinterwordspacing
S.~Bastiaens, W.~Raes, N.~Stevens, L.~Martens, W.~Joseph, and D.~Plets, ``{Impact of a Photodiode's Angular Characteristics on RSS-Based VLP Accuracy},'' \emph{IEEE Access}, vol.~8, pp. 83\,116--83\,130, 2020. [Online]. Available: \url{http://dx.doi.org/10.1109/ACCESS.2020.2991298}
\BIBentrySTDinterwordspacing

\bibitem{phdthesis}
K.~Dividis, ``\BIBforeignlanguage{English}{Design and prototyping of a visible light indoor positioning system},'' EngD Thesis, 2007, eindverslag.

\bibitem{AppliedVLP}
\BIBentryALTinterwordspacing
R.~Amsters, ``{Applied Visible Light Positioning},'' Ph.D. dissertation, KU Leuven, 2021, accessed: 2025-05-15. [Online]. Available: \url{https://lirias.kuleuven.be/retrieve/635434}
\BIBentrySTDinterwordspacing

\bibitem{JMLR:v9:vandermaaten08a}
\BIBentryALTinterwordspacing
{L. V. D. Maaten and G. Hinton}, ``Visualizing data using t-sne,'' \emph{Journal of Machine Learning Research}, vol.~9, no.~86, pp. 2579--2605, 2008. [Online]. Available: \url{http://jmlr.org/papers/v9/vandermaaten08a.html}
\BIBentrySTDinterwordspacing

\bibitem{Hu2024}
\BIBentryALTinterwordspacing
J.~Hu, Y.~Wang, H.~Jia, C.~Jiang, M.~Hassan, B.~Kusy, and W.~Hu, ``{LiDARSpectra: Synthetic Indoor Spectral Mapping with Low-cost LiDARs},'' in \emph{2024 23rd ACM/IEEE International Conference on Information Processing in Sensor Networks (IPSN)}, vol.~35.\hskip 1em plus 0.5em minus 0.4em\relax IEEE, May 2024, p. 75–87. [Online]. Available: \url{http://dx.doi.org/10.1109/IPSN61024.2024.00011}
\BIBentrySTDinterwordspacing

\bibitem{Hu2023}
\BIBentryALTinterwordspacing
J.~Hu, Y.~Wang, H.~Jia, W.~Hu, M.~Hassan, B.~Kusy, A.~Uddin, and M.~Youssef, ``{Iris: Passive Visible Light Positioning Using Light Spectral Information},'' \emph{Proceedings of the ACM on Interactive, Mobile, Wearable and Ubiquitous Technologies}, vol.~7, no.~3, p. 1–27, Sep. 2023. [Online]. Available: \url{http://dx.doi.org/10.1145/3610913}
\BIBentrySTDinterwordspacing

\bibitem{Azizyan2009}
\BIBentryALTinterwordspacing
M.~Azizyan, I.~Constandache, and R.~Roy~Choudhury, ``{SurroundSense: mobile phone localization via ambience fingerprinting},'' in \emph{Proceedings of the 15th annual international conference on Mobile computing and networking}, ser. MobiCom’09.\hskip 1em plus 0.5em minus 0.4em\relax ACM, Sep. 2009. [Online]. Available: \url{http://dx.doi.org/10.1145/1614320.1614350}
\BIBentrySTDinterwordspacing

\bibitem{Rahaim2012}
\BIBentryALTinterwordspacing
M.~Rahaim, G.~B. Prince, and T.~D. Little, ``{State estimation and motion tracking for spatially diverse VLC networks},'' in \emph{2012 IEEE Globecom Workshops}.\hskip 1em plus 0.5em minus 0.4em\relax IEEE, Dec. 2012, p. 1249–1253. [Online]. Available: \url{http://dx.doi.org/10.1109/GLOCOMW.2012.6477760}
\BIBentrySTDinterwordspacing

\bibitem{Saengudomlert2024}
\BIBentryALTinterwordspacing
P.~Saengudomlert and P.~Ubolkosold, ``{Joint Position and Orientation Estimation for Visible Light Positioning Using Extended Kalman Filters},'' in \emph{2024 21st International Conference on Electrical Engineering/Electronics, Computer, Telecommunications and Information Technology (ECTI-CON)}.\hskip 1em plus 0.5em minus 0.4em\relax IEEE, May 2024, p. 1–5. [Online]. Available: \url{http://dx.doi.org/10.1109/ECTI-CON60892.2024.10594863}
\BIBentrySTDinterwordspacing

\bibitem{Li20177}
\BIBentryALTinterwordspacing
Z.~Li, L.~Feng, and A.~Yang, ``{Fusion Based on Visible Light Positioning and Inertial Navigation Using Extended Kalman Filters},'' \emph{Sensors}, vol.~17, no.~5, p. 1093, May 2017. [Online]. Available: \url{http://dx.doi.org/10.3390/s17051093}
\BIBentrySTDinterwordspacing

\bibitem{Hu2015}
\BIBentryALTinterwordspacing
Y.~Hu, Y.~Xiong, W.~Huang, X.-Y. Li, Y.~Zhang, X.~Mao, P.~Yang, and C.~Wang, ``{Lightitude: Indoor Positioning Using Ubiquitous Visible Lights and COTS Devices},'' in \emph{2015 IEEE 35th International Conference on Distributed Computing Systems}.\hskip 1em plus 0.5em minus 0.4em\relax IEEE, Jun. 2015, p. 732–733. [Online]. Available: \url{http://dx.doi.org/10.1109/ICDCS.2015.82}
\BIBentrySTDinterwordspacing

\bibitem{Jimenez2013}
\BIBentryALTinterwordspacing
A.~Jimenez, F.~Zampella, and F.~Seco, ``{Light-matching: A new signal of opportunity for pedestrian indoor navigation},'' in \emph{International Conference on Indoor Positioning and Indoor Navigation}.\hskip 1em plus 0.5em minus 0.4em\relax IEEE, Oct. 2013. [Online]. Available: \url{http://dx.doi.org/10.1109/IPIN.2013.6817843}
\BIBentrySTDinterwordspacing

\bibitem{Fleury1999}
\BIBentryALTinterwordspacing
B.~Fleury, M.~Tschudin, R.~Heddergott, D.~Dahlhaus, and K.~Ingeman~Pedersen, ``{Channel parameter estimation in mobile radio environments using the SAGE algorithm},'' \emph{IEEE Journal on Selected Areas in Communications}, vol.~17, no.~3, p. 434–450, Mar. 1999. [Online]. Available: \url{http://dx.doi.org/10.1109/49.753729}
\BIBentrySTDinterwordspacing

\bibitem{vanderBroeck2007}
\BIBentryALTinterwordspacing
H.~van~der Broeck, G.~Sauerlander, and M.~Wendt, ``{Power driver topologies and control schemes for LEDs},'' in \emph{APEC 07 - Twenty-Second Annual IEEE Applied Power Electronics Conference and Exposition}.\hskip 1em plus 0.5em minus 0.4em\relax IEEE, Feb. 2007. [Online]. Available: \url{http://dx.doi.org/10.1109/APEX.2007.357686}
\BIBentrySTDinterwordspacing

\bibitem{Schmidt1986}
\BIBentryALTinterwordspacing
R.~Schmidt, ``{Multiple emitter location and signal parameter estimation},'' \emph{IEEE Transactions on Antennas and Propagation}, vol.~34, no.~3, p. 276–280, Mar. 1986. [Online]. Available: \url{http://dx.doi.org/10.1109/TAP.1986.1143830}
\BIBentrySTDinterwordspacing

\bibitem{Thomas2005}
\BIBentryALTinterwordspacing
F.~Thomas and L.~Ros, ``{Revisiting trilateration for robot localization},'' \emph{IEEE Transactions on Robotics}, vol.~21, no.~1, p. 93–101, Feb. 2005. [Online]. Available: \url{http://dx.doi.org/10.1109/TRO.2004.833793}
\BIBentrySTDinterwordspacing

\bibitem{WANG1997}
\BIBentryALTinterwordspacing
H.~WANG, T.~ISHIMATSU, and J.~T. MIAN, ``{Self-Localization for an Electric Wheelchair},'' \emph{JSME International Journal Series C}, vol.~40, no.~3, p. 433–438, 1997. [Online]. Available: \url{http://dx.doi.org/10.1299/jsmec.40.433}
\BIBentrySTDinterwordspacing

\bibitem{Wang2005}
\BIBentryALTinterwordspacing
H.~Wang and T.~Ishimatsu, ``{Vision-Based Navigation for an Electric Wheelchair Using Ceiling Light Landmark},'' \emph{Journal of Intelligent and Robotic Systems}, vol.~41, no.~4, p. 283–314, Jan. 2005. [Online]. Available: \url{http://dx.doi.org/10.1007/s10846-005-9902-7}
\BIBentrySTDinterwordspacing

\bibitem{lirias3591348}
\BIBentryALTinterwordspacing
R.~Amsters, P.~Slaets, N.~Stevens, and E.~Demeester, ``\BIBforeignlanguage{eng}{{Applied Visible Light Positioning}},'' 2021-10-28. [Online]. Available: \url{https://kuleuven.limo.libis.be/discovery/search?query=any,contains,LIRIAS3591348&tab=LIRIAS&search_scope=lirias_profile&vid=32KUL_KUL:Lirias&offset=0}
\BIBentrySTDinterwordspacing

\bibitem{Panzieri2005}
\BIBentryALTinterwordspacing
S.~Panzieri, F.~Pascucci, and R.~Setola, ``{Interlaced extended Kalman filter for real time navigation},'' in \emph{2005 IEEE/RSJ International Conference on Intelligent Robots and Systems}.\hskip 1em plus 0.5em minus 0.4em\relax IEEE, 2005, p. 2780–2785. [Online]. Available: \url{http://dx.doi.org/10.1109/IROS.2005.1544979}
\BIBentrySTDinterwordspacing

\bibitem{Folkesson2005}
\BIBentryALTinterwordspacing
J.~Folkesson, P.~Jensfelt, and H.~Christensen, ``{Vision SLAM in the Measurement Subspace},'' in \emph{Proceedings of the 2005 IEEE International Conference on Robotics and Automation}.\hskip 1em plus 0.5em minus 0.4em\relax IEEE, p. 30–35. [Online]. Available: \url{http://dx.doi.org/10.1109/ROBOT.2005.1570092}
\BIBentrySTDinterwordspacing

\bibitem{Hwang2011}
\BIBentryALTinterwordspacing
S.-Y. Hwang and J.-B. Song, ``{Monocular Vision-Based SLAM in Indoor Environment Using Corner, Lamp, and Door Features From Upward-Looking Camera},'' \emph{IEEE Transactions on Industrial Electronics}, vol.~58, no.~10, p. 4804–4812, Oct. 2011. [Online]. Available: \url{http://dx.doi.org/10.1109/TIE.2011.2109333}
\BIBentrySTDinterwordspacing

\bibitem{Zhang20199}
\BIBentryALTinterwordspacing
{H. Zhang, A. Zhou, D. Xu, and S. Xu, X. Zhang, and H. Ma}, ``{Learning to Recognize Unmodified Lights with Invisible Features},'' \emph{Proceedings of the ACM on Interactive, Mobile, Wearable and Ubiquitous Technologies}, vol.~3, no.~2, p. 1–23, Jun. 2019. [Online]. Available: \url{http://dx.doi.org/10.1145/3328938}
\BIBentrySTDinterwordspacing

\bibitem{Zhao2017}
\BIBentryALTinterwordspacing
Z.~Zhao, J.~Wang, X.~Zhao, C.~Peng, Q.~Guo, and B.~Wu, ``{NaviLight: Indoor localization and navigation under arbitrary lights},'' in \emph{IEEE INFOCOM 2017 - IEEE Conference on Computer Communications}.\hskip 1em plus 0.5em minus 0.4em\relax IEEE, May 2017, p. 1–9. [Online]. Available: \url{http://dx.doi.org/10.1109/INFOCOM.2017.8057184}
\BIBentrySTDinterwordspacing

\bibitem{Liu2019Graph}
\BIBentryALTinterwordspacing
W.~Liu, H.~Jiang, G.~Jiang, J.~Liu, X.~Ma, Y.~Jia, and F.~Xiao, ``{Indoor Navigation With Virtual Graph Representation: Exploiting Peak Intensities of Unmodulated Luminaries},'' \emph{IEEE/ACM Transactions on Networking}, vol.~27, no.~1, p. 187–200, Feb. 2019. [Online]. Available: \url{http://dx.doi.org/10.1109/TNET.2018.2884088}
\BIBentrySTDinterwordspacing

\bibitem{Liang2020}
\BIBentryALTinterwordspacing
Q.~Liang and M.~Liu, ``{An Automatic Site Survey Approach for Indoor Localization Using a Smartphone},'' \emph{IEEE Transactions on Automation Science and Engineering}, vol.~17, no.~1, p. 191–206, Jan. 2020. [Online]. Available: \url{http://dx.doi.org/10.1109/TASE.2019.2918030}
\BIBentrySTDinterwordspacing

\bibitem{Umetsu2019}
Y.~Umetsu, Y.~Nakamura, Y.~Arakawa, M.~Fujimoto, and H.~Suwa, ``{EHAAS: Energy Harvesters As A Sensor for Place Recognition on Wearables},'' in \emph{2019 IEEE International Conference on Pervasive Computing and Communications (PerCom)}, 2019, pp. 1--10.

\bibitem{Singh20244}
\BIBentryALTinterwordspacing
J.~Singh, M.~Zuniga, T.~Farnham, and Q.~Wang, ``{HueLoc: Localization Through LEDs’ Hue Spectrum},'' \emph{IEEE Internet of Things Journal}, p. 1–1, 2024. [Online]. Available: \url{http://dx.doi.org/10.1109/JIOT.2024.3512943}
\BIBentrySTDinterwordspacing

\bibitem{Jimnez2014}
\BIBentryALTinterwordspacing
A.~Jiménez, F.~Zampella, and F.~Seco, ``{Improving Inertial Pedestrian Dead-Reckoning by Detecting Unmodified Switched-on Lamps in Buildings},'' \emph{Sensors}, vol.~14, no.~1, p. 731–769, Jan. 2014. [Online]. Available: \url{http://dx.doi.org/10.3390/s140100731}
\BIBentrySTDinterwordspacing

\bibitem{HamidiRad2017}
\BIBentryALTinterwordspacing
S.~Hamidi-Rad, K.~Lyons, and N.~Goela, ``{Infrastructure-less indoor localization using light fingerprints},'' in \emph{2017 IEEE International Conference on Acoustics, Speech and Signal Processing (ICASSP)}.\hskip 1em plus 0.5em minus 0.4em\relax IEEE, Mar. 2017. [Online]. Available: \url{http://dx.doi.org/10.1109/ICASSP.2017.7953307}
\BIBentrySTDinterwordspacing

\bibitem{Bastiaens2020(2)}
\BIBentryALTinterwordspacing
S.~Bastiaens, K.~Deprez, L.~Martens, W.~Joseph, and D.~Plets, ``{Experimental Assessment of the Accuracy of Modulated and Unmodulated Visible Light Positioning},'' in \emph{2020 IEEE International Symposium on Broadband Multimedia Systems and Broadcasting (BMSB)}.\hskip 1em plus 0.5em minus 0.4em\relax IEEE, Oct. 2020, p. 1–5. [Online]. Available: \url{http://dx.doi.org/10.1109/BMSB49480.2020.9379836}
\BIBentrySTDinterwordspacing

\bibitem{Launay}
\BIBentryALTinterwordspacing
F.~Launay, A.~Ohya, and S.~Yuta, ``{A corridors lights based navigation system including path definition using a topologically corrected map for indoor mobile robots},'' in \emph{Proceedings 2002 IEEE International Conference on Robotics and Automation (Cat. No.02CH37292)}, ser. ROBOT-02, vol.~4.\hskip 1em plus 0.5em minus 0.4em\relax IEEE, p. 3918–3923. [Online]. Available: \url{http://dx.doi.org/10.1109/ROBOT.2002.1014338}
\BIBentrySTDinterwordspacing

\bibitem{Wang2018}
\BIBentryALTinterwordspacing
Q.~Wang, X.~Wang, L.~Ye, A.~Men, F.~Zhao, H.~Luo, and Y.~Huang, ``{A Multimode Fusion Visible Light Localization Algorithm using Ambient Lights},'' in \emph{2018 Ubiquitous Positioning, Indoor Navigation and Location-Based Services (UPINLBS)}.\hskip 1em plus 0.5em minus 0.4em\relax IEEE, Mar. 2018, p. 1–9. [Online]. Available: \url{http://dx.doi.org/10.1109/UPINLBS.2018.8559842}
\BIBentrySTDinterwordspacing

\bibitem{Rahman20200}
\BIBentryALTinterwordspacing
M.~H. Rahman, M.~A.~S. Sejan, J.-J. Kim, and W.-Y. Chung, ``{Reduced Tilting Effect of Smartphone CMOS Image Sensor in Visible Light Indoor Positioning},'' \emph{Electronics}, vol.~9, no.~10, p. 1635, Oct. 2020. [Online]. Available: \url{http://dx.doi.org/10.3390/electronics9101635}
\BIBentrySTDinterwordspacing

\bibitem{Moreira1997}
\BIBentryALTinterwordspacing
A.~J. Moreira, R.~T. Valadas, and A.~de~Oliveira~Duarte, ``{Optical interference produced by artificial light},'' \emph{Wireless Networks}, vol.~3, no.~2, p. 131–140, 1997. [Online]. Available: \url{http://dx.doi.org/10.1023/A:1019140814049}
\BIBentrySTDinterwordspacing

\bibitem{Yang2024}
\BIBentryALTinterwordspacing
X.~Yang, Y.~Zhuang, M.~Shi, X.~Sun, X.~Cao, and B.~Zhou, ``{RatioVLP: Ambient Light Noise Evaluation and Suppression in the Visible Light Positioning System},'' \emph{IEEE Transactions on Mobile Computing}, vol.~23, no.~5, p. 5755–5769, May 2024. [Online]. Available: \url{http://dx.doi.org/10.1109/TMC.2023.3312550}
\BIBentrySTDinterwordspacing

\bibitem{deCock2022}
\BIBentryALTinterwordspacing
C.~de~Cock, S.~Coene, B.~van Herbruggen, L.~Martens, W.~Joseph, and D.~Plets, ``{IMU-aided detection and mitigation of Human Body Shadowing for UWB positioning},'' in \emph{2022 IEEE 12th International Conference on Indoor Positioning and Indoor Navigation (IPIN)}.\hskip 1em plus 0.5em minus 0.4em\relax IEEE, Sep. 2022, p. 1–8. [Online]. Available: \url{http://dx.doi.org/10.1109/IPIN54987.2022.9918139}
\BIBentrySTDinterwordspacing

\bibitem{DeCock2023}
\BIBentryALTinterwordspacing
C.~De~Cock, E.~Tanghe, W.~Joseph, and D.~Plets, ``{Robust IMU-Based Mitigation of Human Body Shadowing in UWB Indoor Positioning},'' \emph{Sensors}, vol.~23, no.~19, p. 8289, Oct. 2023. [Online]. Available: \url{http://dx.doi.org/10.3390/s23198289}
\BIBentrySTDinterwordspacing

\bibitem{Cock2021}
\BIBentryALTinterwordspacing
C.~D. Cock, W.~Joseph, L.~Martens, and D.~Plets, ``{Floor Number Detection for Smartphone-based Pedestrian Dead Reckoning Applications},'' in \emph{2021 International Conference on Indoor Positioning and Indoor Navigation (IPIN)}.\hskip 1em plus 0.5em minus 0.4em\relax IEEE, Nov. 2021, p. 1–6. [Online]. Available: \url{http://dx.doi.org/10.1109/IPIN51156.2021.9662470}
\BIBentrySTDinterwordspacing

\bibitem{boqi_pwm_dimming}
\BIBentryALTinterwordspacing
{BOQI LED}, ``{Exploring PWM Dimming in LED Drivers: A Comprehensive Guide},'' 2023, accessed: 2025-04-04. [Online]. Available: \url{https://boqiled.com/exploring-pwm-dimming-in-led-drivers-a-comprehensive-guide/}
\BIBentrySTDinterwordspacing

\bibitem{Kamath2022}
\BIBentryALTinterwordspacing
U.~Kamath, K.~L. Graham, and W.~Emara, \emph{{Transformers for Machine Learning: A Deep Dive}}.\hskip 1em plus 0.5em minus 0.4em\relax Chapman and Hall/CRC, Apr. 2022. [Online]. Available: \url{http://dx.doi.org/10.1201/9781003170082}
\BIBentrySTDinterwordspacing

\bibitem{Zhang2025}
\BIBentryALTinterwordspacing
L.~Zhang, Y.~Chen, X.~Gao, J.~Yang, and C.~Zhang, ``{Indoor visible light fingerprint positioning based on dynamic weighted RSSI and LGSHHO-LSSVM},'' \emph{Optics Communications}, vol. 583, p. 131778, Jun. 2025. [Online]. Available: \url{http://dx.doi.org/10.1016/j.optcom.2025.131778}
\BIBentrySTDinterwordspacing

\bibitem{Aboagye2023}
\BIBentryALTinterwordspacing
S.~Aboagye, A.~R. Ndjiongue, T.~M.~N. Ngatched, O.~A. Dobre, and H.~V. Poor, ``{RIS-Assisted Visible Light Communication Systems: A Tutorial},'' \emph{IEEE Communications Surveys \& Tutorials}, vol.~25, no.~1, p. 251–288, 2023. [Online]. Available: \url{http://dx.doi.org/10.1109/COMST.2022.3225859}
\BIBentrySTDinterwordspacing

\bibitem{Abdelhady2021}
\BIBentryALTinterwordspacing
A.~M. Abdelhady, A.~K.~S. Salem, O.~Amin, B.~Shihada, and M.-S. Alouini, ``{Visible Light Communications via Intelligent Reflecting Surfaces: Metasurfaces vs Mirror Arrays},'' \emph{IEEE Open Journal of the Communications Society}, vol.~2, p. 1–20, 2021. [Online]. Available: \url{http://dx.doi.org/10.1109/OJCOMS.2020.3041930}
\BIBentrySTDinterwordspacing

\bibitem{Sun20222}
\BIBentryALTinterwordspacing
S.~Sun, T.~Wang, F.~Yang, J.~Song, and Z.~Han, ``{Intelligent Reflecting Surface-Aided Visible Light Communications: Potentials and Challenges},'' \emph{IEEE Vehicular Technology Magazine}, vol.~17, no.~1, p. 47–56, Mar. 2022. [Online]. Available: \url{http://dx.doi.org/10.1109/MVT.2021.3127869}
\BIBentrySTDinterwordspacing

\bibitem{Aboagye2022}
\BIBentryALTinterwordspacing
S.~Aboagye, A.~R. Ndjiongue, T.~M.~N. Ngatched, and O.~A. Dobre, ``{Design and Optimization of Liquid Crystal RIS-Based Visible Light Communication Receivers},'' \emph{IEEE Photonics Journal}, vol.~14, no.~6, p. 1–7, Dec. 2022. [Online]. Available: \url{http://dx.doi.org/10.1109/JPHOT.2022.3211730}
\BIBentrySTDinterwordspacing

\end{thebibliography}
\begin{IEEEbiography}
  [{\includegraphics[width=1in,height=1.25in,clip,keepaspectratio]{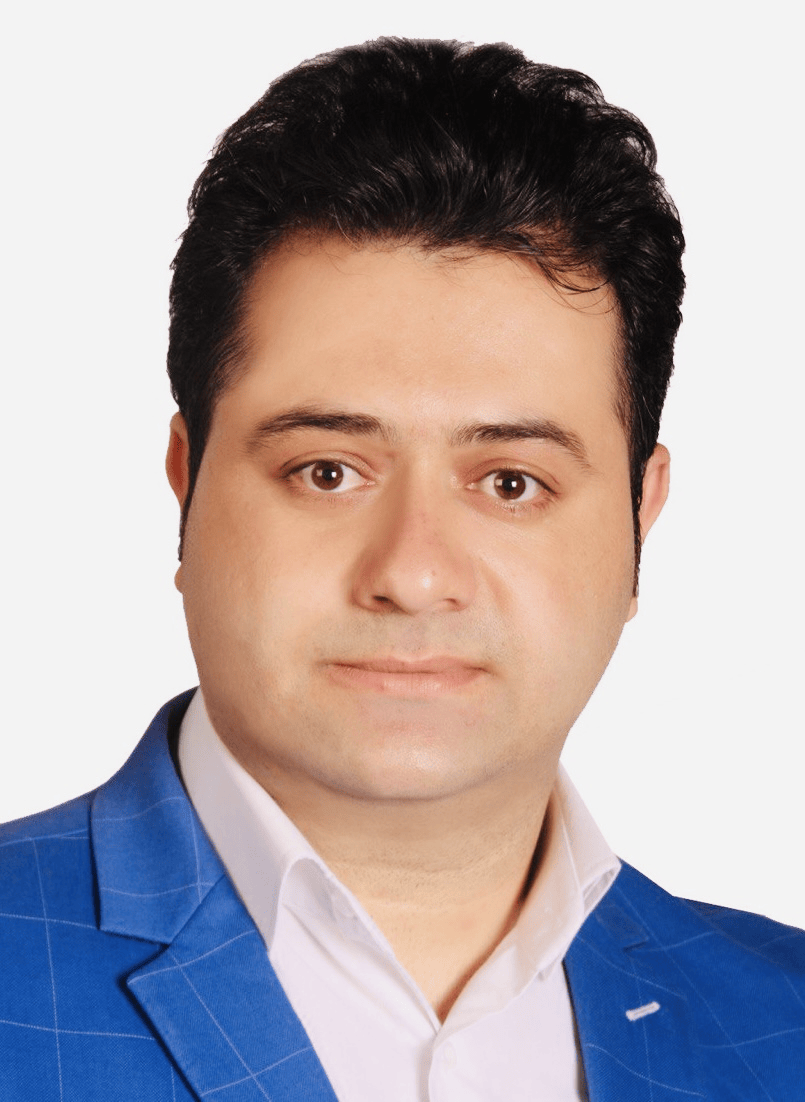}}]{Morteza Alijani} received his first M.Sc. degree in Electrical Engineering with a specialization in Communications in 2014. He obtained his second M.Sc. degree in Mechatronics Engineering, with a focus on robotics and electronics, from the University of Trento, Italy, in 2022. From December 2021 to September 2022, he was a Research Assistant at Centro Ricerche Fiat, Trento, Italy. In September 2022, he joined the imec-WAVES Research Group, Department of Information Technology, Ghent University, Belgium, where he is currently pursuing the Ph.D. degree in Electrical Engineering. His research interests include visible light positioning and sensing, as well as Internet of Things (IoT)-based localization and sensing applications in both indoor and outdoor environments.
  \end{IEEEbiography}

\begin{IEEEbiography}[{\includegraphics[width=1in,height=1.25in,clip,keepaspectratio]{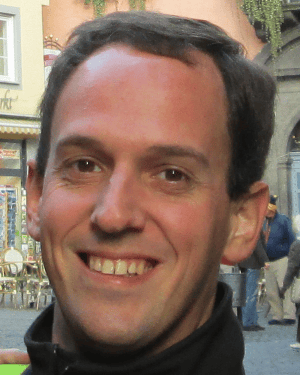}}]{Wout Joseph}
(Senior Member, IEEE) was born in Ostend, Belgium, in 21 October 1977. He received the M.Sc. degree in electrical engineering from Ghent University, Belgium, in July 2000, and the Ph.D. degree in March 2005; this work dealt with measuring and modeling of electromagnetic fields around base stations for mobile communications related to the health effects of the exposure to electromagnetic radiation. He was a Postdoctoral Fellow with the FWO-V (Research Foundation—Flanders) from 2007 to 2012. Since October 2009, he has been a Professor in the domain of “Experimental Characterization of Wireless Communication Systems” with Ghent University. He has been a IMEC PI since 2017. His professional interests are electromagnetic field exposure assessment, propagation for wireless communication systems, antennas, and calibration. Furthermore, he specializes in wireless performance analysis and quality of experience.
\end{IEEEbiography}

\begin{IEEEbiography}
  [{\includegraphics[width=1in,height=0.125\textheight,clip]{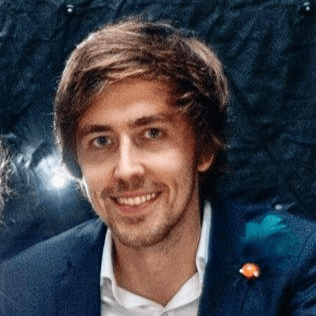}}]{David Plets}
  (Member, IEEE) has been a member of the imec-WAVES Group, Department of Information Technology, Ghent University, since 2006, where he has also been a Professor since 2016. His current research interests include localization techniques and the IoT, for both industry- and health-related applications. He is also involved in the optimization of wireless communication and broadcast networks, with a focus on coverage, exposure, and interference.
\end{IEEEbiography}
\vfill
\end{document}